# *In silico* Modeling of Itk Activation Kinetics in Thymocytes Suggests Competing Positive and Negative IP$_4$ Mediated Feedbacks Increase Robustness

Sayak Mukherjee[1], Stephanie Rigaud[7], Sang-Cheol Seok[1], Guo Fu[7], Agnieszka Prochenka[1,8], Michael Dworkin[1,5], Nicholas R. J. Gascoigne[7], Veronica J. Vieland[1,2,4], Karsten Sauer[7*] and Jayajit Das[1,2,3,6*]

[1]Battelle Center for Mathematical Medicine, The Research Institute at the Nationwide Children's Hospital and Departments of [2]Pediatrics, [3]Physics, [4]Statistics, [5]Mathematics and [6]Biophysics Graduate Program, The Ohio State University, 700 Children's Drive, Columbus, OH 43205. [7] Department of Immunology and Microbial Science, The Scripps Research Institute, La Jolla, CA 92037.[8]Institute of Computer Science, Polish Academy of Sciences, Warsaw, Poland.

One-sentence summary:

Competing positive and negative IP$_4$ feedbacks increase robustness of PI3K signaling via Itk in thymocytes against variations at the single cell level.

* Corresponding authors

**Abstract**

The inositol-phosphate messenger inositol(1,3,4,5)tetrakisphosphate ($IP_4$) is essential for thymocyte positive selection by regulating plasma-membrane association of the protein tyrosine kinase Itk downstream of the T cell receptor (TCR). $IP_4$ can act as a soluble analog of the phosphoinositide 3-kinase (PI3K) membrane lipid product phosphatidylinositol(3,4,5)trisphosphate ($PIP_3$). $PIP_3$ recruits signaling proteins such as Itk to cellular membranes by binding to PH and other domains. In thymocytes, low-dose $IP_4$ binding to the Itk PH domain surprisingly promoted and high-dose $IP_4$ inhibited $PIP_3$ binding of Itk PH domains. However, the mechanisms that underlie the regulation of membrane recruitment of Itk by $IP_4$ and $PIP_3$ remain unclear. The distinct Itk PH domain ability to oligomerize is consistent with a cooperative-allosteric mode of $IP_4$ action. However, other possibilities cannot be ruled out due to difficulties in quantitatively measuring the interactions between Itk, $IP_4$ and $PIP_3$, and in generating non-oligomerizing Itk PH domain mutants. This has hindered a full mechanistic understanding of how $IP_4$ controls Itk function. By combining experimentally measured kinetics of PLCγ1 phosphorylation by Itk with *in silico* modeling of multiple Itk signaling circuits and a maximum entropy (MaxEnt) based computational approach, we show that those *in silico* models which are most robust against variations of protein and lipid expression levels and kinetic rates at the single cell level share a cooperative-allosteric mode of Itk regulation by $IP_4$ involving oligomeric Itk PH domains at the plasma membrane. This identifies MaxEnt as an excellent tool for quantifying robustness for complex TCR signaling circuits and provides testable predictions to further elucidate a controversial mechanism of $PIP_3$ signaling.

**Author Summary:**

Inositol phosphate second messengers can regulate interactions between receptor signaling and lipid metabolic networks, critically affecting cell decision processes. However, the molecular mechanisms underlying such cross-regulation are poorly understood. Pairing mathematical modeling and experiments, we elucidate these mechanisms in the activation of T cells developing in the thymus (thymocytes), which is carefully controlled by TCR induced production of the membrane lipid $PIP_3$, soluble $IP_4$, and activation of the kinase Itk. T cells are major orchestrators of our adaptive immune system. Their development in the thymus is critically monitored to produce functional but self-tolerant T cells in the periphery. We combine experimentally measured kinetics of Itk mediated PLCγ1 phosphorylation with Maximum Entropy (MaxEnt) based computational simulations into a novel approach to analyze the robustness of Itk activation kinetics. Our results provide testable predictions to further elucidate a controversial mechanism of $PIP_3$ signaling.

**Introduction**

Hydrolysis of plasma membrane phospholipids generates various cellular messengers [1]. Among these, ,multiple isomeric inositol phosphates (IP) [1-4] form an "IP code"[5] whose members can regulate critical decision processes downstream of many receptors in diverse cell types. However, the specific mechanisms and precise molecular circuitries that underlie the regulation of cell functions by soluble IPs are poorly understood. We and others previously reported an essential role for inositol(1,3,4,5) tetrakisphosphate ($IP_4$) in regulating T cell development [2,3,6,7].

T cells are key mediators of adaptive immune responses. Through a plasma-membrane anchored TCR, they recognize pathogen-derived peptides bound to Major Histocompatibility Complex proteins (pMHC) on the surface of antigen-presenting cells. TCR engagement triggers activation, proliferation and effector functions in peripheral T cells that then kill pathogen-infected cells and control immune responses. During T cell development in the thymus, somatic mutation of the antigen-binding TCR α/β subunit genes creates a thymocyte repertoire with random TCR specificities. However, many of these TCRs are non-functional or interact with the body's self-antigens with high affinity, causing autoimmune disorders if the respective T cells were allowed to mature. To prevent this, thymic selection processes eliminate thymocytes carrying TCRs that fail to interact with, or interact with too strong affinity with self-peptide-MHC (pMHC) complexes. The latter process is known as negative selection, a key mechanism of central tolerance. Only those thymocytes whose TCR generates mild signals are positively selected to mature into T cells, which then populate peripheral organs. Balanced positive and negative selection are critical for generating a diverse but self-tolerant T cell repertoire [8-10]. Recent experiments provided a more complex picture of thymic selection, where certain high affinity peptides can 'agonist select' distinct regulatory T cell types [11,12].

TCR-pMHC binding triggers a series of signaling reactions, resulting in the formation of a plasma membrane-proximal signalosome containing Src (Lck, Fyn) and Syk family protein tyrosine kinases (Zap70), cytosolic (such as SLP-76, Gads, Grb-2), and transmembrane adapter proteins (such as LAT). TCR-activation of phosphoinositide 3-kinase (PI3K) converts the abundant membrane phospholipid phosphatidylinositol(4,5) bisphosphate ($PIP_2$) into phosphatidylinositol(3,4,5) trisphosphate ($PIP_3$). By binding to pleckstrin homology (PH) or other protein-domains, $PIP_3$ recruits key effectors such as the Tec family protein tyrosine kinase Itk (IL-2 inducing T cell activation kinase). Itk also contains SH2 and SH3 domains that bind to signalosome components. The Src kinase Lck phosphorylates $Y_{511}$ in the A-loop of the murine ($Y_{512}$ in the human) Itk kinase domain [13]. Subsequently, Itk propagates TCR signals by phosphorylating and activating signalosome-recruited phospholipase Cγ1 (PLCγ1). PLCγ1 then hydrolyzes $PIP_2$ into the second messenger molecules diacylglycerol (DAG) and inositol(1,4,5) trisphosphate ($IP_3$). The membrane lipid DAG further recruits and activates Rasgrp1 and PKCs that in turn activate the GTPase Ras and the Bcl-10/CARMA1/MALT complex, ultimately triggering thymocyte positive and negative selection, or peripheral T cell responses [14,15]. Soluble $IP_3$ mobilizes $Ca^{2+}$ from the endoplasmic reticulum (ER). Moreover, $IP_3$ 3-kinases such as ItpkB can phosphorylate $IP_3$ at its 3-position into $IP_4$ [2,6,7,14,16]. $IP_4$ chemically resembles the PH domain binding $PIP_3$ tetraphoinositol headgroup [14,17].

We and others identified ItpkB as essential for thymocyte positive selection [2,6,7]. *ItpkB$^{-/-}$* DP thymocytes show intact proximal TCR signaling but defective $IP_4$ production, Itk $PIP_3$-binding, signalosome recruitment and activation with ensuing reduced PLCγ1 activation, DAG production, and, Ras/Erk activation [2]. The ability of soluble $IP_4$ to bind to the Itk PH domain and in low μM doses promote $PIP_3$ binding, and the ability of the Itk PH domain to oligomerize suggested that $IP_4$ might promote Itk recruitment to membrane-$PIP_3$ through a cooperative-allosteric mechanism. In this model, $IP_4$-binding to one PH domain in an oligomer allosterically increases the ligand affinities of the other PH domains in the same oligomer [2]. $IP_4$ promoted Itk activation appears to be required for sufficient Itk activation to ensure positive selection, because an exogenous DAG-analog restored positive selection of *ItpkB$^{-/-}$* thymocytes [2]. However, high-dose $IP_4$ inhibited Itk PH domain binding to $PIP_3$ *in vitro* [2]. Whether it does so *in vivo* is unknown [14]. In neutrophils, NK cells and myeloid progenitors, $IP_4$ competitively limits Akt PH domain binding to membrane $PIP_3$[18-20]. Which PH domains are positively versus negatively controlled by $IP_4$, and what determines whether $IP_4$ promotes or inhibits PH domain binding to $PIP_3$ or leaves it unaffected are important open questions [14,21]. In particular, the Itk PH domain might be bi-modally regulated by $IP_4$. However, the detailed molecular interactions between Itk, $PIP_3$ and $IP_4$ *in vivo* are not well characterized. This leaves room for multiple alternate hypotheses/mechanisms. For example, one could also propose that the binding affinity of $PIP_3$ and $IP_4$ for Itk changes from a low to a fixed high value above a threshold $IP_4$ concentration. Such a mechanism implies that the interaction of Itk with $IP_4$ and $PIP_3$ after the threshold $IP_4$ concentration is reached does not involve a positive feedback. The situation is further confounded by elusive results from experiments probing Itk oligomerization [2,22-28].

The current lack of a mechanistic understanding of how $IP_4$ controls Itk $PIP_3$-interactions and whether Itk PH domain oligomerization is physiologically relevant arises from difficulties in quantitatively measuring the interactions between Itk, $IP_4$ and $PIP_3$, and in generating soluble Itk PH domain preparations for biophysical studies and non-oligomerizing Itk PH domain mutants for genetic analyses. Additional limitations arise from difficulties in measuring membrane recruitment of Itk in cell population based assays. It is also difficult to measure $PIP_3$ bound Itk or phosphorylation of PLC-γ1, a substrate of $PIP_3$ bound Itk, in large numbers of individual cells using flow cytometry techniques due to limited antibody quality. *In vitro* and cell-based studies based on ectopic Itk expression suggest the existence of several different monomeric and oligomeric Itk species, including head-to-head and head-to-tail dimers [2,22-28]. Andreotti and colleagues[22] showed that Itk molecules can self associate via their SH2-SH3 domains into auto-inhibitory oligomers. This is hindered by SLP-76 interactions with the Itk SH2-SH3 domains. It was suggested that Itk molecules might exist as auto-inhibited multimers in the cytosol, but after plasma membrane recruitment, Itk monomers might mediate downstream activation [22,26]. Other experiments [27,28] employing fluorescence complementation showed that formation of Itk head-to-head and head-to-tail dimers requires the PH domain and may primarily occur at the plasma membrane, although low-abundance cytoplasmic dimers have not been excluded. Here, monomeric Itk was proposed to be primarily cytoplasmic and autoinhibited [27]. At least head-to-head dimerization is unaffected by mutations in the other (SH2/SH3) domains [28]. We found that the Itk PH domain can oligomerize with other Itk PH domains or full length Itk [2]. Thus, the PH domain is well suited to contribute to at least certain modes of Itk oligomerization, some of which could have positive or a combination of positive and negative functions regulated by $IP_4$/$PIP_3$. This could account for the limited activity-enhancing effect of disrupting SH3/SH2-domain mediated Itk dimerization [26].

Altogether, whether Itk PH domain dimerization has a physiological function, whether it promotes or inhibits Itk activation, whether $IP_4$ controls Itk function through positive or negative feedback, or both, and whether $IP_4$ has additional unrelated functions in thymocytes, are all contentious questions in the field. Resolving them is very important, because PI3K is a paramount regulator of signaling from many receptors in most cells. $PIP_3$ hyperactivity is a major contributor to immune, metabolic and other diseases including cancers [29,30]. $IP_3$ 3-kinases are broadly expressed and $IP_4$ has been found in many cell types. Thus, $IP_4$ regulation of $PIP_3$ function could be broadly important and elucidating the precise molecular mechanisms through which $IP_4$ controls $PIP_3$ signaling improves our understanding of a very fundamental and important signaling pathway with great therapeutic relevance [14].

To further explore how the presence or absence of Itk PH domain oligomerization, of positive or negative $IP_4$ feedback or both, or of specific molecular modes of association of Itk, $PIP_3$ and $IP_4$ impact TCR signaling, we constructed seven different molecular models (Table I and Fig. S1B). We used a Maximum Entropy (MaxEnt) [31-33] based approach to quantify the robustness of each model against variations in rate constants and protein expression levels at the single cell level. Each model was constrained to reproduce the Itk activation kinetics of an entire cell population measured in biochemical experiments. We found that those models involving dimeric Itk molecules with $IP_4$ mediated competing positive and negative feedbacks are most robust. As in many other cell signaling systems [34], the actual signaling kinetics in thymocytes are likely to be robust against such variations, while retaining their sensitivity to small variations in antigen affinity or dose. On this basis, our simulations best support biphasic Itk regulation by $IP_4$ in thymocytes. Future testing of this exciting hypothesis will require the so far unsuccessful generation of non-oligomerizing Itk PH domain mutants and their expression in $Itk^{-/-}$ mice, along with currently impossible single-cell measurements of $IP_4$ levels in large cell populations.

**RESULTS**

**Multiple molecular models can be constructed to probe Itk, $IP_4$, and $PIP_3$ interactions *in silico***

We constructed seven different molecular models (Table I, Fig. S1B) based on available details about interactions between Itk, $PIP_3$ and $IP_4$ from the biochemical studies described above. Including Itk kinase domain activation by Lck only caused qualitative changes in the relative robustness of the models (Fig. S17, Tables S9-S15). Therefore, for simplicity, we considered models that do not contain Itk activation by Lck explicitly. We also did not consider Itk autophosphorylation explicitly in the models as it does not affect Itk catalytic activity. In addition, the role of Itk autophosphorylation in PLCγ1 activation remains unclear [22]. Since we aimed to elucidate general characteristics of the kinetics of $PIP_3$ binding to Itk, we used a simplified modeling scheme (Fig. 1) and did not consider the detailed molecular composition of the TCR and the LAT associated signalsome. The models also do not investigate different mechanisms for formation of Itk oligomers. Rather, they probe the functional consequences of having Itk PH domain dimers versus monomers and how these can affect interactions between Itk, $PIP_3$ and $IP_4$ in the presence or absence of $IP_4$ mediated positive feedback. The kinetics of $PIP_3$ production due to signal-dependent recruitment of PI3K are not considered explicitly as $PIP_3$ is produced at a much faster time scale (in seconds, [35][36][37]) than the time scales of PLCγ1 activation (up to 60 min, Figs. 3, S18) analyzed in this study. The concentrations of LAT bound Itk and of $PIP_3$ were considered approximate markers for the strength of the stimulation generated by antigen-TCR interactions. Therefore, we considered fixed initial concentrations of Itk and $PIP_3$ in the models. We approximated the production of $IP_4$ from $PIP_2$ by a single one-step reaction to simplify the models further .

The models can be broadly classified into two types: (i) Models M1-M4 and M7 containing $IP_4$ mediated positive feedbacks. (ii) Models M5 and M6 lacking $IP_4$ positive feedback. In each type, we further considered models that contained Itk dimers (models M1-M3, M5, M7), or monomers (models M4, M6). In models M1-M3, each of the two PH domains in the Itk dimer can independently bind to either $IP_4$ or $PIP_3$ with a weak affinity when the other PH domain is unoccupied. However, once a PH domain is bound to an $IP_4$ molecule, it allosterically increases the affinity of the other PH domain for $PIP_3$ and $IP_4$. Models M1-M3 differ from each other in the relative increase in the affinities of one PH domain in the Itk dimer toward $IP_4$ vs. $PIP_3$ caused by $IP_4$ or $PIP_3$ binding to the other PH domain in the dimer. In contrast, in M7, binding of $PIP_3$ to one PH domain in a dimer increases the affinity of the other PH domain for $PIP_3$ but not for $IP_4$. These models probed potential secondary interactions between Itk dimers and the membrane lipids. In the monomeric model, M4, $IP_4$ binds the single Itk PH domain with a weak affinity and induces a conformational change that increases the affinity of this PH domain for both $PIP_3$ and $IP_4$. Models M5 and M6 lack positive $IP_4$ feedback. Instead, the Itk PH domain binds to $IP_4$ and $PIP_3$ with equal affinity. These models probed a mechanism where the Itk PH domain interacts with $IP_4$ and $PIP_3$ once a small threshold $IP_4$ concentration is generated. We assumed that the small threshold level of $IP_4$ is generated at a time scale much smaller than the timescale (min) of robust Itk activation and did not consider the kinetics generating the threshold level of $IP_4$ explicitly in M5 and M6. The models are summarized in Table I, Fig. S1B, and tables S1-S8.

**The shape of transient Itk activation kinetics depends on specific molecular wirings and feedbacks in the different models**

We studied the kinetics of Itk binding to $PIP_3$ using deterministic mass-action kinetic rate equations described by ordinary differential equations (ODE) for all the models, ignoring stochastic fluctuations in the copy numbers of signaling proteins occurring due to the intrinsic random nature of biochemical reactions [38]. Including such fluctuations did not change the kinetics qualitatively (Figs. S2-S3). In all seven models, the kinetics of $PIP_3$ bound Itk showed a transient behavior (Fig. 2A); $PIP_3$ bound Itk started with a low concentration, reached a peak value at an intermediate time, and then fell back to a small concentration at later times. We found that initially few Itk molecules were bound to $PIP_3$. With increasing time, more Itk molecules became associated with $PIP_3$ molecules due to the binding reactions between Itk and $PIP_3$. This produced the rise in the Itk-$PIP_3$ concentration. However, as the concentrations of $PIP_3$ bound Itk molecules increased, they also induced increased production of $IP_4$ molecules. $IP_4$ competed with $PIP_3$ for binding to the Itk PH domain, and when the number of $IP_4$ molecules exceeded that of $PIP_3$ molecules, most of the Itk molecules were sequestered to the cytosol by forming stable complexes only with $IP_4$. This reduced the rate of $PIP_3$ association of Itk and eventually resulted in the decrease of the $PIP_3$ bound Itk molecules. $IP_4$ outnumbered $PIP_3$ at later times because the number of $PIP_2$ molecules, the source of $IP_3$ and $IP_4$ in a cell, is considered not limiting in contrast to $PIP_3$ [37,39]. We emphasize that the results of our models do not depend on the cytosolic nature of Itk-$IP_4$ complexes, but on the model assumption that Itk (or Itk oligomers) bound to $IP_4$ at every PH domain does not induce any PLCγ1 activation.

We characterized the 'shape' of the temporal profile of $PIP_3$ bound Itk in terms of (i) the largest concentration of $PIP_3$ bound Itk in the entire temporal profile (peak amplitude value $A$); (ii) the time taken for $PIP_3$ bound Itk to reach the peak value (peak time $\tau_p$); and (iii) the time interval during which the $PIP_3$ bound Itk concentration is greater than or equal to half of the peak value (peak duration $\tau_w$, Fig. 2B). A dimensionless variable quantifying the asymmetry in the shape of the kinetics, denoted as the asymmetry ratio R = $\tau_w/\tau_p$ (Fig. 2B), turned out to be a useful indicator for differentiating temporal profiles of concentrations of $PIP_3$ bound Itk in simulations and experiments. R also quantifies if the time scale for the decay of the concentration of $PIP_3$ bound Itk after the peak value is reached is larger than or comparable to $\tau_p$ (the timescale for producing the peak value $A$). E.g., when $R \approx 1$, it implies that the $\tau_p$ is comparable to decay time. $R \gg 1$ indicates a more persistent signal with long decay times. Differences (transient vs. persistent) in the shapes of kinetic profiles of signals downstream of Itk activation have been observed to influence thymocyte decision outcomes [2,40]. Therefore, R, which characterizes the persistent or transient nature of Itk activation, also contains details directly relevant for thymic selection outcomes. We found that the shape of the transient kinetics of $PIP_3$ binding to Itk varied substantially depending on the feedbacks and the molecular wiring of the networks. Since the reaction rates used in the models are difficult to measure *in vivo* for thymocytes, we estimated the rates based on interaction strengths measured *in vitro* between PH domains and inositol phosphates in other cells, and from temporal profiles of PLCγ1 activation measured in experiments with T cells (tables S1-S7). Previous work demonstrated the essential role of phosphorylated PLCγ1 and its kinetics in regulating thymocyte positive, negative and agonist selection [41,42]. Phospho-PLCγ1 is also known to mirror other indicators of T cell activation such as TCRζ- or LAT-phosphorylation, or Erk-activation [2,40]. Therefore phospho-PLCγ1 is a relevant marker for functional T cell responses.

We studied variations in the kinetics of $PIP_3$ bound Itk for different initial concentrations of Itk and $PIP_3$. This probed how different ligand doses or affinities affected the $PIP_3$ binding of Itk. We found that the peak concentration of $PIP_3$ bound Itk increased in a graded manner with increasing initial Itk and $PIP_3$ concentrations in all models (Fig. S4). However, the peak time $\tau_p$ (Fig. S5), and the asymmetry ratio R (Fig. 2C), were affected differently in different models. Among the feedback models, M1-M3 and M7 containing Itk dimers generated smaller values (varied between 2 to 6) of R compared to monomeric model M4 which produced a much larger range of R (~20 -120) (Fig. 2C). The models lacking positive feedbacks (M5 and M6) generated large values (~100-700) of *R* compared to feedback models with Itk dimers (Fig. 2C). In the feedback models, the initial low affinity binding-unbinding interactions between Itk and $PIP_3$/$IP_4$ are converted into high affinity interactions due to the positive feedback. Therefore, a large part of $\tau_p$ is spent in building up the positive feedback interactions controlled primarily by the weak affinity binding-unbinding rates (or $K_D$). The small values of R in models M1-M3 and M7 occurred because stronger negative feedbacks resulted in much smaller timescales for substantially reducing the concentration of $PIP_3$ bound Itk after it reached its peak value compared to the other models. In the models lacking positive feedback (M5, M6), concentrations of $PIP_3$ bound Itk decreased at a much slower rate than the peak time leading to large values of R. In the monomeric model, the relatively weaker strength of positive and negative feedbacks resulted in larger decay time scales for the $PIP_3$ bound Itk, producing large values of R. These results are analyzed in detail in the web supplement and Figs. S6-S11. We will show below how the ability of feedback models with Itk dimers to produce R values within a small range leads to higher robustness of these models against parameter variations at the single cell level.

## Models containing dimeric Itk and $IP_4$ mediated dueling positive and negative feedbacks are the most robust models

*Quantification of robustness in* in silico *models:* The reaction rates describing non-covalent primary and secondary interactions between Itk, $PIP_3$ and $IP_4$ can depend on specific properties of the local cellular environment, such as local membrane curvature [43], molecular crowding [44,45], and the presence of different lipid molecules in the proximity [46]. Since these factors can vary from cell to cell, the reaction rates can vary at the single cell level. In addition, protein expression levels can vary between cells. Such variations are also known as extrinsic noise fluctuations [47,48]. The $IP_4$ production rate depends on the concentrations of ItpkB, Calmodulin (CaM), and released calcium [3]. Hence, the $IP_4$ production rates in our models which approximate all such dependencies with a one-step reaction will vary between individual cells as well. The above variations are capable of producing differences in the shapes of temporal profiles of activation of signaling proteins in individual cells [49]. In the coarse-grained or approximate models we have constructed, many molecular details have been approximated. For example, multiple phosphorylation sites or SH2/SH3 binding sites of Itk, LAT, SLP-76 and their regulation via TCR induced signaling are not considered explicitly [3,22,50,51]. These detailed molecular signaling events can depend on the concentrations of proteins, enzymes, and lipids, and can thus be regulated differently in different cells due to extrinsic noise fluctuations. Consequently, the rates in our *in silico* models that effectively describe those detailed signaling events can vary from cell to cell. Consistent with this view, our simulations with the ODE models showed that the shape of the kinetics of PIP$_3$ bound Itk, characterized by, A, $\tau_p$, and R, changed significantly as the rate constants and initial concentrations in a model were varied (Figs. S12-S14). Thus, activation kinetics of a marker molecule (e.g. PLCγ1) measured in experiments (e.g., immunoblots) assaying a large cell population represent averages over a range of temporal profiles with different shapes occurring at the single cell level.

We found that for some ranges of the reaction rates, multiple different *in silico* models can produce the same values of A, $\tau_p$, and R (Fig. S12-S14). This implied that more than one *in silico* model *could* reproduce the mean temporal profile measured in cell population assays. However, it is possible that each model could show a different degree of robustness to variations in reaction rates and initial concentrations at the single cell level. Robustness of time dependent responses in a cell population against variations at the single cell levels has been observed in several systems, e.g., oscillations in adenosine 3',5' cyclic monophosphate (cAMP) concentrations in a population of *Dyctostelium* [52,53], or damped oscillations of protein 53 (p53) in a population of human breast cancer epithelial cells [54]. Robustness of cellular functions against variations in external conditions and cell-to-cell variability has been proposed as a required design principle for a wide range of biochemical networks [55-58]. We therefore decided to ask: Which model(s) can accommodate the largest variation in reaction rates and initial concentrations, while reproducing the mean temporal profile of PIP$_3$ bound Itk measured as generation of phosphorylated PLCγ1 in cell population experiments? We postulate that the answer to this question will point us to the molecular circuitry most likely to be the relevant model, in the sense that it robustly produces a specific temporal response at the cell population level despite variations in the kinetics in individual cells.

To identify the most robust model(s), we quantified robustness using a method based on the principle of Maximum Entropy (MaxEnt) [31-33]. MaxEnt provides a mechanism for estimating the probability distribution of the rate constants and initial Itk and PIP$_3$ concentrations under constraints derived from experimental data (Fig. 3A-B, 4A-B). Here, we used the experimentally obtained values $\tau_p^{expt}$ and $R^{expt}$ as the constraints. It is difficult to directly relate the amplitude (in units of number of molecules in the simulation box) in the *in silico* models to experiments, where amplitudes are calculated from the fold change of the immunoblot intensities upon stimulation. Therefore, the experimental values of A can be related to the number of activated molecules, at best, through a proportionality constant dependent on specific protocols used in an assay. Because of these issues we chose a value of A$^{expt}$, representing A in experiments, where every *in silico* model produced amplitudes at A$^{expt}$ for a set of parameters within the range of variations considered here. We then varied A$^{expt}$ to investigate the change in robustness of the models and address the arbitrariness in the choice of A$^{expt}$. We constructed a relative entropy measure (Kullback-Leibler distance, D$_{KL}$, calculated on the log$_{10}$ scale) [59] that measures the deviation of the constrained MaxEnt distribution from the unconstrained MaxEnt distribution, in which all values of the rate constants and initial concentrations are equally likely (uniform distribution). Thus D$_{KL}$ is being used as a measure of how "close" each model can get to one which is completely indifferent to the values of the rate constants, given the experimental constraints. We then compared D$_{KL}$ across our models in order to find the most robust model compatible with experimental results. Note that the minimum value of D$_{KL}$ is 0, with smaller values indicating greater robustness. We have also analyzed $D_{KL}$ for different models when $\tau_p^{expt}$ and $R^{expt}$ were constrained but the amplitude $A^{expt}$ was not constrained (Fig. S23). The results are qualitatively similar to that of the case when $\tau_p^{expt}$, $R^{expt}$, and $A^{expt}$ were constrained. This indicates that the robustness of the temporal shape of Itk membrane recruitment kinetics rather than the amplitude contributes substantially toward the increased robustness of the feedback models with Itk dimers.

*Experimental analysis of Itk activation kinetics in mouse thymocytes:* To determine which Itk activation profile predicted by models M1-M7 produces maximum robustness while reproducing experimental data, we analyzed Itk activation kinetics in mouse $CD4^+CD8^+$ double-positive (DP) thymocytes, the developmental stage where positive selection occurs [9,60]. To generate a homogeneous cell population in which every cell expresses the same TCR and in which the TCR has not been stimulated by endogenous ligands prior to *in vitro* stimulation, we used *OT1 TCR-transgenic, RAG2$^{-/-}$, MHCI(β2m)$^{-/-}$* mice. Their DP cells express exclusively the transgenic OT1 TCR, which recognizes the ovalbumin-derived peptide ligand OVA and recently identified endogenous peptide ligands presented by MHCI molecules [40,61]. In *MHCI$^{-/-}$* mice, no endogenous ligands are presented to OT1 TCR-transgenic T cells and their development is blocked at the DP stage due to impaired positive selection. *In vitro*, OT1 TCR transgenic DP cells can be stimulated with MHCI tetramers loaded with OVA peptide [40,61]. Due to its high affinity for the OT1 TCR, OVA stimulation generates strong TCR signals and induces DP cell deletion. A number of OVA-derived altered peptide ligands (APL) have been generated which carry single or multiple amino acid substitutions compared to OVA. In the peptide series OVA>Q4R7>Q4H7>G4, such substitutions progressively reduce OT1 TCR affinity and signaling capacity [40]. Consequently, OVA and Q4R7 cause *OT1 TCR-transgenic* DP cell negative selection, whereas Q4H7 and G4 trigger positive selection.

We used MHCI tetramers presenting either one of these four peptides to stimulate *RAG2$^{-/-}$MHCI$^{-/-}$ OT1 TCR-transgenic* DP cells for various time points. We analyzed PLCγ1 phosphorylation at $Y_{783}$, normalized to total PLCγ1 protein levels, as a measure for Itk activation [2] (Fig. 3A). Stimulation by all peptides induced fast PLCγ1 phosphorylation already at 1 min which peaked at 2 min and then decreased over the next 60 min to very low levels which, however, were still above background levels in unstimulated cells. The decrease was fastest between 2 and 5 min and then progressively slowed down. As expected, overall levels of PLCγ1 phosphorylation progressively decreased with decreasing peptide affinity/signaling capacity in the order OVA>Q4R7>Q4H7>G4. An asymmetric peak shape with an extended right flank was preserved across all signal intensities. We calculated the peak durations ($\tau_w$), peak times ($\tau_p$) and asymmetry ratios $R = \tau_w/\tau_p$ in Table II for stimulation with OVA, Q4R7, Q4H7 and G4, respectively. Consistent with preserved peak asymmetry, all ratios R were >1.

*Comparison between experiments and conclusions:* The phospho-PLCγ1 levels (representing active Itk) for different affinity peptides peaked at $\tau_p$=2 mins with R values from 1.9-4.3 (Table II). Therefore, we fixed $\tau_p^{expt}=\bar{\tau}=2$ mins (the bar indicates average over the cell population) for quantifying robustness in the *in silico* modeling. The low, medium and large initial Itk and $PIP_3$ concentrations represent stimulation by weak (G4), moderate (Q4R7, Q4H7) and high affinity (OVA) ligands, respectively. Analyzing $D_{KL}$ (Fig. 3C) showed that for large initial $PIP_3$ and Itk concentrations (representing OVA stimulation) the feedback models incorporating Itk PH domain dimers (M1-M3, M7) were substantially more robust (Smaller $D_{KL}$ values) than the models lacking feedbacks (M5, M6) for small values of R (<3). Monomeric feedback model M4 produced large $D_{KL}$ values (1.5-5). M5, M6 and M4 produced much larger ranges of R (Fig. S14) as the parameters were varied compared to the feedback models with Itk dimers where the values of $R$ were clustered around $R^{expt}$ ~2. This behavior contributed substantially to the increased robustness of the feedback models with dimers as these models could accommodate for larger ranges of parameter variations while being able to maintain the constraint imposed by $R^{expt}$. The relative robustness of the feedback versus feedback-free models showed similar qualitative trends for the other ligands, Q4R7, Q4H7, and G4 (Fig. S15-S16). This suggests that the models containing feedbacks and Itk dimers are substantially more robust than models with Itk monomers or lacking feedbacks.

*Evaluation of robustness in polyclonal thymocytes:* The molecular wiring of Itk, $PIP_3$ and $IP_4$ interactions is unlikely to depend on the clonal nature of the T cells. Thus, the feedback models with Itk dimers should also be more robust than the other models when used to describe the kinetics of PLCγ1 activation in polyclonal DP thymocytes expressing many different TCRs with different ligand specificities, stimulated by antibodies against the common TCR subunit CD3 alone or with co-ligation of the common coreceptor CD4. Stimulation of non TCR-transgenic $MHC^{-/-}$ DP cells with 1µg/ml or 5µg/ml of αCD3 or combined αCD3/αCD4 antibodies produced different $R^{expt}$ and $\tau_p^{expt}$ values than the OT1 system above (Fig. 4A-4B, Fig S18, table. S16). Calculation of the robustness constrained by $R^{expt}$, $\tau_p^{expt}$ and A showed that feedback models M1, M2, M3 and M7 are again substantially more robust than the other models (Fig. 4C-4F, Fig. S19, S20). Large variations of R in M4, M5 and M6 as parameters were varied again made these models substantially less robust than the feedback models with Itk dimers.

**DISCUSSION**

Here, we used *in silico* simulations combined with a novel Maximum Entropy (MaxEnt) based method and cell population averaged measurements of PLCγ1 activation kinetics to distinguish between multiple models constructed to elucidate different mechanisms of Itk activation in TCR signaling. Our analysis quantified the robustness of seven different models employing monomeric or dimeric Itk PH domains with or without positive and negative $IP_4$ feedback against variations of parameters (rates and concentrations) at the single cell level. MaxEnt has been widely used in diverse disciplines ranging from physics [62] via information theory [63] to biology [64-67] to estimate probability distributions of variables subject to constraints imposed by experimental data [33,65]. However, to our knowledge these methods have not been used for evaluating the robustness of dynamic models in cell signaling or gene regulatory systems. Using thymocyte positive selection as a physiologically important model process, our results show the usefulness of MaxEnt methods for such studies. We are currently working on extending the methods to include additional information from experiments (such as variances), and also evaluating their performance in comparison with closely related approaches such as Bayesian analysis [68].

Our simulations predict that the models containing $IP_4$ feedbacks and Itk dimers are most robust. This is consistent with our previously proposed model of cooperative-allosteric regulation of Itk-$PIP_3$ interactions via $IP_4$-binding to oligomeric Itk PH domains [2]. Thymocyte selection critically depends on TCR induced signals. Small differences in antigen peptide concentration or affinities for the same TCR can produce opposite (negative vs. positive) selection outcomes [40]. Thus, we consider it plausible that for a fixed antigen dose and affinity (or average initial concentrations of Itk and $PIP_3$ in our models), TCR signaling in thymocytes should be robust against cell-to-cell variations of protein/lipid concentrations, rate constants and local environment. But TCR signaling should retain sensitivity to small variations in antigen affinity or dose. A direct experimental validation of this assumption will require to test the probability distributions of $\tau_p$, R, and A in cell populations where PLCγ1 activation kinetics are measured in individual cells. However, we were unable to perform such single cell comparisons due to the insensitivity of FACS-based PLCγ1 signaling assays. This indicates the importance of studying the effects of network architecture, rate constants, protein and lipid concentrations on system robustness in DP thymocyte selection in detail in the future. Thymocytes are an excellent *in vivo* model to probe the exquisite dependency of cell fate decisions on the affinity of TCR ligands with important physiological and pathological implications. This provides a valuable addition to the experimental and theoretical investigations of robustness in synthetic systems or transformed tissue culture cells *in vitro*.

On the basis of robustness, our simulations support bimodal positive and negative Itk regulation by $IP_4$ in thymocytes. They make a supportive argument that Itk PH domain oligomerization and $IP_4$ feedback are physiologically important, consistent with the severely defective TCR signaling, $IP_4$ production, Itk/PLCγ1 activation, positive selection and resulting immunodeficiency in *ItpkB*[-/-] mice, the ability of $IP_4$ to bimodally control Itk PH domain binding to $PIP_3$ *in vitro*, and the reported Itk PH domain oligomerization [2,6,7,28]. They do, however, not exclude the possibility that $IP_4$ also has additional, unknown functions in DP cells [14].

Testing this exciting hypothesis will require currently impossible single-cell measurements of $IP_4$ levels in large cell populations. Moreover, the physiological roles and modes of Itk oligomerization, the specific PH domain contributions to Itk oligomerization, whether Itk oligomerization occurs in the cytoplasm or at the plasma membrane or both, whether it exclusively inhibits or can also promote Itk activation, and whether $IP_4$ promotes or inhibits Itk PH domain binding to $PIP_3$ or does both depending on its local concentration are all matters of active debate [2,22-28]. Their conclusive elucidation requires quantitative biophysical studies of full length Itk with or without mutational perturbation of individual and combined interactions among the different Itk domains implicated in its monomeric and oligomeric self-association, and the reconstitution of *Itk*[-/-] mice with these mutants at endogenous expression levels. Unfortunately, difficulties to produce sufficient quantities of soluble full-length Itk or Itk PH domain protein, and a tendency of Itk and its PH domain to aggregate *in vitro* have precluded more quantitative analyses of Itk PH domain oligomerization and $IP_4$/$PIP_3$ interactions, as well as the generation of non-oligomerizing Itk PH domain mutants. Despite progress regarding SH2/SH3/proline-rich domain interactions [22-26] and some evidence for PH domain involvement [2,27,28], formation of several different homotypic Itk dimers with differing subcellular localization and functions further complicates such analyses and their interpretation. Our *in silico* results suggest that by enabling competing positive and negative $IP_4$ induced feedback, Itk PH domain oligomerization could render Itk signaling in DP thymocytes much more robust to parameter fluctuation between individual cells than could be achieved without Itk dimers, or without $IP_4$ feedback. Models M1-M3 and M7 involving Itk dimers and $IP_4$ feedbacks showed substantially larger robustness than models lacking feedbacks (M5-M6) or containing only monomeric Itk (M4). M1-M3 and M7 can describe the experimentally observed PLCγ1 kinetics with similar robustness. They differ only at the level of secondary Itk/$IP_4$/$PIP_3$ interactions. Similar robustness and the inherent variability of experimental data preclude the identification of one of these dimeric Itk feedback models as the only one operative *in vivo* thus far.

**METHOD AND MATERIALS**

**Signaling kinetics in the *in silico* models**

We constructed ODE based models. The ODEs described kinetics of concentrations of proteins and lipids in two well-mixed compartments representing plasma membrane and cytosol (Fig. S1A). The biochemical signaling reactions for each model are shown in tables S1-S7. The details regarding the construction of the ODEs and the parameters are given in the web supplement and Fig. S1. We use the rule based modeling software package BioNetGen [69] to generate time courses for the species kinetics for the signaling networks described by models M1-M7. This program produces a set of ODEs corresponding to the mass-action kinetics describing biochemical reactions in the networks and solves them numerically using the CVODE solver [70]. The ODEs for each model are listed in the supplementary material.

**Quantification of robustness based on the maximum entropy principle**

When a variable $x$ can assume multiple values and is distributed according to a probability distribution $p(x)$, then the uncertainty associated with the distribution can be quantified by the entropy (S) defined as,

$$S = -\sum_x p(x) \ln p(x). \tag{1}$$

S is non-negative and is maximized when $x$ is distributed according to a uniform distribution (i.e., $x$ can take any value within a range with equal probability). Suppose $p(x)$ is unknown, but we do know the average value of a variable, $f$, that is a function of $x$, i.e., $f \equiv f(x)$. We can then maximize S under the constraint

$$\sum_x f(x) p(x) = \overline{f} . \tag{2}$$

The constrained MaxEnt distribution is given by $p(x) \propto \exp(-\lambda f(x))$, where the constant $\lambda$, also known as the Lagrange multiplier, is determined by solving Eq. (2) for $\lambda$ when the above MaxEnt distribution for p($x$) is used in Eq. (2). The method can be easily generalized to accommodate multiple variables and constraints. We used the constraints imposed by $\tau_p^{expt}$, $R^{expt}$, and $A^{expt}$, or, $\tau_p^{expt}$ and $R^{expt}$ that are measured over a cell population. Therefore, the MaxEnt distribution of the parameters in our calculation is given by, $p(\{k_i\}) \propto \exp(-\lambda_1 \tau_p(\{k_i\}) - \lambda_2 R(\{k_i\}) - \lambda_3 A(\{k_i\}))$, where $\lambda_1, \lambda_2$, and, $\lambda_3$ denote the Lagrange's multipliers, and $\{k_i\}$ denote the values of rate constants and initial concentrations in individual cells. The Lagrange multipliers can be calculated from the constraint equations,

$$\sum_{\{k_i\}} \tau_p(\{k_i\}) \, p(\{k_i\}) = \tau_p^{expt}$$

$$\sum_{\{k_i\}} R(\{k_i\}) \, p(\{k_i\}) = R^{expt} \qquad (3)$$

$$\sum_{\{k_i\}} A(\{k_i\}) \, p(\{k_i\}) = A^{expt}.$$

The MaxEnt distribution thus describes how $\tau_p$, R, and A, in individual cells are distributed over a cell population. The distribution also produces an estimation of the probability distributions for the rate constants and initial concentrations that regulate $\tau_p$, R, and A, through the functions $\tau_p(\{k_i\})$, $R(\{k_i\})$, and, $A(\{k_i\})$, respectively. The specific relationship between the parameters, $\{k_i\}$, and the observables ($\tau_p$, R, and A) is dependent on the molecular details of the models, M1-M7. In all the models prior to the MaxEnt calculation, the rate constants were chosen from a uniform distribution with lower and upper bounds equal to 1/10 and 10 times, respectively, of the base values shown in tables S1-S7. Similarly, the initial concentrations of proteins (e.g., Itk) and lipids (such as PIP3) were varied within a 35% [71] range from uniform distributions centered at the base values shown in Table S8. The joint uniform distribution in the parameters is given by q($\{k_i\}$). We then used these MaxEnt distributions to quantify relative robustness of the models by calculating the Kullback-Leibler distance[59]

$$D_{KL} = \sum_{\{k_i\}} p(\{k_i\}) \ln(p(\{k_i\})/q(\{k_i\})) \qquad (4)$$

That is, for each model, we first find the probability distribution for the rate constants and initial concentrations that maximizes the entropy (robustness) for that model under the experimental constraints, giving the model a kind of "maximum benefit of the doubt." We then compare the resulting MaxEnt models with one another to evaluate their relative robustness to variation in the rate constants, in order to select the model(s) most likely to correctly represent the actual kinetics. When $p(\{k_i\})$ is equal to $q(\{k_i\})$, $D_{KL}$ assumes the minimum value 0; as the distribution $p(\{k_i\})$ starts deviating from the uniform distribution, say by becoming sharply peaked around a particular value, $D_{KL}$ increases. Thus maximizing the entropy S, is equivalent to minimizing $D_{KL}$ in Eq. (4). The calculations of $D_{KL}$ were done at a specific antigen dose which fixed the average values of initial concentrations of Itk and PIP3. Therefore, the robustness calculations did not exclude the sensitivity of PLCγ1 activation to changes in PIP3 concentrations resulting from antigen dose variations. We calculated $p(\{k_i\})$ by by minimizing the $D_{KL}$ subject to the constraints imposed by Eq. (3). We used $D_{KL}$ to rank order the models for a particular measured value of $\tau_p^{expt}$, $R^{expt}$, and, $A_{avg}$. All the calculations were carried out using MATLAB. Additional details can be found in the supplementary material (Figs. S12-S15). Note that $D_{KL}$ is unaffected by inclusion of additional parameters that do not influence the experimentally measured variables (Fig. S21, Table S17). Thus having extra variables in a model does not in and of itself affect the relative robustness of models with variable numbers of parameters. We have used 100,000 sample points, which we have shown to be statistically sufficient in Fig. S22 for the faithful calculation of $D_{KL}$.

**Thymocyte stimulation and immunoblot analysis**

All mice were housed in the TSRI specific pathogen-free vivarium monitored by the TSRI Department of Animal Resources. All animal studies were approved by the TSRI IACUC and conform to all relevant regulatory standards.

DP cells were prepared as in [2] and rested at 37°C for 3 hours. Then, $10^7$ DP cells per sample were incubated on ice for 15 min with 2.4 µM MHCI tetramers pre-loaded with either one of the altered peptide ligands OVA, Q4R7, Q4H7 or G4 [40], stimulated by rapidly adding 37°C warm PBS for the indicated times and quickly lysed in 100 mM Tris, pH 7.5, 600 mM NaCl, 240 mM n-octyl-β-D-glucoside, 4% Triton, 4 mM EDTA and a protease/phosphatase inhibitor cocktail (Roche). Lysates were cleared by centrifugation at 14000 rpm for 10 minutes at 4°C, resolved by SDS-PAGE and analyzed via immunoblot as previously described [2]. Band intensities were quantified via densitometry using NIH *ImageJ* software, and phosphoY$_{783}$-PLCγ1 intensities normalized to total PLCγ1 amounts.

**ACKNOWLEDGEMENTS**

We thank our lab members for valuable discussions, Luise Sternberg and Lyn'Al Nosaka for mouse genotyping, and the TSRI vivarium for expert mouse husbandry. S.M thanks Susmita Basak for help with MATLAB.

**Figure Captions:**

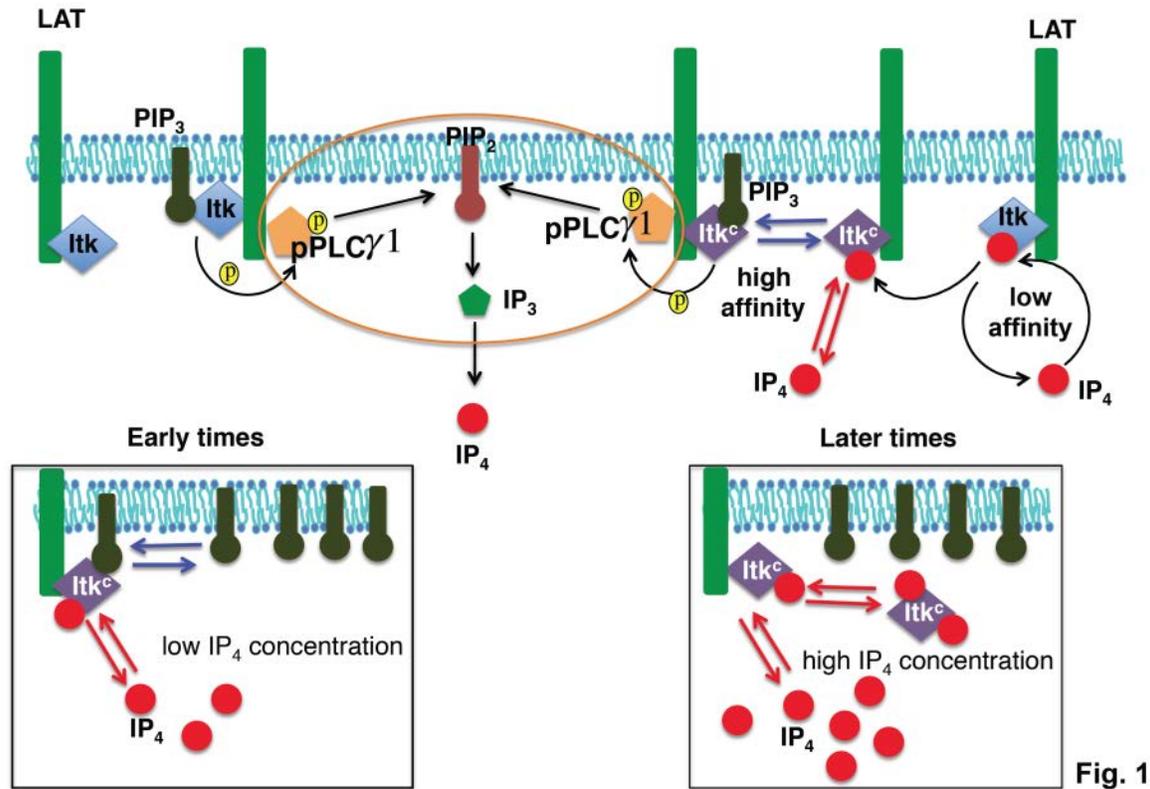

**Fig. 1. Relevant basic interactions between Itk, PIP$_3$ and IP$_4$.** Following TCR-pMHC binding, Itk molecules are bound by the LAT signalosome via SLP-76 (not shown). Itk molecules (monomers or dimers, blue diamonds), bind the membrane lipid PIP$_3$ with low affinity through their PH domains. PIP$_3$ bound Itk phosphorylates and thereby activates LAT-bound PLCγ1. Activated PLCγ1 then hydrolyzes the membrane lipid PIP$_2$ into the soluble second messenger IP$_3$, a key mediator of Ca$^{2+}$ mobilization. IP$_3$ 3-kinase B (ItpkB) converts IP$_3$ into IP$_4$ (red filled circle). For our *in silico* models, we simplified this series of reactions, encircled by the orange oval, into a single second order reaction where PIP$_3$ bound Itk converts PIP$_2$ into IP$_4$. In models M1-M4 and M7, IP$_4$ modifies the Itk PH domain (denoted as Itk$^C$, purple diamonds) to promote PIP$_3$ and IP$_4$ binding to the Itk PH domain. At the onset of the signaling, when the concentration of IP$_4$ is smaller than that of PIP$_3$, IP$_4$ helps Itk$^C$ to bind to PIP$_3$ (left lower panel). However, as the concentration of IP$_4$ is increased at later times, IP$_4$ outcompetes PIP$_3$ for binding to Itk$^C$ and sequesters Itk$^C$ to the cytosol (right lower panel). In models M5/M6, IP$_4$ and PIP$_3$ do not augment each other's binding to Itk. However, IP$_4$ still outcompetes PIP$_3$ for Itk PH domain binding when the number of IP$_4$ molecules becomes much larger than that of PIP$_3$ molecules at later times.

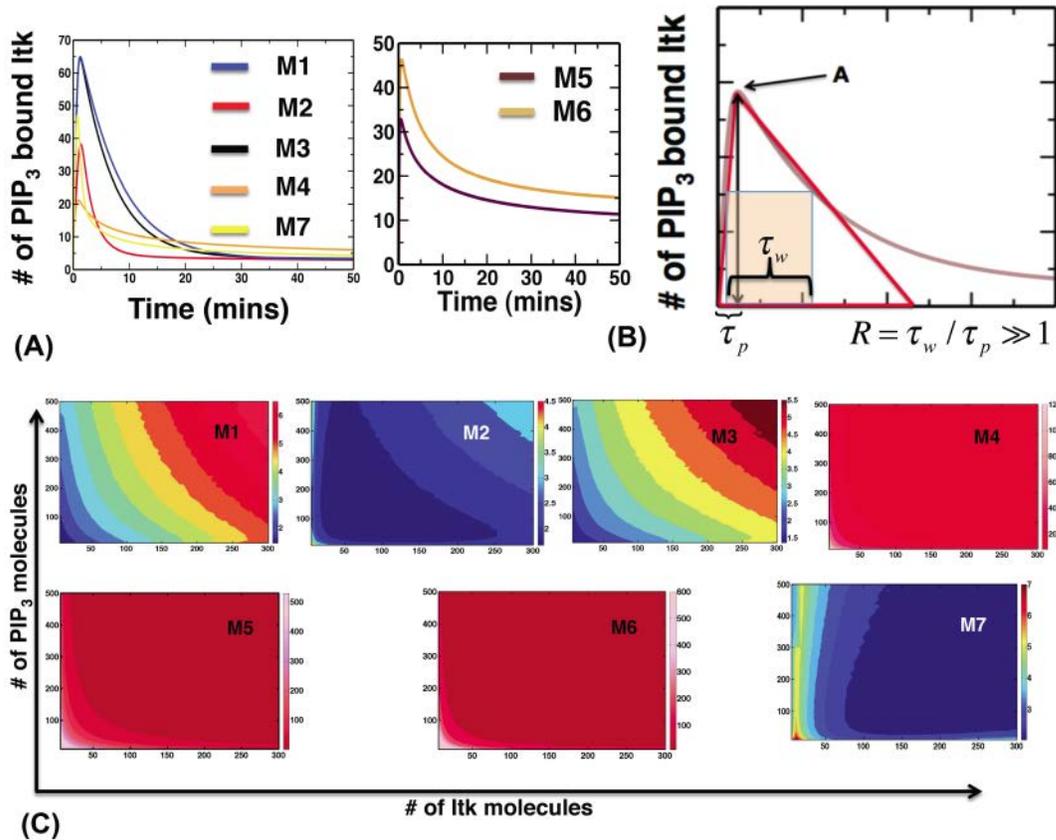

**Fig. 2. Different molecular interactions in models M1-M7 produce different temporal profiles of PIP$_3$ binding to Itk. (A)** Kinetics of PIP$_3$ association of Itk for fixed initial PIP$_3$ and Itk concentrations (100 and 370 molecules, respectively) in models with feedbacks (M1-M4, and M7, left panel) and no feedbacks (M5-M6, right panel). **(B)** The shapes of the temporal profiles can be characterized by the parameters peak time ($\tau_p$), peak width ($\tau_w$), and peak value or amplitude (A). The dimensionless asymmetry ratio R=$\tau_w/\tau_p$ quantifies how symmetric the shape of the time profile is. A larger R value indicates larger asymmetry. **(C)** Variations in R in models M1-M7 for different initial concentrations of Itk and PIP$_3$. Color scales for R values are shown on the right of each panel.

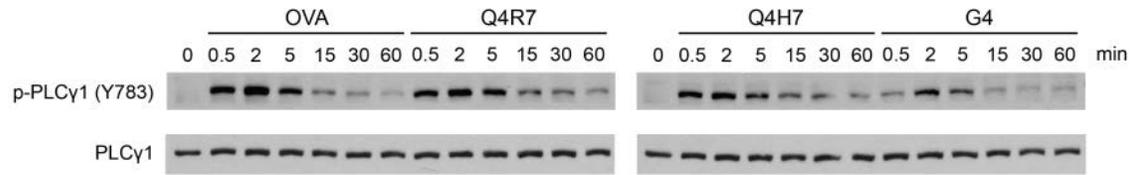

(A)

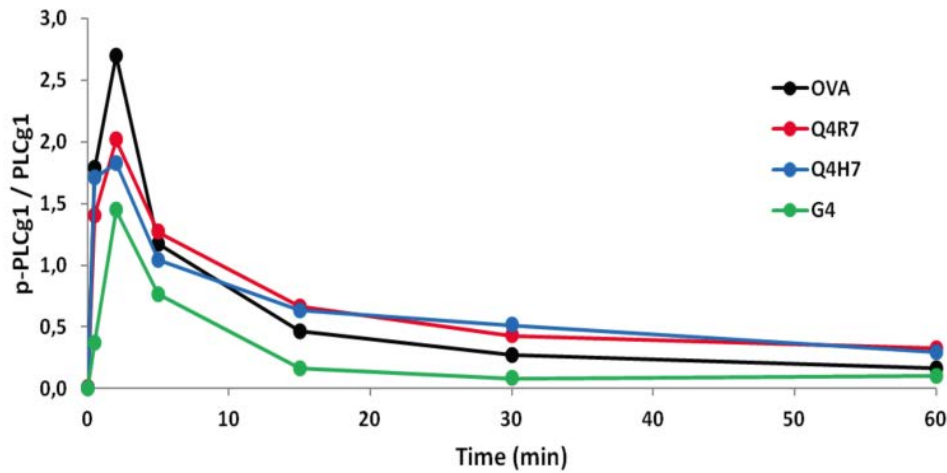

(B)

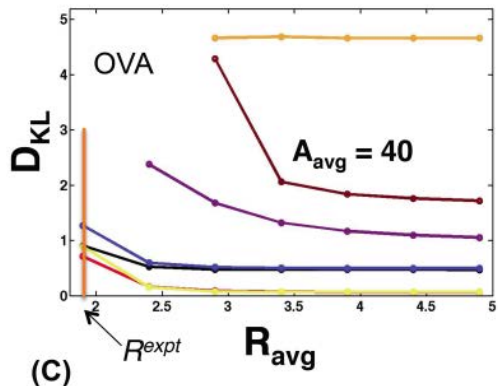

(C)

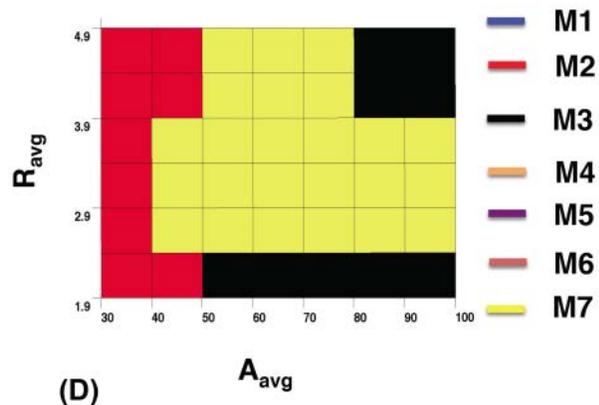

(D)

**Fig. 3. Experimentally measured PLCγ1 activation kinetics in DP thymocytes stimulated with TCR ligands of different affinities and robustness of *in silico* models.** (A) Immunoblots showing $Y_{783}$-phosphorylated (upper panels) and total (lower panels) PLCγ1 protein amounts in *RAG2$^{-/-}$MHC$^{-/-}$ OT1 TCR-transgenic* DP thymocytes stimulated for the indicated times with MHCI tetramers presenting the indicated altered peptide ligands (APL). (B) Phospho-PLCγ1 levels normalized to total PLCγ1 protein amounts plotted over time for the indicated APLs. Their TCR affinity decreases in the order OVA (black) > Q4R7 (red) > Q4H7 (blue) > G4 (green). Band intensities were quantified via scanning and analysis with *ImageJ* software. Representative of several independent experiments. (C) Variation of the Kulback-Leibler distance $D_{KL}$ with *R* for models M1-M3 (blue, red and black, respectively), M7 (yellow), and M4-M6 (orange, purple, and maroon, respectively) at high initial Itk (Itk$^0$=140 molecules) and PIP$_3$ concentrations (PIP$_3^0$=530 molecules), representing high-affinity OVA stimulation for

$\tau_p$=2 min and A (shown as $A_{avg}$)=40 molecules. Note we use A to represent the amplitude $A^{expt}$ in experiments measuring fold change in Itk phosphorylation (see the main text for further details). The vertical orange bar indicates $R^{expt}$ for OVA. Color legend in (D). (D) The color map shows which model is most robust (has the lowest $D_{KL}$) as $R^{expt}$ and A (shown as $A_{avg}$) are varied for the same parameters as in (C). The color legend is depicted on the right.

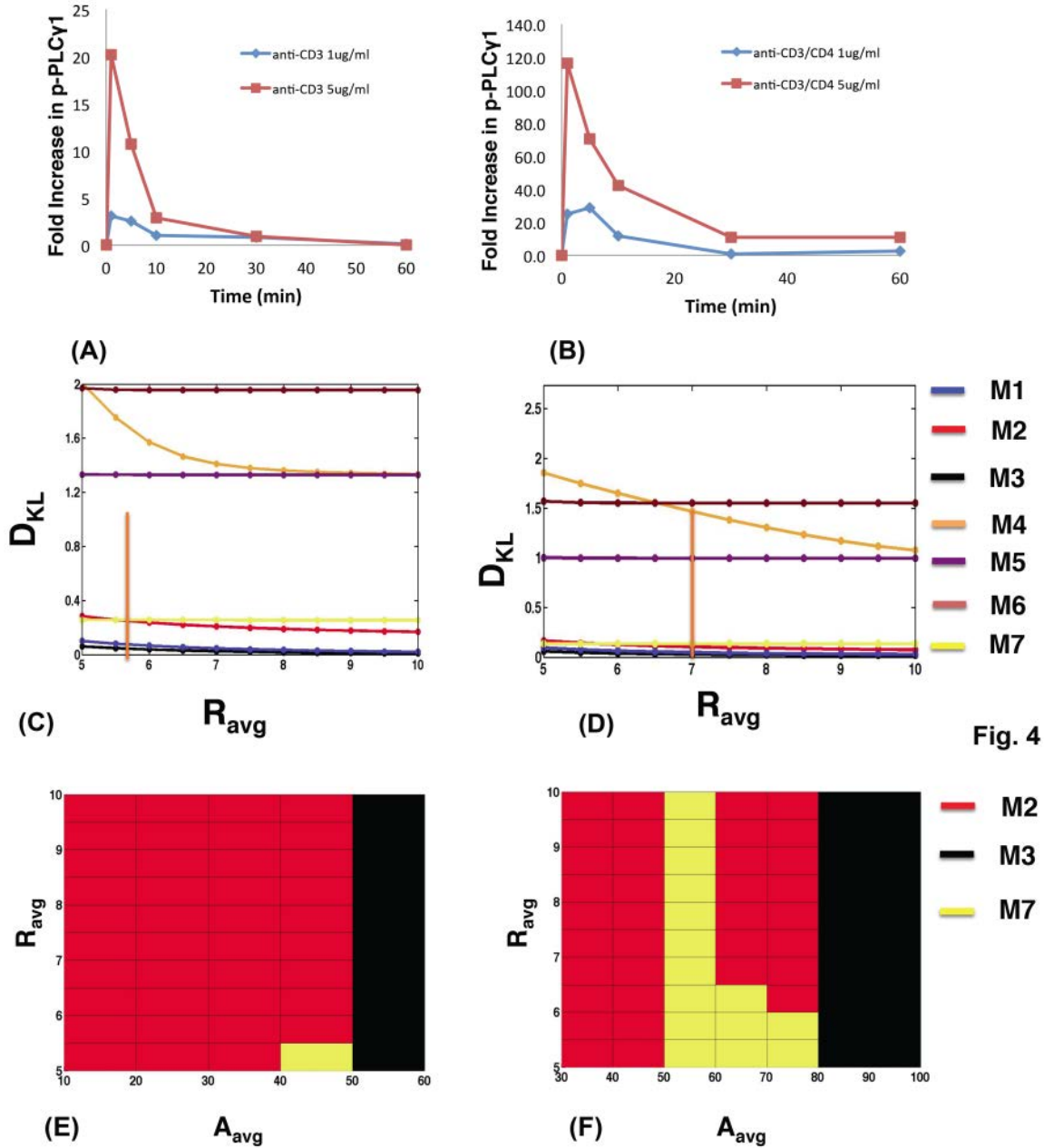

**Fig. 4. Models containing Itk dimers and dueling feedbacks also show higher robustness for polyclonal T cells stimulated by anti-CD3 antibodies**. PLCγ1 phosphorylation kinetics in $MHC^{-/-}$ T cells stimulated by antibodies against (A) CD3 or (B) CD3 and CD4 at 1 µg/ml versus 5 µg/ml. (C) Variation of $D_{KL}$ with R for the *in silico*

*models* M1-M3 (blue, red and black, respectively), M7 (yellow), and M5-M6 (purple and maroon, respectively) at initial Itk ($Itk^0$= 100 molecules) and $PIP_3$ concentrations ($PIP_3^0$= 370 molecules) at $\tau_p$=1 min and A=60 molecules, representing anti-CD3 stimulation at 5 µg/ml. The orange bar indicates $R^{expt}$. Note we use A to represent the amplitude $A^{expt}$ in experiments measuring fold change in Itk phosphorylation (see the main text for further details). (D) Variation of $D_{KL}$ with R for anti-CD3/CD4 stimulation at 5 µg/ml at $\tau_p$=1min and A=80 molecules. The initial Itk ($Itk^0$= 140 molecules) and $PIP_3$ concentrations ($PIP_3^0$= 530 molecules) were used. The orange bar indicates $R^{expt}$. (E) and (F) show maps of the most robust models (with the lowest $D_{KL}$) as $R^{expt}$ and A (shown as $A_{avg}$) are varied for the same parameters as in (C) and (D), respectively.

**Table I: Molecular models describing interactions between Itk, IP$_4$ and PIP$_3$.**

| | M1 | M2 | M3 | M7 | M4 | M5 | M6 |
|---|---|---|---|---|---|---|---|
| | **Contain IP$_4$ induced +ve feedback** | | | | | **No +ve feedback** | |
| | **Contain Itk dimers** | | | | **Contains Itk monomers** | **Contains Itk dimers** | **Contains Itk monomers** |
| **Effect of IP$_4$ binding to one PH domain of an Itk dimer** | Increases affinity of the other PH domain toward IP$_4$ and PIP$_3$. | Same as in M1. | Same as in M1. | Increases affinity of the other PH domain toward PIP$_3$ and IP$_4$. | IP$_4$ and PIP$_3$ bind to the Itk PH domain with weak affinities. However, IP$_4$ bound to Itk gets replaced by PIP$_3$ with high affinity, and then the PIP$_3$ bound to Itk can get replaced by IP$_4$ with high affinity. | No change in affinity | The monomeric PH domain binds IP$_4$ and PIP$_3$ with equal but always low affinity. |
| **Effect of PIP$_3$ binding to one PH domain of an Itk dimer** | Increases affinity of the other PH domain for IP$_4$ *and* PIP$_3$. | *Does not* increase the affinity of the other PH domain for IP$_4$ or PIP$_3$. | Increases affinity of the other PH domain only for IP$_4$ but *not* for PIP$_3$. | Increases affinity of the other PH domain for PIP$_3$ but *not* for IP$_4$. | | No change in affinity | |
| **Number of parameters (Rate constants + initial concentrations)** | 5 + 3 | 5 + 3 | 5 + 3 | 5 + 3 | 4 + 3 | 3 + 3 | 3 + 3 |

**Table II: Values of peak time, peak width, and asymmetry ratio R calculated from the PLCγ1 activation kinetics in Fig. 3 for different ligands.**

| Ligand | Peak time ($\tau_p$) (min) | Peak width ($\tau_w$) (min) | R |
|---|---|---|---|
| OVA | 2.0 | 3.9 | 1.9 |
| Q4R7 | 2.0 | 8.6 | 4.3 |
| Q4H7 | 2.0 | 7.5 | 3.8 |
| G4 | 2.0 | 4.3 | 2.1 |

**Supplementary Information for "*In silico* Modeling of Itk Activation Kinetics in Thymocytes Suggests Competing Positive and Negative IP$_4$ Mediated Feedbacks Increase Robustness "**

**Details of the models and simulations:**

We used a set of ODEs to describe the signaling kinetics of concentrations of proteins and lipids in the system. The descriptions of the kinetics using ODEs neglect stochastic fluctuations arising from intrinsic noise fluctuations and assume a spatially well-mixed system. This is a good approximation when diffusion time scales are fast compared to the reaction time scales. Since the time scales for diffusion depend on the spatial extent of the system, the system can be assumed to be spatially homogeneous in a region of volume V (Fig. S1A), which is small compared the to the total volume of a T cell. In addition, the signaling reactions in our system can take place between a pair of molecules, where both species reside in the plasma membrane. For these reactions, we convert the values of binding rates given in three dimensions to approximate two dimensional rates. We provide details regarding the above approximations below. The reactions are shown in Tables S1-S7, and the corresponding ODEs are shown below the tables. The model reactions are also shown graphically in Fig. S1B. The software package, BIONETGEN, was used to construct and solve the ODEs. The BIONETGEN codes are available at http://planetx.nationwidechildrens.org/~jayajit/.

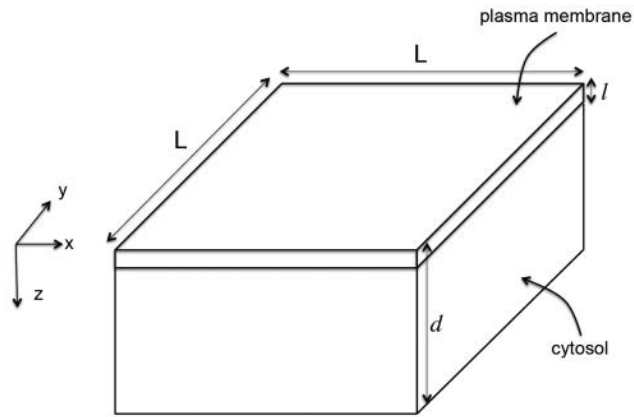

**Fig. S1A. Details of the simulation box.** We used L=2 μm, $l$=2 nm and d=0.02 μm for our simulations.

The simulation box (Fig. S1A) is divided into two compartments, plasma membrane and cytosol. The molecules in the cytosol can react with the plasma membrane bound molecules only when they are in a close proximity (~$l$=2 nm, Fig. S1). This length scale is used to convert three dimensional reaction rates to two dimensional reaction rates describing reactions in the plasma membrane. The area of the plasma membrane is taken to be 4 μm$^2$ (LxL in Fig. S1). The volume of the cytosol is 0.08 μm$^3$ (LxLxd in Fig. S1) and the volume of the plasma membrane compartment is 0.008 μm$^3$ (LxLx0.002 μm). All the molecular species were distributed homogenously across the simulation box. This is a good assumption when the time scales of diffusion of the molecules are much faster compared to the fastest reaction time scale. Since the fastest reaction time scale (~10 s, corresponding to unbinding rate of Itk – PIP$_3$, see the tables below) is comparable to the slowest diffusion time scales ($D \sim 0.1$ $\mu m^2/s$ for membrane bound species (1)) on the length scale (2 μm) of the simulation box, we assumed the plasma membrane bound species to be homogeneously distributed in space. The cytosolic molecules diffuse with a much faster time scale ($D \sim 10$ $\mu m^2/s$ (1)) compared to the plasma membrane bound molecules, therefore, we also assumed that those molecules are also homogeneously distributed in the simulation box.

ODEs for Model 1: The reactions for Model 1 are given in the table below. The corresponding ODEs are shown below the table.

**Table S1: Reactions and rate constants for model M1.**

| Reactions | $k_{on}$ ($\mu M^{-1} s^{-1}$) | $k_{off}$ ($s^{-1}$) | $K_D$ ($\mu M$) | $k_{cat}$ ($\mu M^{-1} s^{-1}$) |
|---|---|---|---|---|
| $Itk - Itk + PIP_3 \leftrightarrow Itk - Itk - PIP_3$ | $2.5 \times 10^{-4}$ | 0.1 | 400 * | |
| $PIP_3 - Itk - Itk + PIP_3 \leftrightarrow PIP_3 - Itk - Itk - PIP_3$ | 0.01 | 0.003 | 0.3 * | |
| $Itk - Itk + IP_4 \leftrightarrow Itk - Itk - IP_4$ | $2.5 \times 10^{-3}$ | 0.1 | 40 * | |
| $PIP_3 - Itk - Itk + IP4 \leftrightarrow PIP_3 - Itk - Itk - IP_4$ | 0.1 | 0.003 | 0.03 † (2) | |
| $IP_4 - Itk - Itk + PIP_3 \leftrightarrow IP_4 - Itk - Itk - PIP_3$ | 0.01 | 0.003 | 0.3 † (2) | |
| $IP_4 - Itk - Itk + IP4 \leftrightarrow IP_4 - Itk - Itk - IP_4$ | 0.1 | 0.003 | 0.03 † (2) | |
| $Itk - Itk - PIP_3 + S \leftrightarrow IP_4 + Itk - Itk - PIP_3$ | | | | $1.5 \times 10^{-4}$ ** |
| $IP_4 - Itk - Itk - PIP_3 + S \leftrightarrow IP_4 + IP_4 - Itk - Itk - PIP_3$ | | | | $1.5 \times 10^{-4}$ ** |
| $PIP_3 - Itk - Itk - PIP_3 + S \rightarrow IP_4 + PIP_3 - Itk - Itk - PIP_3$ | | | | $1.5 \times 10^{-4}$ ** |

† The high affinity binding of IP$_4$ to the PH domain is taken to be in the nano-molar range based on the binding affinity of the isolated Btk PH domain for both IP$_4$ and PIP$_3$ (2). The binding affinity of the Itk PH domain for IP$_4$ or PIP$_3$ however is not known. The $K_D$ for PIP$_3$ binding to the isolated Btk PH domain is reported to be 7 times higher than the IP$_4$ binding (2). For convenience we have taken it to be 10 times larger.

* Estimated. The low affinity binding of Itk PH domains to IP$_4$/PIP$_3$ has been assumed to be 1000 times weaker than the high affinity binding.

** Estimated. Hydrolysis of PIP$_2$ and subsequent production of IP$_3$, release of Ca$^{2+}$ from the endoplasmic stores, Ca$^{2+}$/Calmodulin dependent activation of ItpkB, and, finally the enzymatic turnover of IP$_3$ by ItpkB have all been subsumed into the one single step where membrane bound Itk cleaves PIP$_2$ to produce IP$_4$. The rate constant for the above reaction is chosen to match the time scale of PLCγ1 activation reported in the experiments (3). PIP$_2$ is denoted as S for convenience.

*Construction of the ODEs describing the signaling kinetics.*

The molecular species, Itk-Itk (we denote this species by Itk$_D$), representing Itk-Itk dimers bound to the TCR and LAT signalosome, resides at the interface of the plasma membrane and the cytosol. The lipid PIP$_3$, and the Itk-Itk-PIP$_3$ complexes (Itk-Itk-PIP$_3$, IP$_4$-Itk-Itk-PIP$_3$), and PIP$_2$ (denoted as S) also reside in the plasma membrane. The molecules IP$_4$ and IP$_4$-Itk-Itk-IP$_4$ are soluble and dwell in the cytosol. We use two well-mixed compartments representing plasma membrane and cytosol (Fig. S1A) and use ODEs to describe the kinetics for the molecules in those compartments. E.g., concentration of Itk$_D$ or [Itk$_D$] at a time t is given by,

$[Itk_D](t) = \frac{N_{Itk_D}(t)}{Al}\theta(z+l)$ , where, $N_{Itk_D}(t)$ denotes the total number of Itk dimers in the system, and $\theta(z)$ is the Heaveside step function defined as, $\theta(z) = 0$ for $z < 0$, and, $\theta(z) = 1$ for $z \geq 0$. In the simulation box, z ranges from z=0 to z=-d. The region between $0 \geq z \geq -l$ represents the plasma membrane and $-l \geq z \geq -d$ represents the cytosolic part in the simulation. All the plasma membrane bound species concentrations are defined as above. The concentration of cytosolic IP$_4$ is given by,

$[IP_4](t) = \frac{N_{IP_4}(t)}{A(d-l)}\theta(z+d)\theta(z-l) \approx \frac{N_{IP_4}(t)}{Ad}\theta(z+d)\theta(z-l)$. The last approximation follows as $l(= 2nm) \ll d(= 0.02\mu m)$. Concentrations of all the cytosolic species are defined in the same way. Using the scheme described above we write down the ODEs describing kinetics for the total concentrations of the species present in the model. Below we show how the ODEs are constructed from the reactions using an example.

Consider the following reaction
$Itk - Itk + IP_4 \underset{K_{off}}{\overset{K_{on}}{\longleftrightarrow}} Itk - Itk - IP_4$

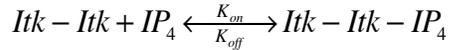

, where $K_{on}$ and $K_{off}$ are the binding and the unbinding rates respectively. Using the law of mass action the time evolution of the Itk-Itk-IP$_4$ complex can be written as

$\frac{d[Itk - Itk - IP_4]}{dt} = K_{on}[Itk - Itk][IP_4] - K_{off}[Itk - Itk - IP_4]$.

The species Itk-Itk-IP$_4$ and Itk-Itk reside in the plasma membrane whereas IP$_4$ resides in the cytosol. Therefore, as described above, $[Itk - Itk - IP_4](t) = \frac{N_{Itk-Itk-IP_4}(t)}{Al}\theta(z+l)$,

$[Itk - Itk](t) = \frac{N_{Itk-Itk}(t)}{Al}\theta(z+l)$ , and [IP$_4$](t) is related to N$_{IP4}$ as described previously.

We denote, V$_{membrane}$=$Al$ and Ad=V$_{cytosol}$=$Ad$. Multiplying the above rate equation by V$_{membrane}$, we get,

$$\frac{dN_{Itk-Itk-IP_4}}{dt} = \frac{K_{on}}{V_{cytosol}} N_{Itk-Itk} N_{IP_4} - K_{off} N_{Itk-Itk-IP_4}$$ Note that the rate, $K_{on}$, for a binding reaction between a plasma membrane bound molecule and a cytosolic molecule, is scaled by the volume of the cytosol whereas the $K_{off}$ remains the same. For a reaction like $Itk - Itk + PIP_3 \underset{K_{off}}{\overset{K_{on}}{\longleftrightarrow}} Itk - Itk - PIP_3$, where all the molecular species reside in the plasma membrane we find,

$$\frac{dN_{Itk-Itk-PIP_3}}{dt} = \frac{K_{on}}{V_{membrane}} N_{Itk-Itk} N_{PIP_3} - K_{off} N_{Itk-Itk-PIP_3}.$$

The on rate, $K_{on}$, is scaled by the volume of the membrane $V_{membrane}$ instead of the $V_{cytosol}$= Ad. Following the above procedure we write the ODEs for the model M1. For convenience we denote the Itk-Itk dimer as $Itk_D$. We do not show the volume scaling factors explicitly, all the rates shown in the ODEs have the unit of [time]$^{-1}$. These rates are calculated using the values given in Table S1 and then scaling those values with appropriate volume factors.

$$\frac{dN_{Itk_D}}{dt} = k_{-1}N_{Itk_D-PIP_3} - k_1 N_{Itk_D} N_{PIP_3} + k_{-2} N_{Itk_D-IP_4} - k_2 N_{Itk_D} N_{IP_4}$$

$$\frac{dN_{Itk_D-PIP_3}}{dt} = k_1 N_{Itk_D} N_{PIP_3} - k_{-1} N_{Itk_D-PIP_3} + 2\tilde{k}_{-1} N_{PIP_3-Itk_D-PIP_3}$$
$$- \tilde{k}_1 N_{Itk_D-PIP_3} N_{PIP_3} + \tilde{k}_{-2} N_{IP_4-Itk_D-PIP_3} - \tilde{k}_2 N_{Itk_D-PIP_3} N_{IP_4}$$

$$\frac{dN_{Itk_D-IP_4}}{dt} = k_2 N_{Itk_D} N_{IP_4} - k_{-2} N_{Itk_D-IP_4} + 2\tilde{k}_{-2} N_{IP_4-Itk_D-IP_4} - \tilde{k}_2 N_{Itk_D-IP_4} N_{IP_4}$$
$$+ \tilde{k}_{-1} N_{IP_4-Itk_D-PIP_3} - \tilde{k}_1 N_{Itk-IP_4} N_{PIP_3}$$

$$\frac{dN_{PIP_3}}{dt} = k_{-1} N_{Itk_D-PIP_3} - k_1 N_{Itk_D} N_{PIP_3} + 2\tilde{k}_{-1} N_{PIP_3-Itk_D-PIP_3} - \tilde{k}_1 N_{Itk_D-PIP_3} N_{PIP_3}$$
$$+ \tilde{k}_{-1} N_{IP_4-Itk_D-PIP_3} - \tilde{k}_1 N_{Itk-IP_4} N_{PIP_3}$$

$$\frac{dN_{PIP_3-Itk_D-PIP_3}}{dt} = \tilde{k}_1 N_{Itk_D-PIP_3} N_{PIP_3} - 2\tilde{k}_{-1} N_{PIP_3-Itk_D-PIP_3}$$

$$\frac{dN_{IP_4-Itk_D-PIP_3}}{dt} = \tilde{k}_1 N_{Itk-IP_4} N_{PIP_3} - \tilde{k}_{-1} N_{IP_4-Itk_D-PIP_3} + \tilde{k}_2 N_{Itk_D-PIP_3} N_{IP_4}$$
$$- \tilde{k}_{-2} N_{IP_4-Itk_D-PIP_3}$$

$$\frac{dN_{IP_4-Itk_D-IP_4}}{dt} = \tilde{k}_2 N_{Itk_D-IP_4} N_{IP_4} - 2\tilde{k}_{-2} N_{IP_4-Itk_D-IP_4}$$

$$\frac{dN_{IP_4}}{dt} = 2\tilde{k}_{-2} N_{IP_4-Itk_D-IP_4} - \tilde{k}_2 N_{Itk_D-IP_4} N_{IP_4} + \tilde{k}_{-2} N_{IP_4-Itk_D-PIP_3} - \tilde{k}_2 N_{Itk_D-PIP_3} N_{IP_4}$$
$$+ k_{-2} N_{Itk_D-IP_4} - k_2 N_{Itk_D} N_{IP_4}$$
$$+ k_3 \left\{ N_{IP_4-Itk_D-PIP_3} + N_{Itk_D-PIP_3} + N_{PIP_3-Itk_D-PIP_3} \right\} N_S$$

$$\frac{dN_S}{dt} = -k_3 \left\{ N_{IP_4-Itk_D-PIP_3} + N_{Itk_D-PIP_3} + N_{PIP_3-Itk_D-PIP_3} \right\} N_S$$

There are three conservation laws, namely

$$N_{Itk_D} + N_{Itk_D-PIP_3} + N_{PIP_3-Itk_D-PIP_3} + N_{IP_4-Itk_D-PIP_3} + N_{IP_4-Itk_D-IP_4} = N_{Itk_D}(t=0)$$
$$N_{PIP_3} + N_{Itk_D-PIP_3} + N_{IP_4-Itk_D-PIP_3} + 2*N_{PIP_3-Itk_D-PIP_3} = N_{PIP_3}(t=0)$$
$$N_{IP_4} + N_{Itk_D-IP_4} + N_{IP_4-Itk_D-PIP_3} + 2*N_{IP_4-Itk_D-IP_4} + N_S = N_S(t=0)$$

the equations above give the time evolution of all the proteins and the protein complexes.

ODEs for Model 2: The reactions for Model 2 are given in the table below. The corresponding ODEs are shown below the table.

**Table S2: Reactions and rate constants for model M2.**

| Reactions | $k_{on}$ ($\mu M^{-1}s^{-1}$) | $k_{off}$ ($s^{-1}$) | $K_D$ ($\mu M$) | $k_{cat}$ ($\mu M^{-1}s^{-1}$) |
|---|---|---|---|---|
| $Itk-Itk + PIP_3 \leftrightarrow Itk-Itk-PIP_3$ | $2.5 \times 10^{-4}$ | 0.1 | 400 | |
| $Itk-Itk + IP_4 \leftrightarrow Itk-Itk-IP_4$ | $2.5 \times 10^{-3}$ | 0.1 | 40 | |
| $PIP_3-Itk-Itk + IP4 \leftrightarrow PIP_3-Itk-Itk-IP_4$ | $2.5 \times 10^{-3}$ | 0.1 | 40 | |
| $IP_4-Itk-Itk + PIP_3 \leftrightarrow IP_4-Itk-Itk-PIP_3$ | 0.01 | 0.003 | 0.3 | |
| $IP_4-Itk-Itk + IP4 \leftrightarrow IP_4-Itk-Itk-IP_4$ | 0.1 | 0.003 | 0.03 | |
| $Itk-Itk-PIP_3 + S \leftrightarrow IP_4 + Itk-Itk-PIP_3$ | | | | $1.5 \times 10^{-4}$ |
| $IP_4-Itk-Itk-PIP_3 + S \leftrightarrow IP_4 + IP_4-Itk-Itk-PIP_3$ | | | | $1.5 \times 10^{-4}$ |

All the high and low affinity binding of Itk PH domains with $PIP_3$ and $IP_4$, and, the $IP_4$ production rate are taken to be the same as that shown for model M1 (Table S1).

$$\frac{dN_{Itk_D}}{dt} = k_{-1} N_{Itk_D-PIP_3} - k_1 N_{Itk_D} N_{PIP_3} + k_{-2} N_{Itk_D-IP_4} - k_2 N_{Itk_D} N_{IP_4}$$

$$\frac{dN_{Itk_D-PIP_3}}{dt} = k_1 N_{Itk_D} N_{PIP_3} - k_{-1} N_{Itk_D-PIP_3} + \tilde{k}_{-2} N_{IP_4-Itk_D-PIP_3} - \tilde{k}_2 N_{Itk_D-PIP_3} N_{IP_4}$$

$$\frac{dN_{Itk_D-IP_4}}{dt} = k_2 N_{Itk_D} N_{IP_4} - k_{-2} N_{Itk_D-IP_4} + 2\tilde{k}_{-2} N_{IP_4-Itk_D-IP_4} - \tilde{k}_2 N_{Itk_D-IP_4} N_{IP_4}$$
$$+ \tilde{k}_{-1} N_{IP_4-Itk_D-PIP_3} - \tilde{k}_1 N_{Itk-IP_4} N_{PIP_3}$$

$$\frac{dN_{PIP_3}}{dt} = k_{-1} N_{Itk_D-PIP_3} - k_1 N_{Itk_D} N_{PIP_3} + \tilde{k}_{-1} N_{IP_4-Itk_D-PIP_3} - \tilde{k}_1 N_{Itk-IP_4} N_{PIP_3}$$

$$\frac{dN_{IP_4-Itk_D-PIP_3}}{dt} = \tilde{k}_1 N_{Itk-IP_4} N_{PIP_3} - \tilde{k}_{-1} N_{IP_4-Itk_D-PIP_3} + \tilde{k}_2 N_{Itk_D-PIP_3} N_{IP_4}$$
$$- \tilde{k}_{-2} N_{IP_4-Itk_D-PIP_3}$$

$$\frac{dN_{IP_4-Itk_D-IP_4}}{dt} = \tilde{k}_2 N_{Itk_D-IP_4} N_{IP_4} - 2\tilde{k}_{-2} N_{IP_4-Itk_D-IP_4}$$

$$\frac{dN_{IP_4}}{dt} = 2\tilde{k}_{-2} N_{IP_4-Itk_D-IP_4} - \tilde{k}_2 N_{Itk_D-IP_4} N_{IP_4} + \tilde{k}_{-2} N_{IP_4-Itk_D-PIP_3} - \tilde{k}_2 N_{Itk_D-PIP_3} N_{IP_4}$$
$$+ k_{-2} N_{Itk_D-IP_4} - k_2 N_{Itk_D} N_{IP_4}$$
$$+ k_3 \left\{ N_{IP_4-Itk_D-PIP_3} + N_{Itk_D-PIP_3} \right\} N_S$$

$$\frac{dN_S}{dt} = -k_3 \left\{ N_{IP_4-Itk_D-PIP_3} + N_{Itk_D-PIP_3} \right\} N_S$$

The conservation laws are,

$$N_{Itk_D} + N_{Itk_D-PIP_3} + N_{IP_4-Itk_D-PIP_3} + N_{IP_4-Itk_D-IP_4} = N_{Itk_D}(t=0)$$
$$N_{PIP_3} + N_{Itk_D-PIP_3} + N_{IP_4-Itk_D-PIP_3} = N_{PIP_3}(t=0)$$
$$N_{IP_4} + N_{Itk_D-IP_4} + N_{IP_4-Itk_D-PIP_3} + 2*N_{IP_4-Itk_D-IP_4} + N_S = N_S(t=0)$$

ODEs for Model 3: The reactions for Model 3 are given in the table below. The corresponding ODEs are shown below the table.

**Table S3: Reactions and rate constants for model M3.**

| Reactions | $k_{on}$ ($\mu M^{-1}s^{-1}$) | $k_{off}$ ($s^{-1}$) | $K_D$ ($\mu M$) | $k_{cat}$ ($\mu M^{-1}s^{-1}$) |
|---|---|---|---|---|
| $Itk-Itk + PIP_3 \leftrightarrow Itk-Itk-PIP_3$ | $2.5 \times 10^{-4}$ | 0.1 | 400 | |
| $Itk-Itk + IP_4 \leftrightarrow Itk-Itk-IP_4$ | $2.5 \times 10^{-3}$ | 0.1 | 40 | |
| $PIP_3-Itk-Itk + IP4 \leftrightarrow PIP_3-Itk-Itk-IP_4$ | 0.1 | 0.003 | 0.03 | |
| $IP_4-Itk-Itk + PIP_3 \leftrightarrow IP_4-Itk-Itk-PIP_3$ | 0.01 | 0.003 | 0.3 | |
| $IP_4-Itk-Itk + IP4 \leftrightarrow IP_4-Itk-Itk-IP_4$ | 0.1 | 0.003 | 0.03 | |
| $Itk-Itk-PIP_3 + S \leftrightarrow IP_4 + Itk-Itk-PIP_3$ | | | | $1.5 \times 10^{-4}$ |
| $IP_4-Itk-Itk-PIP_3 + S \leftrightarrow IP_4 + IP_4-Itk-Itk-PIP_3$ | | | | $1.5 \times 10^{-4}$ |

All the high and low affinity binding of Itk PH domains with $PIP_3$ and $IP_4$, and, the $IP_4$ production rate are taken to be the same as that shown for model M1 (Table S1).

The ODEs and the conservation laws are same as Model 2.

ODEs for Model 4: The reactions for Model 4 are given in the table below. The corresponding ODEs are shown below the table.

**Table S4: Reactions and rate constants for model M4**

| Reactions | $k_{on}$ ($\mu M^{-1}s^{-1}$) | $k_{off}$ ($s^{-1}$) | $K_D$ ($\mu M$) | $k_{cat}$ ($\mu M^{-1}s^{-1}$) |
|---|---|---|---|---|
| $Itk + PIP_3 \leftrightarrow Itk - PIP_3$ | $2.5 \times 10^{-4}$ | 0.1 | 400 | |
| $Itk + IP_4 \leftrightarrow Itk^* - IP_4$ | $2.5 \times 10^{-3}$ | 0.1 | 40 | |
| $Itk^* - IP_4 + PIP_3 \leftrightarrow Itk^* - PIP_3 + IP_4$ | 10 (we used this value to get similar kinetics as given by Ref. (3)) | | | |
| $Itk^* - PIP_3 + S \rightarrow IP_4 + Itk^* - PIP_3$ | | | | $1.5 \times 10^{-4}$ |
| $Itk-PIP_3 + S \rightarrow IP_4 + Itk-PIP_3$ | | | | $1.5 \times 10^{-4}$ |

All the low affinity binding of Itk PH domains with $PIP_3$ and $IP_4$, and, the $IP_4$ production rate are taken to be the same as that shown for model M1 (Table S1).

$$\frac{dN_{Itk}}{dt} = k_{-1}N_{Itk-PIP_3} - k_1 N_{Itk} N_{PIP_3} + k_{-2} N_{Itk^*-IP_4} - k_2 N_{Itk} N_{IP_4}$$

$$\frac{dN_{Itk-PIP_3}}{dt} = k_1 N_{Itk} N_{PIP_3} - k_{-1} N_{Itk-PIP_3}$$

$$\frac{dN_{Itk^*-IP_4}}{dt} = k_2 N_{Itk} N_{IP_4} - k_{-2} N_{Itk^*-IP_4} + k_3 N_{Itk^*-PIP_3} N_{IP_4} - k_3 N_{Itk^*-IP_4} N_{PIP_3}$$

$$\frac{dN_{Itk^*-PIP_3}}{dt} = k_3 N_{Itk^*-IP_4} N_{PIP_3} - k_3 N_{Itk^*-PIP_3} N_{IP_4}$$

$$\frac{dN_{PIP_3}}{dt} = k_{-1}N_{Itk-PIP_3} - k_1 N_{Itk} N_{PIP_3} + k_3 N_{Itk^*-PIP_3} N_{IP_4} - k_3 N_{Itk^*-IP_4} N_{PIP_3}$$

$$\frac{dN_{IP_4}}{dt} = k_{-2}N_{Itk^*-IP_4} - k_2 N_{Itk} N_{IP_4} + k_3 N_{Itk^*-IP_4} N_{PIP_3} - k_3 N_{Itk^*-PIP_3} N_{IP_4}$$
$$+ k_4 \{N_{Itk-PIP_3} + N_{Itk^*-PIP_3}\} N_S$$

$$\frac{dN_S}{dt} = -k_4 \{N_{Itk-PIP_3} + N_{Itk^*-PIP_3}\} N_S$$

Here we have used Itk not Itk$_D$ to denote the monomeric Itk molecules. The conservation laws are,

$$N_{Itk} + N_{Itk-PIP_3} + N_{Itk^*-IP_4} + N_{Itk^*-PIP_3} = N_{Itk}(t=0)$$

$$N_{PIP_3} + N_{Itk^*-PIP_3} + N_{Itk-PIP_3} = N_{PIP_3}(t=0)$$

$$N_{IP_4} + N_{Itk^*-IP_4} + N_S = N_S(t=0)$$

ODEs for Model 5: The reactions for Model 5 are given in the table below. The corresponding ODEs are shown below the table.

**Table S5: Reactions and rate constants for model M5.**

| Reactions | $k_{on}$ ($\mu M^{-1} s^{-1}$) | $k_{off}$ ($s^{-1}$) | $K_D$ ($\mu M$) | $k_{cat}$ ($\mu M^{-1} s^{-1}$) |
|---|---|---|---|---|
| $Itk-Itk+PIP_3 \leftrightarrow Itk-Itk-PIP_3$ | $1.25 \times 10^{-4}$ | 0.05 | 400 | |
| $PIP_3-Itk-Itk+PIP_3 \leftrightarrow PIP_3-Itk-Itk-PIP_3$ | $1.25 \times 10^{-4}$ | 0.05 | 400 | |
| $Itk-Itk+IP_4 \leftrightarrow Itk-Itk-IP_4$ | $1.25 \times 10^{-3}$ | 0.05 | 40 | |
| $PIP_3-Itk-Itk+IP4 \leftrightarrow PIP_3-Itk-Itk-IP_4$ | $1.25 \times 10^{-3}$ | 0.05 | 40 | |

| Reactions | $k_{on}$ ($\mu M^{-1}s^{-1}$) | $k_{off}$ ($s^{-1}$) | $K_D$ ($\mu M$) | $k_{cat}$ ($\mu M^{-1}s^{-1}$) |
|---|---|---|---|---|
| $IP_4 - Itk - Itk + PIP_3 \leftrightarrow IP_4 - Itk - Itk - PIP_3$ | $1.25 \times 10^{-4}$ | 0.05 | 400 | |
| $IP_4 - Itk - Itk + IP4 \leftrightarrow IP_4 - Itk - Itk - IP_4$ | $1.25 \times 10^{-3}$ | 0.05 | 40 | |
| $Itk - Itk - PIP_3 + S \leftrightarrow IP_4 + Itk - Itk - PIP_3$ | | | | $1.5 \times 10^{-4}$ |
| $IP_4 - Itk - Itk - PIP_3 + S \leftrightarrow IP_4 + IP_4 - Itk - Itk - PIP_3$ | | | | $1.5 \times 10^{-4}$ |
| $PIP_3 - Itk - Itk - PIP_3 + S \leftrightarrow IP_4 + PIP_3 - Itk - Itk - PIP_3$ | | | | $1.5 \times 10^{-4}$ |

All the low affinity binding of Itk PH domains with $PIP_3$ and $IP_4$, and, the $IP_4$ production rate are taken to be the same as that shown for model M1 (Table S1).

The ODEs are same as model 1 but the rate constants are different.

ODEs for Model 6: The reactions for Model 6 are given in the table below. The corresponding ODEs are shown below the table.

**Table S6: Reactions and rate constants for model M6.**

| Reactions | $k_{on}$ ($\mu M^{-1}s^{-1}$) | $k_{off}$ ($s^{-1}$) | $K_D$ ($\mu M$) | $k_{cat}$ ($\mu M^{-1}s^{-1}$) |
|---|---|---|---|---|
| $Itk + PIP_3 \leftrightarrow Itk - PIP_3$ | $1.25 \times 10^{-4}$ | 0.05 | 400 | |
| $Itk + IP_4 \longleftrightarrow Itk - IP_4$ | $1.25 \times 10^{-3}$ | 0.05 | 40 | |
| $Itk - PIP_3 + S \rightarrow IP_4 + Itk - PIP_3$ | | | | $1.5 \times 10^{-4}$ |

All the low affinity binding of Itk PH domains with $PIP_3$ and $IP_4$, and, the $IP_4$ production rate are taken to be the same as that shown for model M1 (Table S1).

$$\frac{dN_{Itk}}{dt} = k_{-1}N_{Itk-PIP_3} - k_1 N_{Itk} N_{PIP_3} + k_{-2}N_{Itk-IP_4} - k_2 N_{Itk} N_{IP_4}$$

$$\frac{dN_{Itk-PIP_3}}{dt} = k_1 N_{Itk} N_{PIP_3} - k_{-1}N_{Itk-PIP_3}$$

$$\frac{dN_{Itk-IP_4}}{dt} = k_2 N_{Itk} N_{IP_4} - k_{-2}N_{Itk-IP_4}$$

$$\frac{dN_{PIP_3}}{dt} = k_{-1}N_{Itk-PIP_3} - k_1 N_{Itk} N_{PIP_3}$$

$$\frac{dN_{IP_4}}{dt} = k_{-2}N_{Itk-IP_4} - k_2 N_{Itk} N_{IP_4} + k_4 N_{Itk-PIP_3} N_S$$

$$\frac{dN_S}{dt} = -k_4 N_{Itk-PIP_3} N_S$$

The conservation laws are

$$N_{Itk} + N_{Itk-PIP_3} + N_{Itk-IP_4} = N_{Itk}(t=0)$$
$$N_{PIP_3} + N_{Itk-PIP_3} = N_{PIP_3}(t=0)$$
$$N_{IP_4} + N_{Itk-IP_4} + N_S = N_S(t=0)$$

ODEs for Model 7: The reactions for Model 7 are given in the table below. The corresponding ODEs are shown below the table.

**Table S7: Reactions and rate constants for model M7.**

| Reactions | $k_{on}$ ($\mu M^{-1}s^{-1}$) | $k_{off}$ ($s^{-1}$) | $K_D$ ($\mu M$) | $k_{cat}$ ($\mu M^{-1}s^{-1}$) |
|---|---|---|---|---|
| $Itk - Itk + PIP_3 \leftrightarrow Itk - Itk - PIP_3$ | $2.5 \times 10^{-4}$ | 0.01 | 40 | |
| $PIP_3 - Itk - Itk + PIP_3 \leftrightarrow PIP_3 - Itk - Itk - PIP_3$ | 0.01 | 0.1 | 10 | |
| $Itk - Itk + IP_4 \leftrightarrow Itk - Itk - IP_4$ | $2.5 \times 10^{-3}$ | 0.01 | 4 | |
| $PIP_3 - Itk - Itk + IP4 \leftrightarrow PIP_3 - Itk - Itk - IP_4$ | $2.5 \times 10^{-3}$ | 0.01 | 4 | |
| $IP_4 - Itk - Itk + PIP_3 \leftrightarrow IP_4 - Itk - Itk - PIP_3$ | 0.01 | 0.1 | 10 | |
| $IP_4 - Itk - Itk + IP4 \leftrightarrow IP_4 - Itk - Itk - IP_4$ | 0.1 | 0.1 | 1 | |
| $Itk - Itk - PIP_3 + S \leftrightarrow IP_4 + Itk - Itk - PIP_3$ | | | | $1.5 \times 10^{-4}$ |
| $IP_4 - Itk - Itk - PIP_3 + S \leftrightarrow IP_4 + IP_4 - Itk - Itk - PIP_3$ | | | | $1.5 \times 10^{-4}$ |
| $PIP_3 - Itk - Itk - PIP_3 + S \rightarrow IP_4 + PIP_3 - Itk - Itk - PIP_3$ | | | | $1.5 \times 10^{-4}$ |

All the high and low affinity binding of Itk PH domains with $PIP_3$ and $IP_4$, and, the $IP_4$ production rate are estimated to match the PLCg1 kinetics in Ref. (3).

The ODEs are same as model 1 but the rate constants (shown above) are different.
**Description of the models:**

Models M1-M4 and M7 contain $IP_4$ mediated feedbacks. Models M5-M6 lack those feedbacks. Models M4 and M6 contain monomeric Itk molecules. Models M1-M3, M5 and M7 contain dimeric Itk molecules. In model M1 (Fig. S1B), $PIP_3$ binding to one of the PH domains increases the affinity of the other PH domain for both $IP_4$ and $PIP_3$. M2, M3 and M7 (Fig. S1B) are variants of this model where $PIP_3$ binding to one PH domain in a dimer does not change the affinity of the other PH domain for either $IP_4$ or $PIP_3$ (M2), or allosterically increases the affinity of the other PH domain for $IP_4$ but _not_ $PIP_3$ (M3), or allosterically increases the affinity of the other PH domain for $PIP_3$ but _not_ $IP_4$ (M7). These models probed potential secondary interactions between Itk dimers and the membrane lipids. In the monomeric model, M4 (Fig. S1B), $IP_4$ binds the single Itk PH domain with a weak affinity and induces a conformational change (denoted by Itk* in table S4) that increases the affinity of this PH domain for both $PIP_3$ and $IP_4$. In the models lacking $IP_4$ mediated positive feedbacks, the PH domains of dimeric (model M5) or monomeric (model M6) Itk molecules bind both $PIP_3$ and $IP_4$ without an allosteric modification of the binding affinities. The models are summarized in Table I in the main text. We describe the reactions used for each of the models in the tables S1-S7. In all models we make the following assumptions to make the models simpler while preserving important general features.

**Assumptions (M1-M7).**

1. The initial concentration of the molecular species LAT bound Itk, is referred to as $Itk^0$, and, has been used as a surrogate of signal strength given by antigen dose or antigens of different affinities. The stronger the stimulation, the larger is the value of $Itk^0$.
2. The details of signal dependent recruitment of PI3K and consequent production of $PIP_3$ by PI3K have not been considered. $PIP_3$ is produced within seconds after stimulation (4-6). Hence, we started our simulations with an initial concentration of $PIP_3$, referred to as $PIP_3^0$ from now on. For a given strength of stimulation (i.e. binding of a peptide of a particular affinity), the concentration of total $PIP_3$ (bound plus unbound) has been held fixed to the initial $PIP_3$ concentration for the entirety of the simulation. The number of $PIP_3^0$ is decreased as the peptide affinities are decreased in order to mimic decreasing PI3K-mediated $PIP_3$ production with decreasing intensity of TCR engagement. For example, we have used $PIP_3^0$ = 530 molecules for OVA stimulation but only 50 molecules of $PIP_3^0$ for G4 stimulation.
3. As PLCγ1 is the immediate downstream effector of $PIP_3$ bound Itk, we have assumed that the kinetics of the concentration of $PIP_3$ bound Itk directly represents PLCγ1 activation kinetics.
4. A 5-phosphatase, responsible for the turnover of $IP_4$, is absent in our models. Following TCR ligation, the experimentally determined $IP_4$ level in thymocytes/ T cells increases until it reaches its peak at around 10 mins (7). Then owing to the 5-phosphatase activity, $IP_4$ levels decrease, decaying to half of its peak value at around 30 mins (7). Activation of PLCγ1 on the other hand is much faster, peaking at around 1 to 2 mins (main text, (3)). Based on these data, we believe that the role of 5-phosphatase, if any, in the activation of PLCγ1, is insignificant.

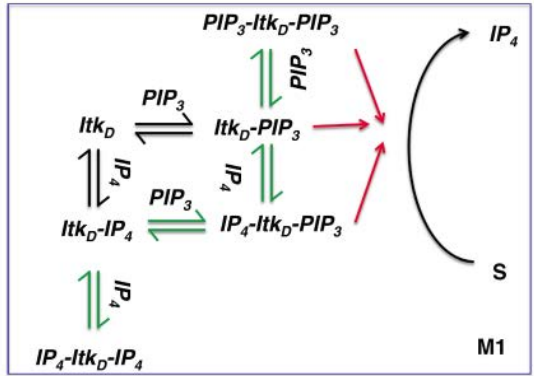
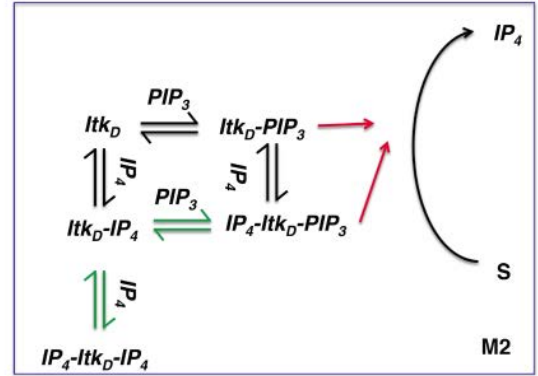
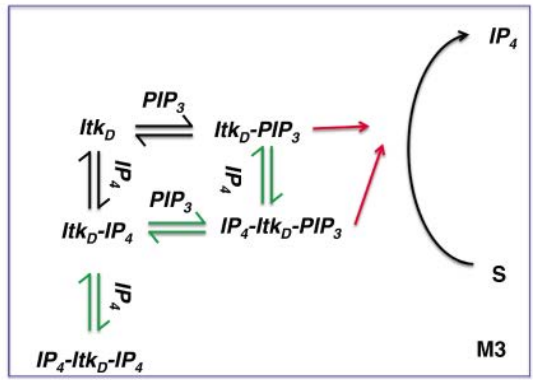
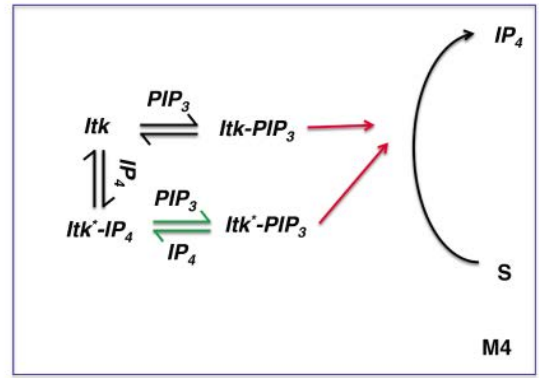
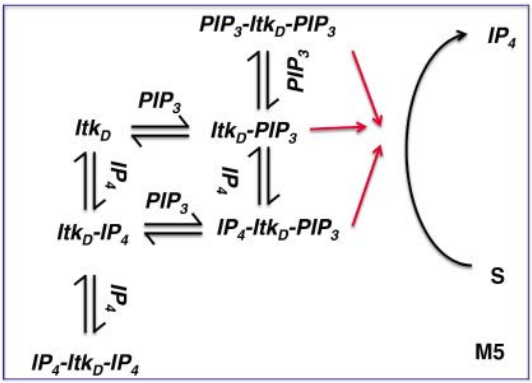
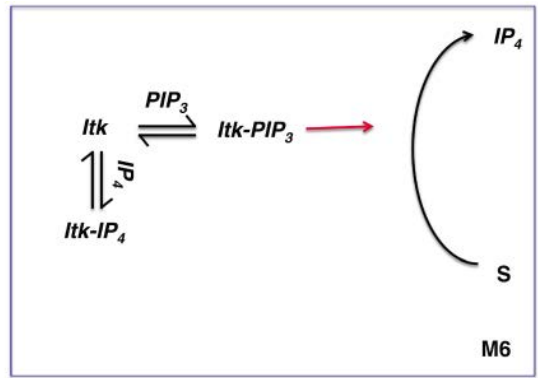
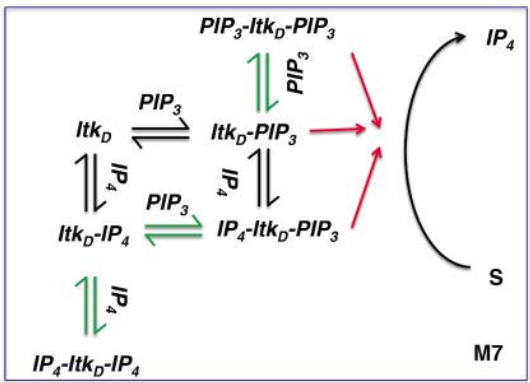

**Fig. S1B: Graphical networks describing the signaling reactions in models M1-M7.** Itk shown in this figure represents an Itk molecule that is bound to the TCR and LAT signalosome (not shown). High affinity binding reactions are shown as green arrows. $PIP_2$ hydrolysis into DAG and $IP_3$ which ultimately produces $IP_4$ (S) is shown as red arrows. **(M1)** In model M1, both $IP_4$ and $PIP_3$ can equally induce allosteric modifications of the PH domains in Itk dimers. **(M2)** Model M2. Similar to M1, however, modification of the PH domains by $PIP_3$ cannot stabilize $IP_4$ or $PIP_3$ binding to the Itk PH domains. **(M3)** Model M3. Similar to M1, however, modification of the PH domains by $PIP_3$ can only stabilize $IP_4$ but not $PIP_3$ binding to the Itk PH domains. **(M4)** Model M4. The Itk PH domains are monomeric and unable to interact allosterically. $IP_4$ or $PIP_3$, upon binding with a weak affinity, instantaneously changes Itk to a high affinity conformation ($Itk^*$) where $IP_4$ (or $PIP_3$) can replace PH domain bound $PIP_3$ (or $IP_4$) with high affinity. **(M5)** Model M5. Both $IP_4$ and $PIP_3$ bind to the PH domains of the Itk dimer with low affinity. No allosteric modification occurs. **(M6)** Model M6. Similar to model M5 but Itk exists only in monomers. **(M7)** Model M7. Similar to M1, however, modification of the PH domains by $PIP_3$ can only stabilize $PIP_3$ but not $IP_4$ binding to the Itk PH domains.

**Table S8: Values of the concentrations of different molecular species used in the models.**

| Molecules | Number | Comments |
|---|---|---|
| $PIP_3^0$ | Varied from 50-530 | Roughly 5% of the available $PIP_2$ pool (8) is taken to be the upper limit of $PIP_3$ concentration. In a separate measurement, the $PIP_3$ concentration reaches to about 150-200 µM in neutrophils, 10 seconds after stimulation (6). |
| Itk-Itk$^0$ (for dimers)/ Itk$^0$(for monomers) | Varied from 20-300 | The upper limit of Itk is assumed to be the upper limit of phosphorylated LAT in thymocytes (9). |
| $S^0$ | 17000 | 3.5 mM, 10 seconds after stimulation in neutrophils (6). |
| $IP_4$ | We do not have any basal level of $IP_4$ in the models. $IP_4$ is generated via the | The $IP_4$ level in Jurkat T-lymphocytes increased to $1125 \pm 125$ pmol/$10^9$ |

| | cleavage of PIP$_2$ (S). | cells after stimulation by anti-CD3 antibody OKT3 (7). This number, when converted to molecules/cell, is roughly two times the number we have used as an upper limit for IP$_4$ (i.e. the initial PIP$_2$ concentration) in our simulations. |
|---|---|---|

**Unit Conversion Table**

$1\mu M = 600$ molecules/$(\mu m)^3$
$k_{on}^{3D} = 1\ (\mu M^{-1} s^{-1}) = 0.16 \times 10^{-2}\ (\mu m)^3/$ (molecules) $s^{-1}$

Note that the reactions involving two plasma membrane bound complexes take place only at the plasma membrane. Hence we have to convert our $k_{on}^{3D}$ to the corresponding $k_{on}^{2D}$ in order to describe a binding reaction on a plane. This is done by dividing the $k_{on}^{3D}$ by the lengths $l = 2$ nm (Fig. S1A).

**Effects of intrinsic noise fluctuations**

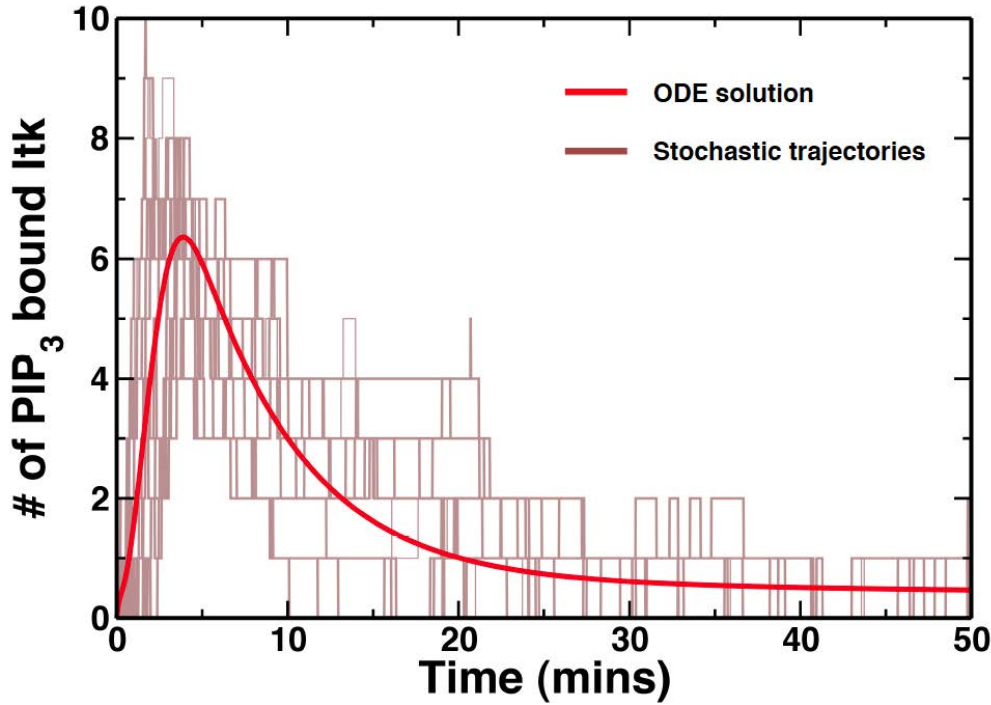

**Fig. S2: Presence of Intrinsic fluctuations does not lead to qualitatively different temporal profiles as compared with the deterministic model.** We show 11 different stochastic trajectories for $Itk^0 = 20$ molecules and $PIP_3^0 = 50$ molecules, the lowest concentration used in our simulations, for model M3. The stochastic trajectories for concentrations of $PIP_3$ bound Itk were obtained by solving the Master equation associated with the signaling reactions (Table S3) using the Gillespie algorithm (10). The curve in red is the solution of the mass action kinetics given by a set of ODEs. We use the same kinetic rates and initial concentrations for the stochastic simulations and the ODEs. The above figure shows that the stochastic trajectories spread around the solution of the ODEs (shown in red). In the next figure (Fig. S3) we show how the ODE solution compares with stochastic trajectories when averaged over a small number of *in silico* "cells". The smaller the difference between the two, the more accurate the ODEs are in describing the kinetics for the cell population, even in the presence of intrinsic stochastic fluctuations.

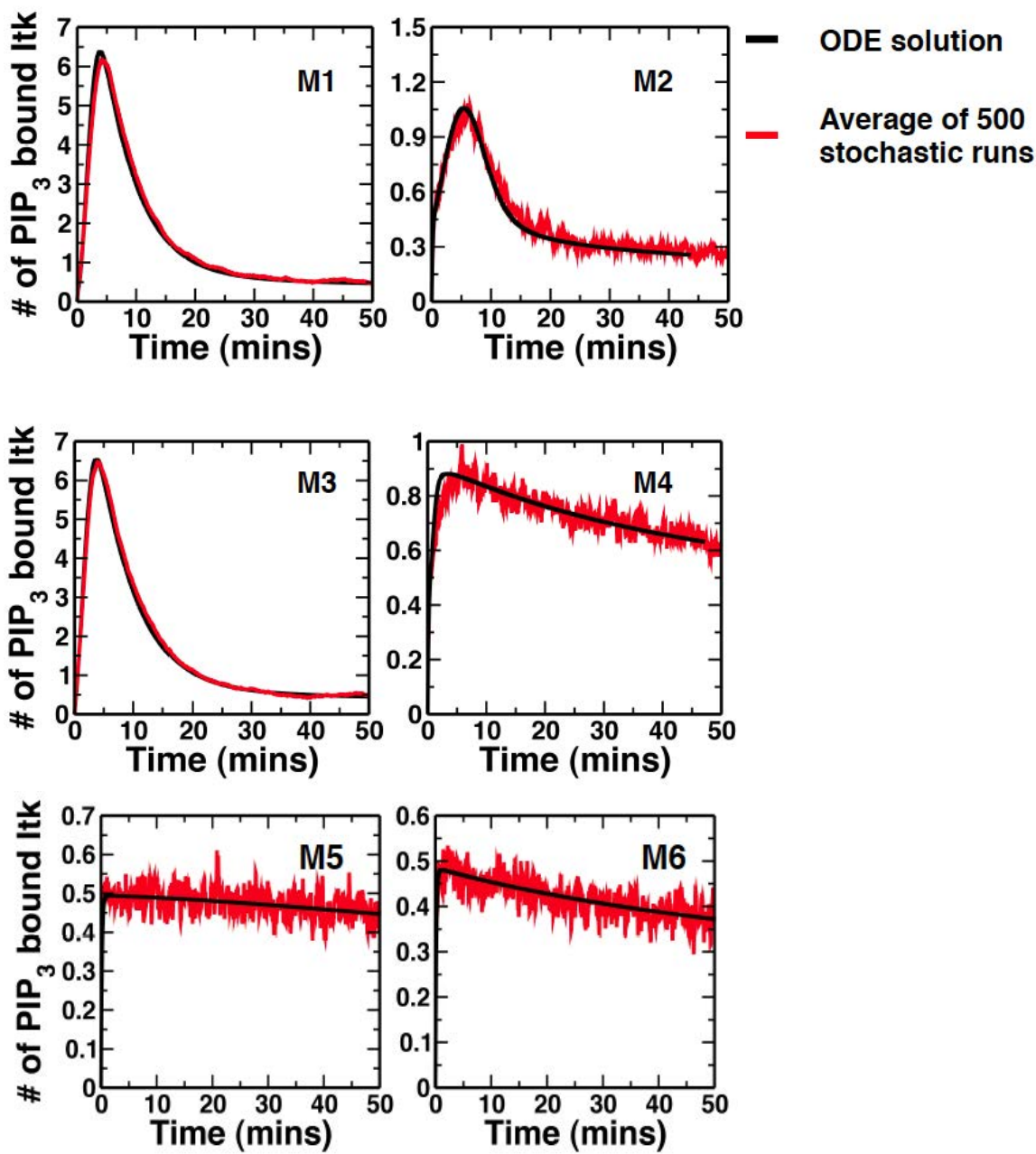

**Fig. S3: Comparison between the ODE solutions and the stochastic trajectories averaged over a small number of cells.** We compared the temporal profiles of concentrations of PIP$_3$ bound Itk obtained in simulations including stochastic copy number variations due to intrinsic noise fluctuations (red) with the solutions of the deterministic mass action reaction kinetics that ignored such fluctuations (solid black lines). The stochastic simulations were carried out by using Gillespie's method (10) which provided exact numerical solution of the Master equations associated with the models. We used the same rate constants and initial concentrations for the stochastic simulations and ODE solutions. The kinetic trajectories were averaged over 500

realizations (or *in silico* "cells") for the stochastic simulations. We show the results for the smallest concentrations of $Itk^0$ (20 molecules) and $PIP_3^0$ (50 molecules) where the effect of the stochastic fluctuations is expected to be the largest. We observed that for all the models the ODE solutions produce qualitatively similar shapes as the average stochastic trajectories. Model M7 is not shown.

**Variation of the peak value (A) as the initial concentrations of Itk and $PIP_3$ were increased in models M1 to M7.**

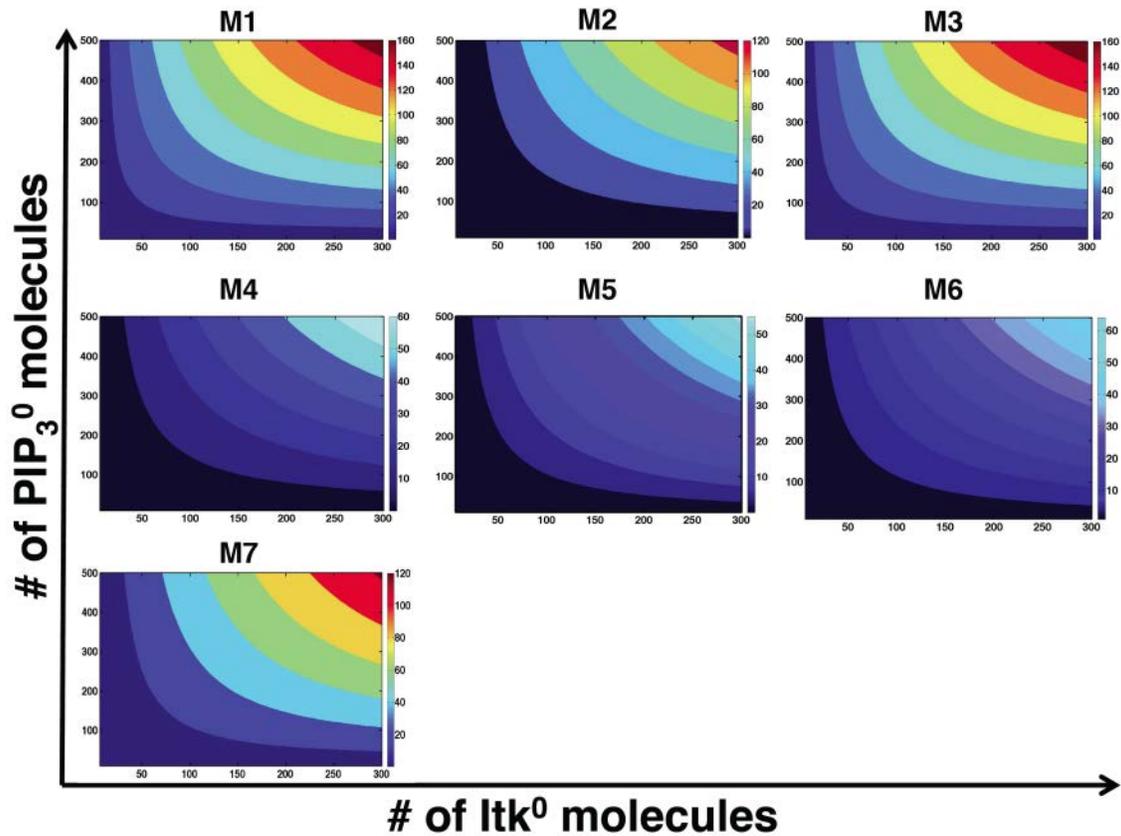

**Fig. S4: Variation of the peak value (A) with $Itk^0$ and $PIP_3^0$ for all seven models.** The peak value of $PIP_3$ bound Itk increased in a graded manner with increasing initial Itk and $PIP_3$ concentrations.

## The variation of peak time ($\tau_p$) as initial concentrations of Itk and PIP$_3$ were increased in models M1 to M7.

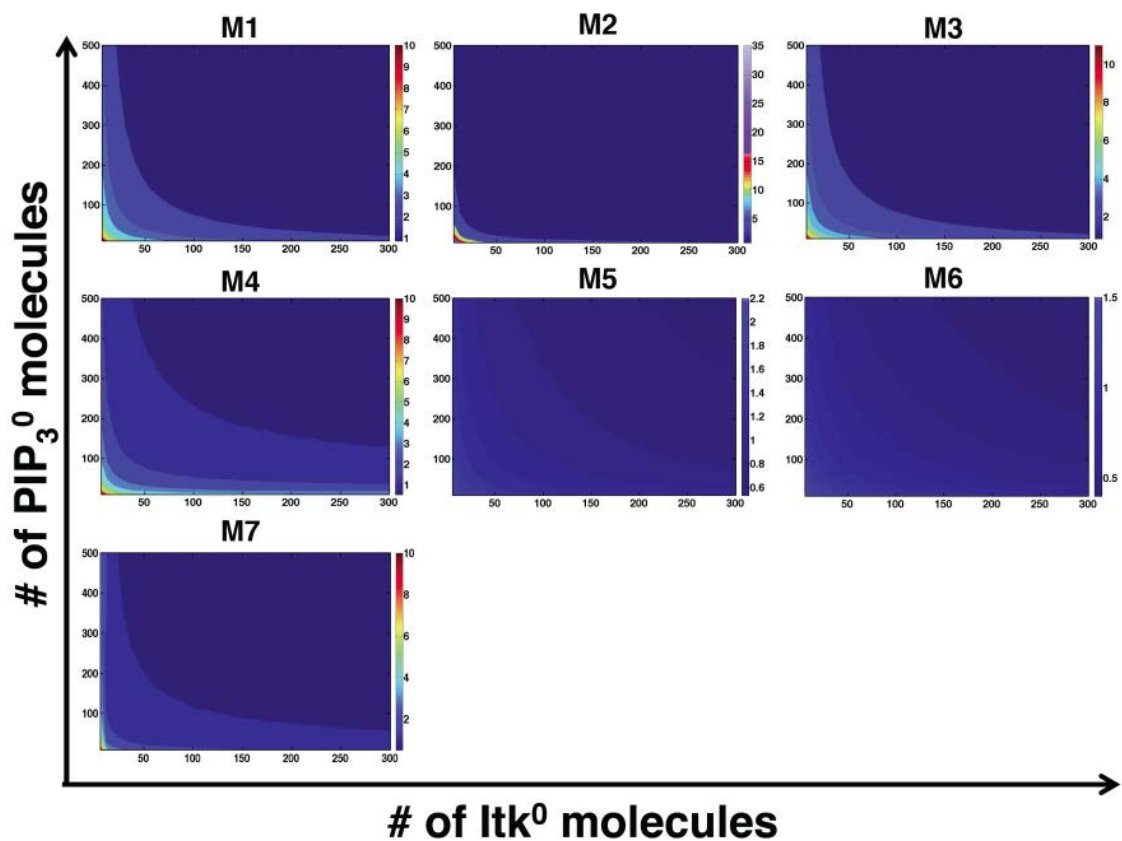

**Fig. S5: Variation of $\tau_p$ with Itk$^0$ and PIP$_3^0$ for all six models.** The peak time ($\tau_p$) of the temporal profile of the concentration of PIP$_3$ bound Itk varied by an order of magnitude (roughly from 1 min to 10 mins) in models M1-M4 and M7, while the peak time did not change appreciably in models M5 and M6 over the entire range of variation. However, $\tau_p$ did not vary appreciably over a large range of initial Itk (>100) and PIP$_3$ concentrations (>150) even in the models M1-M4 and M7. Most of the large variations occurred at small concentrations of Itk and PIP$_3$. We analyzed the above behavior using simple models as described in the following section.

## Understanding mechanisms that regulate of the shape of the temporal profiles of PIP$_3$ bound Itk

We constructed simpler models that effectively describe M1-M4 to analyze the effect of the feedbacks and the binding-unbinding reactions in controlling the shape of the kinetics of PIP$_3$ bound Itk. The simpler models could be analytically tractable under certain conditions which allowed us to characterize the dependence of kinetics on the reaction rates and concentrations. Such calculations are usually very useful in gleaning mechanistic understanding into the system.

Effect of the positive feedback: We aimed to understand how the positive feedback controls $\tau_p$ as the initial concentrations of Itk ($Itk^0$) and PIP$_3$ ($PIP_3^0$) were varied. In models M1-M4, multiple reactions occurring at different time scales work in concert to create the positive feedback which in essence increases the binding affinity of Itk PH domains for PIP$_3$. Therefore, in order to analyze the initial concentration dependence of the peak time, $\tau_p$, we constructed an effective binding unbinding reaction between Itk and PIP$_3$, where, the reaction rates ($k_1$ and $k_{-1}$) are initial concentration dependent.

$$Itk + PIP_3 \underset{k_{-1}}{\overset{k_1}{\longleftrightarrow}} Itk - PIP_3 \qquad \text{Reaction 1}$$

The concentration dependence in the above reaction rates can arise because the effective reaction captures the kinetics of binding of Itk to PIP$_3$ in models M1-M4, where multiple second order reactions associated with different time scales induce positive feedback interactions between Itk and PIP$_3$. In order to compare reaction 1 with the effect of only the positive feedback, we removed the negative feedback interactions from models M1-M4. Therefore, in all the models, the concentration of Itk bound PIP$_3$ reached a non-zero concentration at the steady state (Fig. S6A, black curve). Then we estimated the effective rates, $k_1$ and $k_{-1}$ that will produce similar kinetics (same $\tau_{1/2}$ and the same steady state) (Fig. S6B, blue curve) following the scheme below. The kinetics of $x =$ Itk – PIP$_3$ in Reaction 1 is given by,

$$\frac{dx}{dt} = k_1(Itk^0 - x)(PIP_3^0 - x) - k_{-1}x = k_1\left[(Itk^0 - x)(PIP_3^0 - x) - K_D x\right] \qquad (1)$$

where, $Itk^0$ and $PIP_3^0$ denote the initial concentrations of Itk and PIP$_3$, respectively, and, $K_D = k_{-1}/k_1$. The solution of the above equation is,

$$x(t) = \frac{x_s^+ x_s^- \left(1 - e^{k_1(x_s^+ - x_s^-)t}\right)}{x_s^- - x_s^+ e^{k_1(x_s^+ - x_s^-)t}} \qquad (2)$$

$x_s^+$ and $x_s^-$ being the two steady state (one stable another unstable) solutions given by

$$x_s^\pm = \frac{Itk^0 + PIP_3^0 + k_D \pm \sqrt{\left(Itk^0 + PIP_3^0 + K_D\right)^2 - 4 Itk^0 PIP_3^0}}{2} \qquad (3)$$

The time ($\tau_{1/2}$) taken by $x$, to reach the half of the steady state (the stable fixed point in Eq. (3)) concentration is given by,

$$\tau_{1/2} = \frac{1}{k_1(x_s^+ - x_s^-)} Log\left(\frac{2x_s^+ - x^-}{x_s^+}\right)$$
(4)

For a particular set of initial concentrations, we calculated the steady state concentration of $PIP_3$ bound Itk, and, $\tau_{1/2}$, by numerically solving the corresponding ODEs for models M1-M4 with the negative feedbacks being turned off. Then using Eqns. (3) and (4), we estimated the rate constants, $k_1$ and $K_D$ (or, equivalently, $k_1$ and $k_{-1}$) for each set of initial concentrations (Figs. S7 and S8). Both $k_1$ and $K_D$ varied with initial concentrations of Itk and $PIP_3$. However, $K_D$ did not change appreciably with concentrations for M1 and M3 as compared to M4 or M2 (Fig. S7). M1 and M3 showed qualitatively similar variations in $K_D$ and $k_1$ with increasing initial concentrations. This demonstrates a large degree of similarity between the models. For models M1-M3, the values of $K_D$ in the effective binding-unbinding reaction are substantially smaller (<100 times) to bare Itk and $PIP_3$ interaction ($K_D = 2000$) used in tables S1-S3, in the absence of any $IP_4$ feedback. This again demonstrates that the feedback reactions convert the low affinity interactions between Itk and $PIP_3$ to a high affinity binding unbinding reaction.

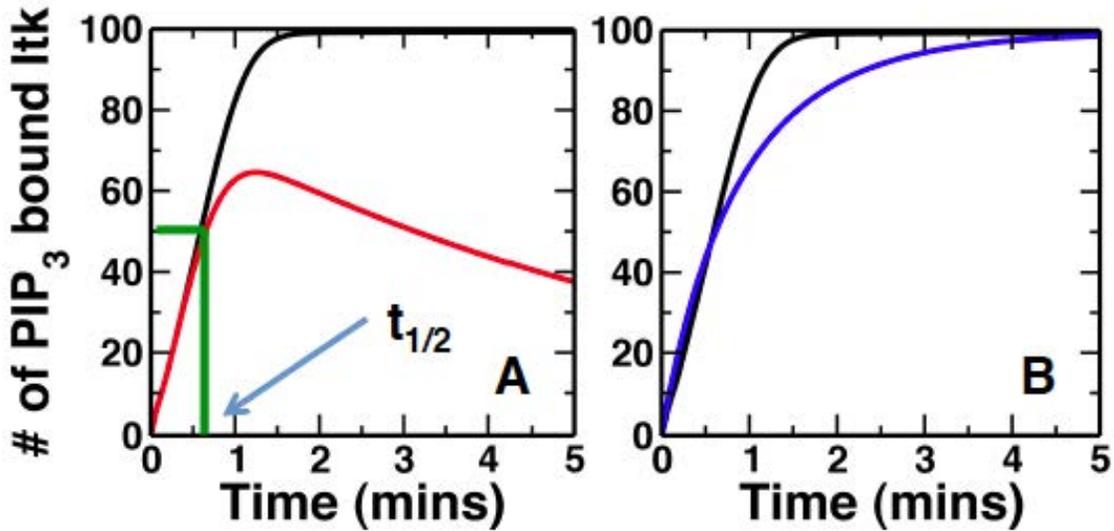

**Fig. S6: Estimation of the reaction rates in the effective binding-unbinding reaction.** A) The transient kinetics of $PIP_3$ bound Itk in M1 (red) is compared with the case when the negative feedback is removed (black). We use $\tau_{1/2}$ and the steady state concentration of the kinetics of $PIP_3$ bound Itk in the absence of the negative feedback to calculate the rates in the effective binding-unbinding reaction 1. B) Kinetics of $PIP_3$ bound Itk in the absence of the negative feedback in model M1 (black). Blue, kinetics of $PIP_3$ bound Itk in the corresponding binding unbinding process (reaction 1) where the $\tau_{1/2}$ and the steady state concentration of $PIP_3$ bound Itk is exactly the same as the black curve.

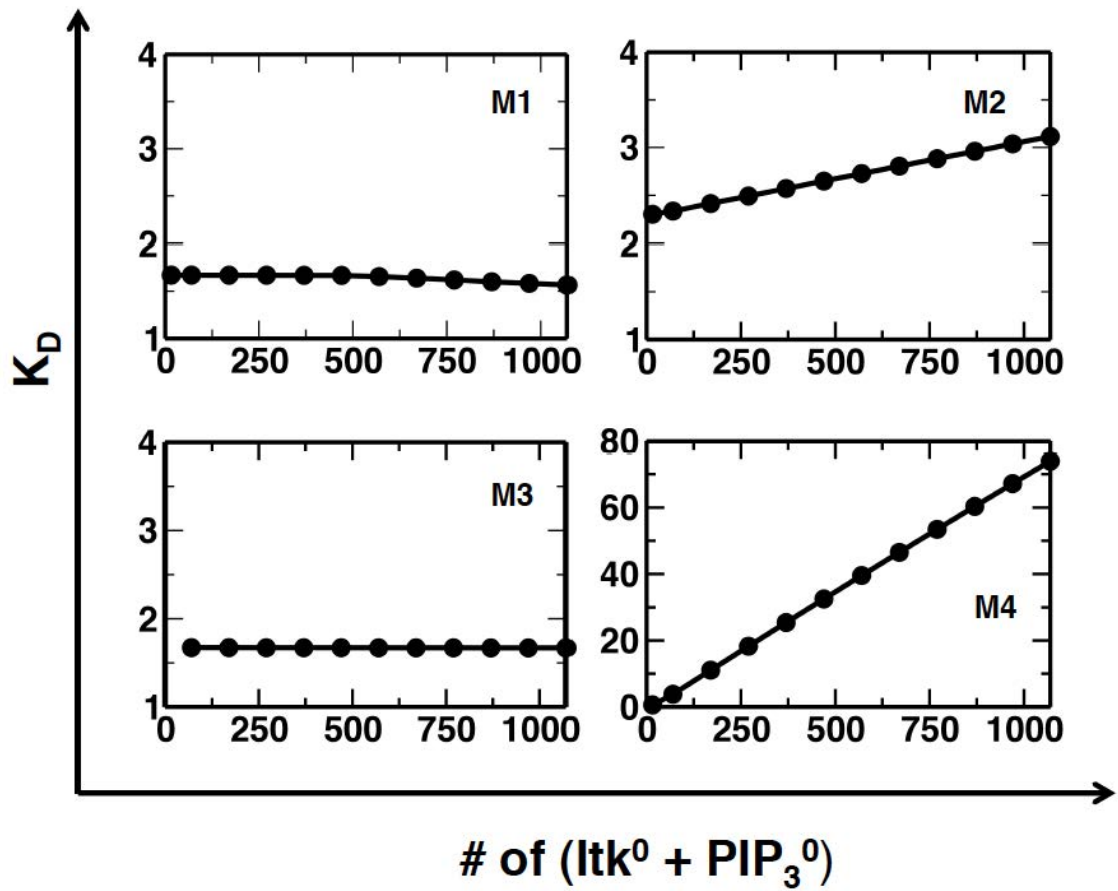

**Fig. S7. Variation of $K_D$ as a function of the sum of $Itk^0$ and $PIP_3^0$ for models M1 to M4.** The $K_D$ for the binding unbinding process has been estimated using the steady state values of the Itk kinetics in presence of the positive but not negative feedback. For models M1-M3, $K_D$ does not change significantly with increasing concentrations of initial Itk and $PIP_3$. The value of $K_D$ is much smaller than the sum of $\left(Itk^0 + PIP_3^0\right)$ as well. For M4 however, $K_D$ increases significantly (by an order of magnitude). The absolute value of the $K_D$ is still a lot less than $\left(Itk^0 + PIP_3^0\right)$.

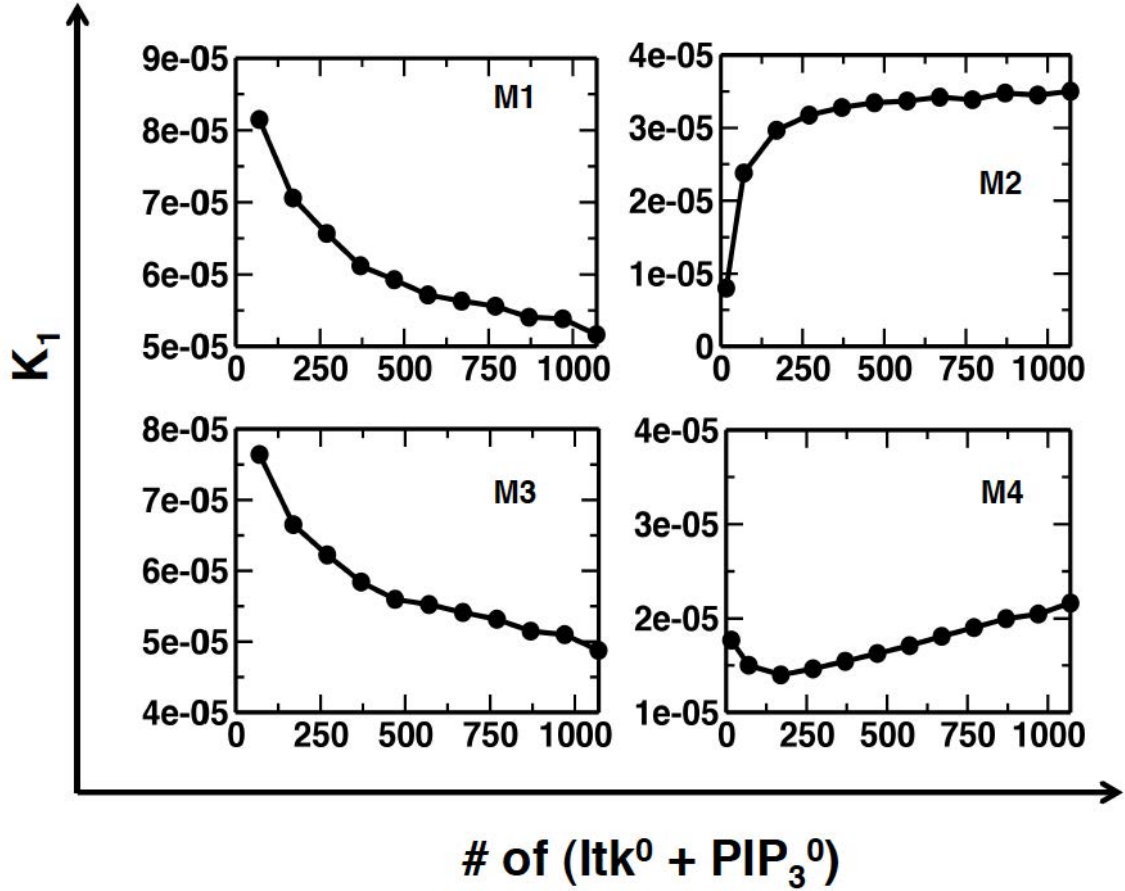

**Fig. S8: Variation of $k_1$ as a function of the sum of $Itk^0$ and $PIP_3^0$ for models M1 to M4.** $k_1$ decreased roughly 2 fold with the increase in $Itk^0$ and $PIP_3^0$ for M1 and M3, while, for model M2, $k_1$ increased 4 times. In M4, $k_1$ did not change appreciably.

Next we analyzed the concentration dependence of $\tau_{1/2}$ in the effective binding-unbinding reaction. This provided us with an estimate of concentration dependence of $\tau_p$ in model M1-M4, when the negative feedbacks do not contribute appreciably to $\tau_p$. When, $K_D \ll Itk^0 + PIP3^0$, as in Fig.S7, the Eqn (4) can be well approximated by,

$$\tau_{1/2} = \frac{1}{k_1(PIP_3^0 - Itk^0)} \log\left(\frac{2PIP_3^0 - Itk^0}{PIP_3^0}\right) \tag{5}$$

For the range of concentrations of $Itk^0$ and $PIP_3^0$ that we have considered in the simulation of our models,
$\log\left(\frac{2PIP_3^0 - Itk^0}{PIP_3^0}\right) \sim constant$, therefore,

$$\tau_{1/2} \simeq \frac{1}{k_1(PIP_3^0 - Itk^0)} \tag{6}$$

When $k_1$ does not depend on concentrations, a tenfold increase in $PIP_3^0 - Itk^0$, will lead to a tenfold decrease in $\tau_{1/2}$. However, our calculations showed that the effective $k_1$ also changes with $Itk^0$ and $PIP_3^0$ (Fig. S8). E.g., $k_1$ decreased roughly two fold in models M1 and M3 (Fig. S8) as $Itk^0$ and $PIP_3^0$ were increased, implying that when $PIP_3^0 - Itk^0$ is increased ten times, the decrease in $\tau_{1/2}$ will be roughly 5 times instead of 10 times. For M2, however, owing to the four times increase in $k_1$, $\tau_{1/2}$ will decrease twenty fold for the same increase in $PIP_3^0 - Itk^0$. This is similar to what we observe in Fig S5 for the variation of $\tau_p$ with initial PIP$_3$ and Itk concentrations. For M1 and M3, the concentration dependence in $k_1$ actually restricts the variation in $\tau_{1/2}$ (~5 times), while in M2 it helps in the variation in $\tau_{1/2}$ (~20 times).

For model M4, $K_D$ showed a monotonic increase with increasing $Itk^0$ and $PIP_3^0$ (Fig. S7). However, values of $K_D$ are much smaller than $(Itk^0 + PIP_3^0)$ for the range of concentrations we considered, therefore, we can still use Eq. (6) to estimate $\tau_{1/2}$. In contrast, for model M4, $k_1$ did not vary appreciably with $Itk^0$ and $PIP_3^0$, therefore, the variation in $\tau_{1/2}$ is largely determined by the change in with $Itk^0$ and $PIP_3^0$ as given by Eq. (6). This is reflected in the dependence of $\tau_p$ on initial Itk and PIP$_3$ concentrations.

Models (M5 and M6) lacking positive feedbacks:
For models M5 and M6, when the negative feedbacks are turned off, the effective binding unbinding reaction can represent the kinetics of Itk – PIP$_3$ with constant $K_D$ and $k_1$ for all concentrations. The estimated $K_D$ values for the effective reaction were much larger ($K_D$~2000) compared to the values of $Itk^0$ and $PIP_3^0$ we considered in the simulations, i,e, $K_D \gg (Itk^0 + PIP_3^0)$. Therefore, in this situation $\tau_{1/2}$ can be approximated by

$$\tau_{1/2} \simeq \frac{1}{k_1 K_D} = \frac{1}{k_{-1}}. \tag{7}$$

where, $\tau_{1/2}$, is solely determined by the unbinding rate and does not show any concentration dependence. This is in agreement with the results (Fig. S5) for models M5-M6.

**Dependence of the decay time $\tau_d$ of PIP$_3$ bound Itk on the initial concentrations of Itk and PIP$_3$ due to the negative feedbacks**

The decay time $\tau_d$ characterizes the time scale for the decay of the concentration of PIP$_3$ bound Itk from its peak value as IP$_4$ molecules outnumber PIP$_3$ molecules. We defined $\tau_d$ as the difference of the time (t$_2$) taken to decay to the half maximum value after the peak value is reached and the peak time, $\tau_p$. The dependence of $\tau_d$ on the initial concentrations of Itk and PIP$_3$ manifests in the variations of the asymmetry ratio R with increasing Itk$^0$ and PIP$_3^0$ as shown in Fig. 2 in the main text. We aim to characterize the concentration dependence of $\tau_d$ for the different models in this section.

Feedback models (M1-M3) with Itk dimers:
Owing to the strong positive feedback, most of the PIP$_3$ bound Itk molecules exist in PIP$_3$ – Itk – Itk – IP$_4$ heterodimers. Itk is sequestered into the cytosol via the reactions inducing negative feedbacks in the system as a result of formation of the soluble complex IP$_4$ – Itk – Itk – IP$_4$. This complex is produced by reactions occurring via two channels:

Channel I

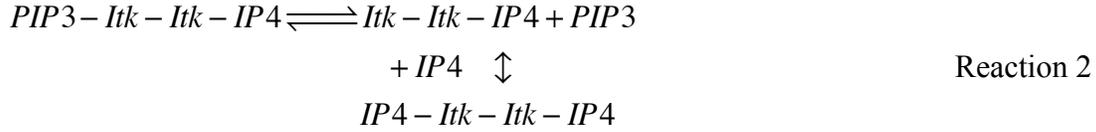

$$PIP3 - Itk - Itk - IP4 \rightleftharpoons Itk - Itk - IP4 + PIP3$$
$$+ IP4 \updownarrow$$
$$IP4 - Itk - Itk - IP4$$

Reaction 2

Channel II

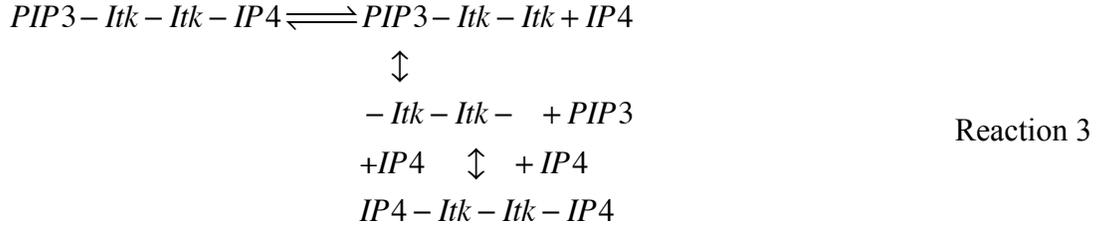

$$PIP3 - Itk - Itk - IP4 \rightleftharpoons PIP3 - Itk - Itk + IP4$$
$$\updownarrow$$
$$- Itk - Itk - \quad + PIP3$$
$$+IP4 \quad \updownarrow \quad + IP4$$
$$IP4 - Itk - Itk - IP4$$

Reaction 3

When the concentration of IP$_4$ is much larger than that of PIP$_3^0$, formation of the soluble IP$_4$ – Itk – Itk – IP$_4$ is more likely to occur through channel I, because, in channel II when IP$_4$ unbinds from PIP$_3$ – Itk – Itk – IP$_4$, the complex, Itk – Itk – PIP$_3$ is quickly transformed back to PIP$_3$ – Itk – Itk – IP$_4$, due to the presence of large number of IP$_4$ molecules. The rate of change of $x = IP_4 - Itk - Itk - PIP_3$ can be described by

$$\frac{dx}{dt} = (\tilde{k}_1.[IP_4].[Itk - Itk - PIP_3] - \tilde{k}_{-1}x) + (\tilde{k}_1.[Itk - Itk - IP_4].[PIP_3] - \tilde{k}_{-1}x) \quad (8)$$

,where, $\tilde{k}_1$ and $\tilde{k}_{-1}$ are high affinity binding unbinding rates of IP$_4$ and PIP$_3$ to the Itk PH domains. The terms in the first and the second parentheses in the right hand side describe the binding unbinding reactions in channel I and II, respectively. We have considered the rates to be the same for IP$_4$ and PIP$_3$, therefore, the above reaction is more appropriate for models M1 and M3. However, the general conclusions drawn in this calculation will apply for M2 as well. As argued above, when concentration of IP$_4$ is much larger than that of PIP$_3$, the first set of binding unbinding reactions in channel II occur at faster time scales, and $\tilde{k}_1.[IP_4].[Itk - Itk - PIP_3] \approx \tilde{k}_{-1}x$. In addition, a large number of IP$_4$ molecules quickly convert the unstable $- Itk - Itk - IP_4$ complexes into stable complexes (reactions in channel I). Therefore, we can write down the following inequality,
$\tilde{k}_1.[Itk - Itk - IP_4][PIP_3] \ll \tilde{k}_1.[IP_4].[Itk - Itk - PIP_3] \approx \tilde{k}_{-1}x$. Keeping in mind
$\tilde{k}_1.[Itk - Itk - IP_4][PIP_3] \ll \tilde{k}_1.[Itk - Itk - IP_4][IP_4]$, $x$ in Eq. (8) can be approximated as,

$$\frac{dx}{dt} = -\tilde{k}_{-1} x \tag{9}$$

Therefore, in this situation, the decay time does not appreciably depend on the initial concentrations of Itk and $PIP_3$, and is determined by the unbinding rate of $PIP_3$ from the $PIP_3 - Itk - Itk - IP_4$ complex (Fig. S9).

The above results change for M2, as $IP_4$ binding does not stabilize binding of $PIP_3$ to the Itk dimers. Therefore, the reactions in channel II play a greater role in the decay of $x$, and consequently, the decay time scales in M2 vary appreciably with the initial concentrations of Itk and $PIP_3$ (Fig. S9).

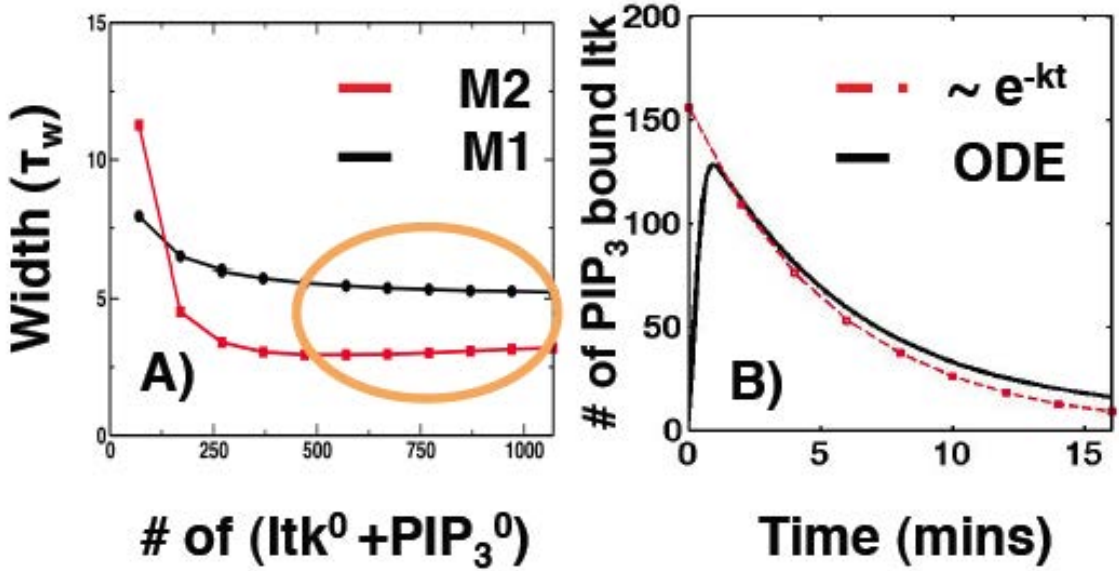

**Fig. S9: The saturation of the width in the feedback models**. A) We have varied both $Itk^0$ and $PIP_3^0$ such that $PIP_3^0 \geq Itk^0$. The plot of the width of $PIP_3$ bound Itk as a function of $(Itk^0 + PIP_3^0)$ is shown for M1 (black line) and M2 (red line). For large values of ($Itk^0+PIP_3^0$) the width saturates (the orange oval) both for M1 and M2. For M2 however the rate of decay of the width of $Itk - PIP_3$ kinetics is much faster than for M1 as can be seen from the fact that the red curve decays from roughly 12 mins to 3 mins where as the black curve goes down from 7 mins to 5 mins. B) The transient activation kinetics of the membrane bound Itk in M3 are shown in black. $PIP_3^0 = 500$, $Itk^0 = 200$. The dotted red curve is the exponential decay curve of the form $e^{-kt}$ with the time constant equal to the inverse of the high affinity $PIP_3$ unbinding rate, which validates the claim of Eqn (9). The same holds true for M1. The saturation of width ($\tau_w$) of the kinetics of $PIP_3$ bound Itk

concentration would be useful to understand the dependence of the ratio R= $\tau_w/\tau_p$ on initial concentrations of Itk and PIP$_3$.

Models lacking feedbacks (M5-M6):

Since both IP$_4$ and PIP$_3$ bind to Itk PH domains with a low affinity in M5 and M6, a large excess of IP$_4$ is required to sequester Itk into the cytosol. We quantify the amount of IP$_4$ that is required to effectively sequester Itk into the cytosol below.

Let us consider model M6. The rates of change of $x = Itk - PIP_3$ and $y = Itk - IP_4$ are given by

$$\frac{dx}{dt} = k_1[Itk^0 - x - y].[PIP_3^0 - x] - k_{-1}x$$
$$\frac{dy}{dt} = k_1[Itk^0 - x - y].[IP_4] - k_{-1}y \qquad (10)$$

, where, $Itk^0$ and $PIP_3^0$ denote the initial concentrations of Itk and PIP$_3$, respectively. Using, $IP_4 + y_s = S^0$, where, $S^0$ denotes the initial concentration of PIP$_2$, we find that at the steady state,

$$\frac{[PIP_3^0 - x_s]}{[S^0 - y_s]} = \frac{x_s}{y_s} \quad \Rightarrow \quad \frac{x_s}{y_s} = \frac{PIP_3^0}{S^0} \qquad (11)$$

$x_s$ can be calculated from the above equations,

$$x_s = \frac{Itk^0 + PIP_3^0 + k_D + S^0 - \sqrt{(Itk^0 + PIP_3^0 + k_D + S^0)^2 - 4.\left(\frac{S^0}{PIP_3^0} + 1\right)Itk^0.PIP_3^0}}{2.\left(\frac{S^0}{PIP_3^0} + 1\right)}$$

(12)

In order to get a finite $\tau_w$, the kinetic of Itk-PIP$_3$ has to decay to half of its peak value (A) i.e., $x_s(S^0) < A/2$, for a given K$_D$, PIP$_3^0$ and Itk$^0$. It is however hard to analytically write down a closed form of A. Instead, we can use $x_s(S^0=0)$ as an approximate upper bound of A (A$_{max}$). The reason being, for S$^0$=0 i.e., when PIP$_2$ concentration is zero, we recover the steady state for the binding unbinding process in absence of any competition from IP$_4$. This will represent the largest value A can ever attain. We used this as our approximation for the peak value (A) of PIP$_3$ recruited Itk in M6. The variation of $x_s$ as a function of $S^0$ for two separate K$_D$ is shown below.

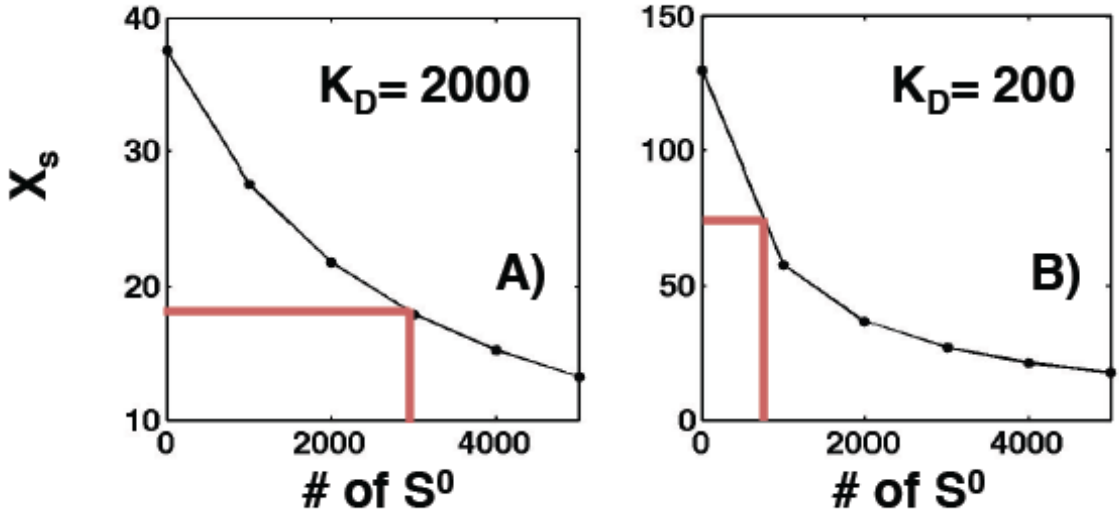

**Fig. S10: A large concentration of IP$_4$ is required to replace PIP$_3$ in models M5-M6.**
A) Variation of the steady state $x_s$ (Itk-PIP$_3$) as a function of initial substrate (PIP$_2$) concentration $S^0$ when the K$_D$ = 2000. B) Variation of the steady state $x_s$ as a function of initial substrate concentration $S^0$ when the K$_D$ = 200.

In Fig S10, the values of $x_s$ at $S_0$=0 denote the values for A$_{max}$. Fig. S10A shows that for high K$_D$ values the system requires a large concentration of $S^0$ (roughly 3000 molecules) to reach a steady state just about the half of A$_{max}$. Moreover, when K$_D$ is high, A$_{max}$ is small (Fig. S10A to Fig. S10B), which in turn slows down the production rate of IP$_4$. We can estimate the initial concentration dependence of this time scale as follows. The fastest time scale for IP$_4$ production is given by $(k_{cat} A_{max})^{-1}$, where, $k_{cat}$ is the rate at which PIP$_2$ is converted into IP$_4$ by Itk – PIP$_3$ by a one step reaction (tables S1-S6). Since, A$_{max}$ grows in a graded fashion with the increase in Itk$^0$ and PIP$_3^0$ (Eqn 12), the production timescales for IP$_4$ generation decrease slowly with increasing Itk$^0$ and PIP$_3^0$. Both these facts play hand in hand to give rise to a larger $\tau_d$ in M5 and M6 as compared to models M1-M3.

Model M4:

In model M4, in addition to the low affinity binding unbinding reactions, there is a bi-directional high affinity augmentation process. Following the same procedure as described above we compute the steady state concentration of PIP$_3$ bound Itk for M4. Denoting concentrations of $Itk - PIP_3$, $Itk^* - PIP_3$ and $Itk\text{-}IP_4$ by $x_1$, $x_2$ and $y$, respectively, the rate equations are given by,

$$\frac{dx_1}{dt} = k_1 [Itk^0 - x - y][PIP_3^0 - x] - k_{-1} x_1$$

$$\frac{dx_2}{dt} = k_2 [PIP_3^0 - x] y - k_2 x_2 [IP_4]$$

$$\frac{dy}{dt} = k_1[Itk^0 - x - y][IP_4] - k_{-1}y - \frac{dx_2}{dt}$$

(13)

where, $k_1$ and $k_{-1}$ are the usual binding unbinding rates while $k_2$ is the high affinity augmentation rate. $x$ is the sum of $x_1$ and $x_2$. In the steady state we have,

$$[PIP_3^0 - x_s]y_s - x_{2s}[S^0 - y_s] = 0$$
$$[Itk^0 - x_s - y_s][S^0 - y_s] - K_D y_s = 0$$
$$[Itk^0 - x_s - y_s][PIP_3^0 - x_s] - K_D x_{1s} = 0$$

(14)

The subscript "s" is used denote steady state concentrations. From Eqn (14) it is apparent that

$$\frac{[PIP_3^0 - x_s]}{[S^0 - y_s]} = \frac{x_{1s}}{y_s}$$
$$[S^0 - y_s] = \frac{y_s[PIP_3^0 - x_s]}{x_{2s}}$$

(15)

which implies $x_{1s} = x_{2s}$ for nonzero values of $y_s$. Making use of this fact we have,

$$K_D x_s (x_s - 2PIP_3^0) - 2(PIP_3^0 - x_s)\{(2PIP_3^0 + S^0 - x_s)x_s + Itk^0(x_s - 2PIP_3^0)\} = 0 .$$  (16)

Eqn(16) is a cubic equation yielding three real roots for the values of $K_D$, $PIP_3^0$ and $Itk^0$ we have used. However, only one root provides physically meaningful result, the other roots create an unphysical situation where $x_s > Itk^0$. We show the variation of $x_s$ with two different values of $K_D$ below (Fig. S11). Following similar analysis as in Fig. S10, we find that a large number of substrate is required to bring down the activation of $PIP_3$ bound Itk. $A_{max}$ in this case was calculated by taking a limit $S^0 \to 0$, note, $x_s$ has a discontinuity at $S^0=0$.

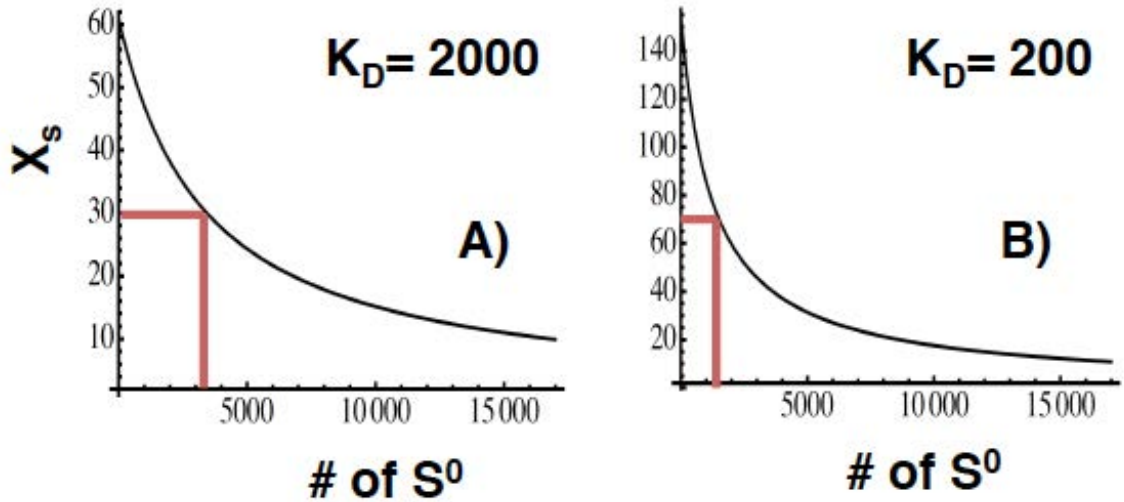

**Fig. S11: A large concentration of *IP₄* is required to replace *PIP₃* in model M4.** A) Variation of the steady state $x_s$ (Itk-PIP$_3$) as a function of initial substrate (PIP$_2$) concentration $S^0$ when the K$_D$ = 2000. B) Same as in A) for K$_D$ = 200.

### Variation of R with increasing initial concentrations of Itk and PIP3 for models M1-M6

The asymmetry ratio, R, is calculated using, $R = \tau_w / \tau_p = [(\tau_p - t_1) + \tau_d]/\tau_p$, where, $\tau_w$ is the width of the temporal profile of the concentration of PIP$_3$ bound Itk. $\tau_w$ can be expressed in terms of the decay time, $\tau_d$, $t_1$ (the time system takes to reach half of the peak concentration) and the peak time $\tau_p$. We aimed to understand the variation of R with increasing Itk$^0$ and PIP$_3{}^0$ (Fig. 2) based on the results described in the last two sections.

Models M1 and M3
The asymmetry ratio, R, increases as Itk$^0$ and PIP$_3{}^0$ are increased. Our calculations show (Fig. S9) that $\tau_w$ does not vary appreciably as Itk$^0$ and PIP$_3{}^0$ are increased in the range of moderate to high values. This occurs because the decay time, $\tau_d$, is primarily determined by the small unbinding rate of PIP$_3$ from the stable IP$_4$ – Itk – Itk–PIP$_3$ complex. Since $\tau_d$ is much larger than the peak time in this range, $\tau_w$ is mainly determined by $\tau_d$ in this range of concentrations. The increase in R, hence, arises from the decrease in $\tau_p$ as Itk$^0$ and PIP$_3{}^0$ are increased. The concentration dependence of this variation is determined by Eq. (5). At lower concentrations of Itk$^0$ and PIP$_3{}^0$, both $\tau_w$ and $\tau_d$ decrease, as Itk$^0$ and PIP$_3{}^0$ are decreased, however, $\tau_p$ decreases at a faster rate compared to $\tau_d$, resulting in increasing R values.

Model M2

In this model, in contrast to models M1 and M3, the asymmetry ratio, R, initially decreases with increasing $Itk^0$ and $PIP_3^0$, and, then at larger values of $Itk^0$ and $PIP_3^0$ starts increasing with increasing $Itk^0$ and $PIP_3^0$. At smaller values of $Itk^0$ and $PIP_3^0$, $\tau_w$, decreases (Fig. S9A) at a much faster rate as compared to $\tau_p$, as $Itk^0$ and $PIP_3^0$ are increased. At higher values of $Itk^0$ and $PIP_3^0$, $\tau_w$ does not change appreciably with increasing initial concentrations because of the same mechanisms as described for M1 and M3. However, in this range of concentrations $\tau_p$ decreases with increasing $Itk^0$ and $PIP_3^0$ resulting in increased R as the initial concentrations increase.

Models M5 and M6

For these models, $\tau_d$ is substantially larger than $\tau_p$, therefore, $\tau_w$ is well approximated by $\tau_d$. Since $\tau_p$ does not change appreciably, but $\tau_d$ decreases with increasing $Itk^0$ and $PIP_3^0$ we see a decrease in the ratio R as the initial concentrations are increased.

Model M4

$\tau_w$ behaves in a very similar manner to models M5 and M6. $\tau_p$ for M4 decreases with increasing $Itk^0$ and $PIP_3^0$ following Eqn (5), however, the rate of the decrease is still smaller than that of $\tau_w$, which results in decrease in the values of R as the initial concentrations are increased.

Model 7 can be analyzed in a similar way.

**Relative Entropy calculation**

The continuous relative entropy is defined as
$$D_{KL} = \int_{x \in K} p(x) \log \frac{p(x)}{q(x)} dx \quad (17)$$
where, $p(x)$ is the distribution of the parameters denoted by $x \in K$, which is subject to constraints imposed by experiments, and, $q(x)$ is a uniform distribution, such that $q(x) = 1/|K|$ for all $x \in K$. Now we seek for $p(x)$ which maximizes Eqn (17) under the constraints:

$$\int_{x \in K} R(x) p(x) dx = R_{avg} = R_{exp}$$

$$\int_{x \in K} \tau(x) p(x) dx = \tau_{avg} = \tau_{exp}$$

$$\int_{x\in K} A(x)p(x)dx = A_{avg} = A_{exp}$$

$$\int_{x\in K} p(x)dx = 1$$

(18)

We use three Lagrange multipliers ($\lambda$, $\lambda_1$, $\lambda_2$ and $\lambda_3$) to incorporate the constraints in Eq. (18) and maximize the following function:

$$G(p,\lambda,\lambda_1,\lambda_2,\lambda_3) = \int_{x\in K} p(x)\log\frac{p(x)}{q(x)}dx + \lambda\left(\int_{x\in K} p(x)-1\right) + \lambda_1\left(\int_{x\in K} R(x)p(x)dx - R_{avg}\right) + \lambda_2\left(\int_{x\in K} \tau(x)p(x)dx - \tau_{avg}\right)$$
$$+ \lambda_3\left(\int_{x\in K} A(x)p(x)dx - A_{avg}\right)$$

$$\frac{\partial G}{\partial p} = 0 \Rightarrow \log\frac{p(x)}{q(x)} + 1 + \lambda + \lambda_1 R(x) + \lambda_2 \tau(x) + \lambda_3 A(x) = 0$$

$$\frac{\partial G}{\partial \lambda} = 0 \Rightarrow \int_{x\in K} p(x)dx = 1$$

$$\frac{\partial G}{\partial \lambda_1} = 0 \Rightarrow \int_{x\in K} R(x)p(x)dx = R_{avg}$$

$$\frac{\partial G}{\partial \lambda_2} = 0 \Rightarrow \int_{x\in K} \tau(x)p(x)dx = \tau_{avg}$$

$$\frac{\partial G}{\partial \lambda_3} = 0 \Rightarrow \int_{x\in K} A(x)p(x)dx = A_{avg}$$

(19)

From Eqn (19) it is clear that

$$p(x) = \frac{q(x)e^{-\lambda_1 R(x) - \lambda_2 \tau(x) - \lambda_3 A(x)}}{\int_{x\in k} q(x)e^{-\lambda_1 R(x) - \lambda_2 \tau(x) - \lambda_3 A(x)}}$$

$$R_{avg} = \frac{\int_{x\in k} R(x)q(x)e^{-\lambda_1 R(x) - \lambda_2 \tau(x) - \lambda_3 A(x)}}{\int_{x\in k} q(x)e^{-\lambda_1 R(x) - \lambda_2 \tau(x) - \lambda_3 A(x)}}$$

$$\tau_{avg} = \frac{\int_{x\in K} \tau(x)q(x)e^{-\lambda_1 R(x) - \lambda_2 \tau(x) - \lambda_3 A(x)}}{\int_{x\in k} q(x)e^{-\lambda_1 R(x) - \lambda_2 \tau(x) - \lambda_3 A(x)}}$$

$$A_{avg} = \frac{\int_{x\in K} A(x)q(x)e^{-\lambda_1 R(x) - \lambda_2 \tau(x) - \lambda_3 A(x)}}{\int_{x\in k} q(x)e^{-\lambda_1 R(x) - \lambda_2 \tau(x) - \lambda_3 A(x)}}$$

We substituted the p(x) from Eq. (19) in the equations for $R_{avg}$, $\tau_{avg}$ and $A_{avg}$ (Eq. 18), and solved for $\lambda_1$, $\lambda_2$ and $\lambda_3$ when the values of $R_{avg}$ and $\tau_{avg}$ were taken from the table II in the main text. Owing to the lack of knowledge about the absolute value of $A_{avg}$, we have used some $A_{avg}$, which all the models can yield. (We have also varied the $A_{avg}$, to study how the dependence of the robustness on the choice of $A_{avg}$). Then we calculate the Kullback-Leibler distance

$$D_{KL} = \int_{x \in K} p(x) \log \frac{p(x)}{q(x)} dx \tag{20}$$

for each model.

*Range of parameter variation:* For the results shown in Fig. S12, 13, 14, 15, 16, 17, 19, 20, 21, 22 and 23, the rate constants were chosen from uniform distributions with lower and upper bounds equal to 1/10 and 10 times, respectively, the base values shown in tables S1-S7. We used 100,000 sample points, each point representing a set of rate constants and initial concentrations for all the models. For models M1-M3 the high affinity binding unbinding rates are drawn from a uniform distribution whilst the low affinity $K_D$ is determined as $K_D^{low} = \alpha K_D^{high}$, where $\alpha$ is drawn from a uniform distribution with lower and upper bound of 1 and 4000 respectively. For M7, while high affinity binding and unbinding rates are drawn from a uniform distribution, $\alpha$ is chosen uniformly from 1 to 50. The initial concentrations of Itk and $PIP_3$ were varied within a 35% (11) range from uniform distributions centered at the base values shown in table S8.

**Histograms of the asymmetry ratio (R) and the peak time $\tau_p$ for low, moderately high and high stimulation.**

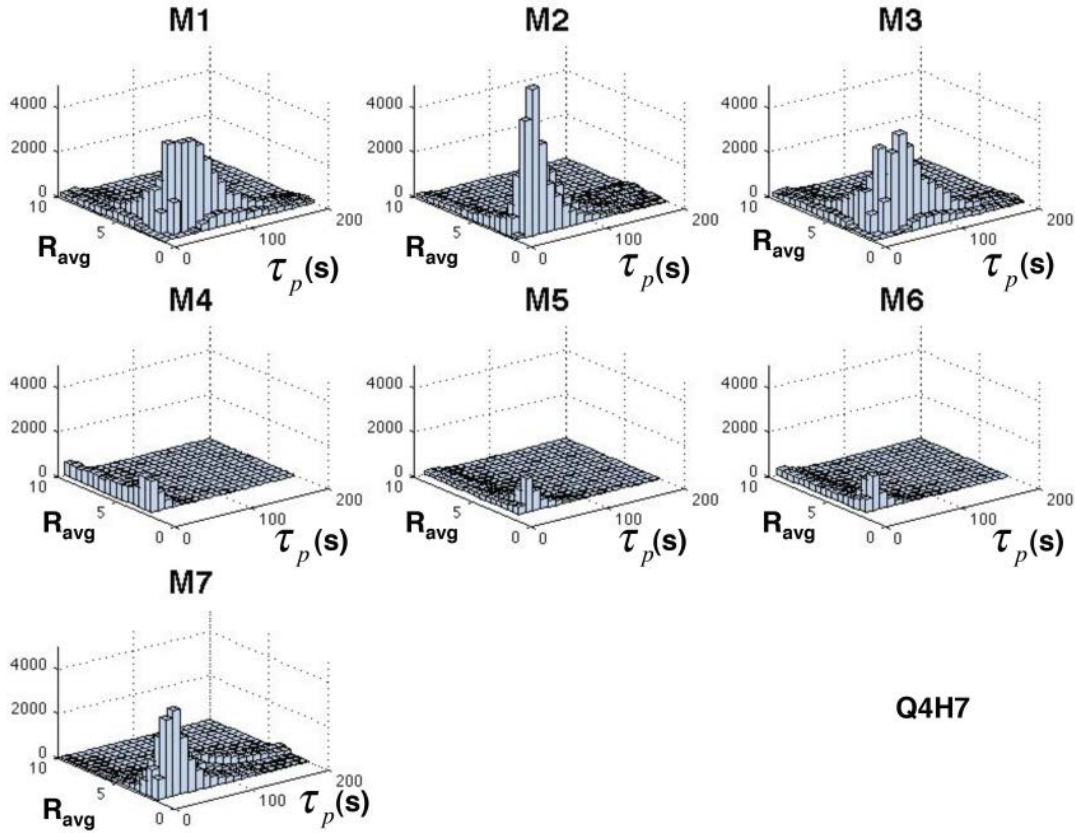

**Fig. S12:** The histograms for R and $\tau$ as the parameters are varied in all 7 models for moderately low initial concentrations of $Itk^0$ and $PIP_3^0$. All the rate constants are varied by two orders of magnitude with the constraint $K_D^{low} = \alpha K_D^{high}$. For M1-M3, $\alpha$ is distributed uniformly over 1 to 4000 while for M7 it is distributed uniformly over 1 to 50. The initial concentrations of species involved are varied in a 35% window about the base value of $Itk^0 = 40$, $PIP_3^0 = 130$ and $PIP_2^0 = 17000$.

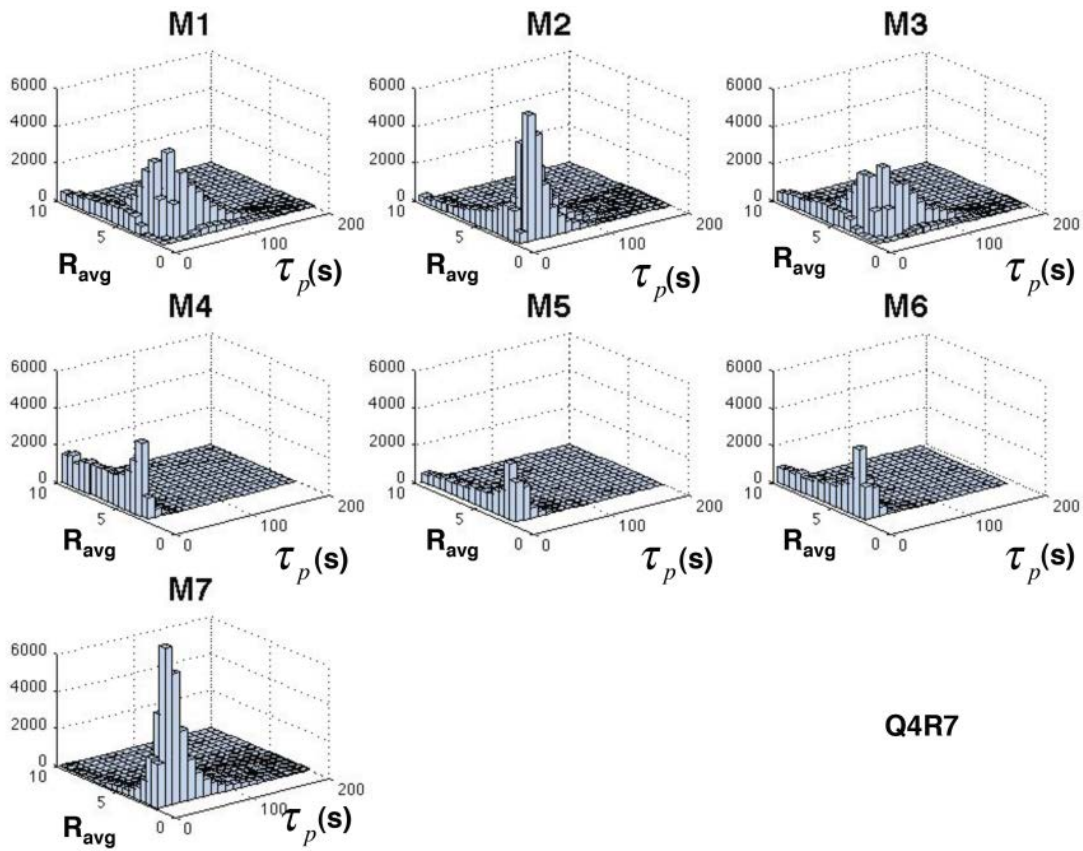

**S13: The histograms for R and $\tau$ as the parameters are varied in all 7 models for moderately high initial concentrations of $Itk^0$ and $PIP_3^0$.** All the rate constants are varied by two orders of magnitude with the constraint $K_D^{low} = \alpha K_D^{high}$. For M1-M3, $\alpha$ is distributed uniformly over 1 to 4000 while for M7 it is distributed uniformly over 1 to 50. The initial concentrations of species involved are varied in a 35% window about the base value of $Itk^0 = 100$, $PIP_3^0 = 370$ and $PIP_2^0 = 17000$.

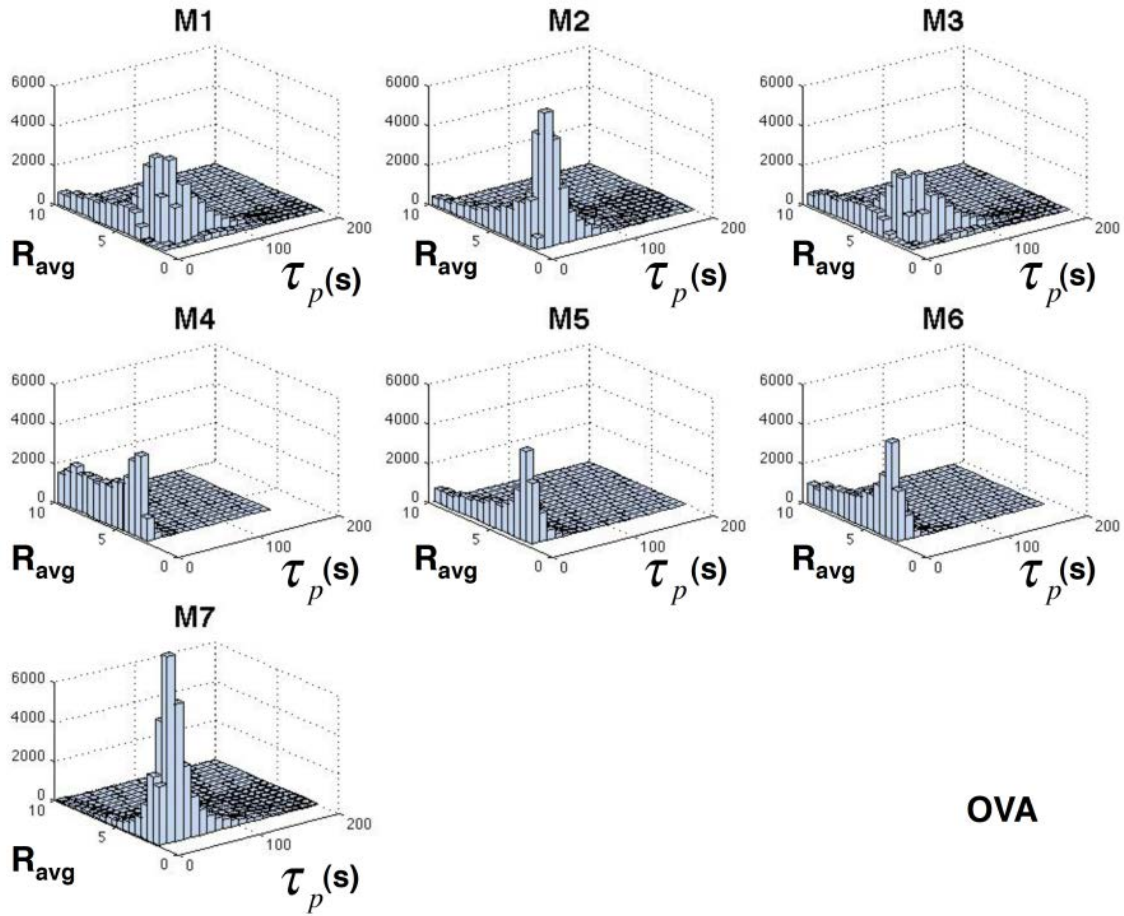

**S14: The histograms for R and $\tau$ as the parameters are varied in all 7 models for high initial concentrations of $Itk^0$ and $PIP_3^0$.** All the rate constants are varied by two orders of magnitude with the constraint $K_D^{low} = \alpha\, K_D^{high}$. For M1-M3, $\alpha$ is distributed uniformly over 1 to 4000 while for M7 it is distributed uniformly over 1 to 50. The initial concentrations of species involved are varied in a 35% window about the base value of $Itk^0 = 140$, $PIP_3^0 = 530$ and $PIP_2^0 = 17000$.

**The checkerboard plot of the most robust models for different peptide affinities as the $R_{avg}$ and $A_{avg}$ are varied.**

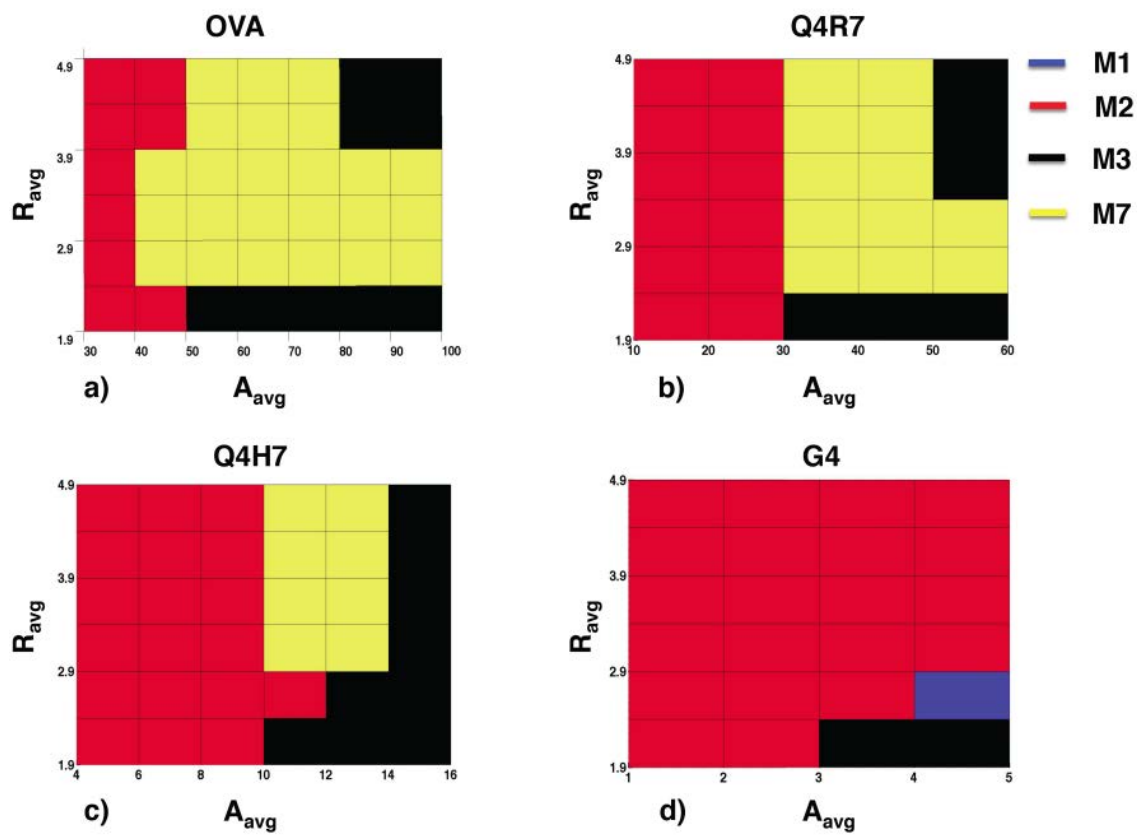

**Fig. S15: Checkerboard plot of the most robust models for different ligand affinities as $R_{avg}$ and $A_{avg}$ are varied for a fixed $\tau_{avg} = 2$ mins**. **a)** Plot of the most robust models for $Itk^0 = 140$ and $PIP_3^0 = 530$ molecules. **b)** The same plot as **a)** for $Itk^0 = 100$ and $PIP_3^0 = 370$ molecules. **c)** Same plot as **a)** for $Itk^0 = 40$ and $PIP_3^0 = 130$ molecules. **d)** The same plot as **a)** for $Itk^0 = 20$ and $PIP_3^0 = 50$ molecules.

**Relative robustness of the seven models for specific amplitude averages as the ligand affinities are varied.**

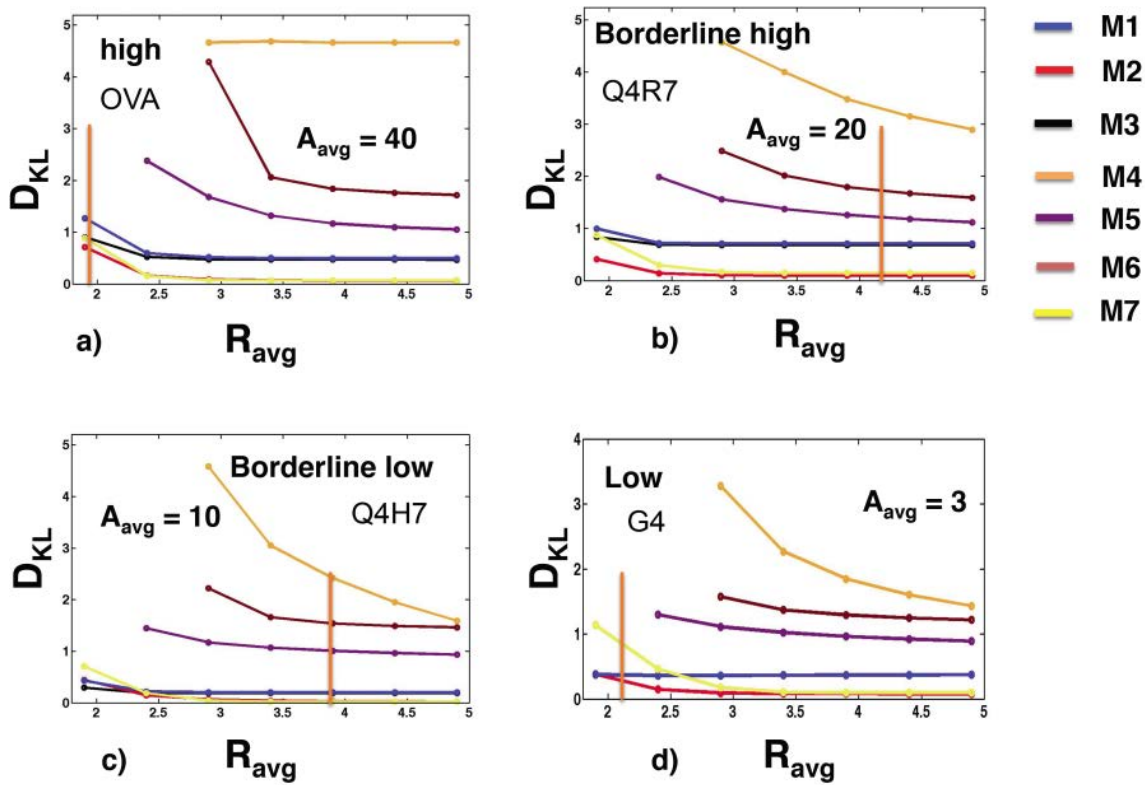

**Fig. S16: Plots of the relative robustness of all the 7 models for a specific $A_{avg}$ for different ligand affinities as $R_{avg}$ is varies for a fixed $\tau_{avg} = 2\,\text{mins}$. a)** For $Itk^0 = 140$ and $PIP_3^0 = 530$ molecules the $D_{KL}$ is shown for an $A_{avg}$ of 40 molecules. **b)** The same plot as **a)** for $Itk^0 = 100$ and $PIP_3^0 = 370$ molecules when the $A_{avg}$ is held fixed at 20 molecules. **c)** Same plot as **a)** for $Itk^0 = 40$ and $PIP_3^0 = 130$ molecules when $A_{avg} = 10$ molecules. **d)** The same plot as **a)** for $Itk^0 = 20$ and $PIP_3^0 = 50$ molecules when $A_{avg} = 3$ moelcules. The orange vertical bar in all the plots show the experimentally observed value of $R_{avg}$ (table I main text).

### The effect of Lck mediated phosphorylation of PIP$_3$ bound Itk complex.

We have studied the effect of Lck mediated phosphorylation of membrane bound Itk at its Y511 residue. Here the membrane bound Itk is phosphorylated by Lck (modeled as a first order reaction). Upon activation it becomes active (denoted by $Itk^{act}$ except for M4 where it is denoted as $Itk_{act}$). Only an active form of Itk is capable of producing IP$_4$ from the hydrolysis of PIP$_2$. The activation and de-activation rates of membrane bound Itk is chosen so that the kinetics of active Itk roughly agrees with the experimental data in the main text and (3). The reactions and the reaction rates are detailed in table S9-S15. We find that adding the Lck mediated phosphorylation of the membrane bound Itk does not alter the rank order of the models.

**Table S9: Reactions and rate constants for model M1$^{lck}$.**

| Reactions | $k_{on}$ ($\mu M^{-1} s^{-1}$) | $k_{off}$ ($s^{-1}$) | $K_D$ ($\mu M$) | $k_{cat}$ ($\mu M^{-1} s^{-1}$) | $k_{act}$ ($s^{-1}$) | $k_{deact}$ ($s^{-1}$) |
|---|---|---|---|---|---|---|
| $Itk-Itk + PIP_3 \leftrightarrow Itk-Itk-PIP_3$ | $2.5 \times 10^{-4}$ | 0.1 | 400 | | | |
| $PIP_3-Itk-Itk + PIP_3 \leftrightarrow PIP_3-Itk-Itk-PIP_3$ | 0.01 | 0.003 | 0.3 | | | |
| $Itk-Itk + IP_4 \leftrightarrow Itk-Itk-IP_4$ | $2.5 \times 10^{-3}$ | 0.1 | 40 | | | |
| $PIP_3-Itk-Itk + IP4 \leftrightarrow PIP_3-Itk-Itk-IP_4$ | 0.1 | 0.003 | 0.03 | | | |
| $IP_4-Itk-Itk + PIP_3 \leftrightarrow IP_4-Itk-Itk-PIP_3$ | 0.01 | 0.003 | 0.3 | | | |
| $IP_4-Itk-Itk + IP4 \leftrightarrow IP_4-Itk-Itk-IP_4$ | 0.1 | 0.003 | 0.03 | | | |
| $IP_4-Itk-Itk^{act}-PIP_3 + S \rightarrow IP_4 + IP_4-Itk-Itk^{act}-PIP_3$ | | | | $1.5 \times 10^{-4}$ | | |
| $Itk-Itk^{act}-PIP_3 + S \rightarrow IP_4 + Itk-Itk^{act}-PIP_3$ | | | | $1.5 \times 10^{-4}$ | | |
| $PIP_3-Itk-Itk^{act}-PIP_3 + S \rightarrow IP_4 + PIP_3-Itk-Itk^{act}-PIP_3$ | | | | $1.5 \times 10^{-4}$ | | |
| $Itk-Itk-PIP_3 \leftrightarrow Itk-Itk^{act}-PIP_3$ | | | | | 0.3 | 0.09 |
| $IP_4-Itk-Itk-PIP_3 \leftrightarrow IP_4-Itk-Itk^{act}-PIP_3$ | | | | | 0.3 | 0.09 |
| $PIP_3-Itk-Itk-PIP_3 \leftrightarrow PIP_3-Itk-Itk^{act}-PIP_3$ | | | | | 0.3 | 0.09 |
| $Itk-Itk^{act}-PIP_3 \rightarrow Itk-Itk + PIP_3$ | | 0.1 | | | | |
| $IP_4-Itk-Itk^{act}-PIP_3 \leftrightarrow Itk-Itk^{act}-PIP_3 + IP_4$ | 0.1 | 0.003 | 0.03 | | | |
| $IP_4-Itk-Itk^{act}-PIP_3 \leftrightarrow Itk-Itk-IP_4 + PIP_3$ | 0.01 | 0.003 | 0.3 | | | |
| $PIP_3-Itk-Itk^{act}-PIP_3 \leftrightarrow Itk-Itk^{act}-PIP_3 + PIP_3$ | 0.01 | 0.003 | 0.3 | | | |

**Table S10: Reactions and rate constants for model M2$^{lck}$.**

| Reactions | $k_{on}$ ($\mu M^{-1} s^{-1}$) | $k_{off}$ ($s^{-1}$) | $K_D$ ($\mu M$) | $k_{cat}$ ($\mu M^{-1} s^{-1}$) | $k_{act}$ ($s^{-1}$) | $k_{deact}$ ($s^{-1}$) |
|---|---|---|---|---|---|---|
| $Itk-Itk + PIP_3 \leftrightarrow Itk-Itk-PIP_3$ | $2.5 \times 10^{-4}$ | 0.1 | 400 | | | |
| $Itk-Itk + IP_4 \leftrightarrow Itk-Itk-IP_4$ | $2.5 \times 10^{-3}$ | 0.1 | 40 | | | |
| $PIP_3-Itk-Itk + IP4 \leftrightarrow PIP_3-Itk-Itk-IP_4$ | $2.5 \times 10^{-3}$ | 0.1 | 40 | | | |
| $IP_4-Itk-Itk + PIP_3 \leftrightarrow IP_4-Itk-Itk-PIP_3$ | 0.01 | 0.003 | 0.3 | | | |
| $IP_4-Itk-Itk + IP4 \leftrightarrow IP_4-Itk-Itk-IP_4$ | 0.1 | 0.003 | 0.03 | | | |
| $IP_4-Itk-Itk^{act}-PIP_3 + S \rightarrow IP_4 + IP_4-Itk-Itk^{act}-PIP_3$ | | | | $1.5 \times 10^{-4}$ | | |
| $Itk-Itk^{act}-PIP_3 + S \rightarrow IP_4 + Itk-Itk^{act}-PIP_3$ | | | | $1.5 \times 10^{-4}$ | | |
| $Itk-Itk-PIP_3 \leftrightarrow Itk-Itk^{act}-PIP_3$ | | | | | 0.3 | 0.09 |
| $IP_4-Itk-Itk-PIP_3 \leftrightarrow IP_4-Itk-Itk^{act}-PIP_3$ | | | | | 0.3 | 0.09 |
| $Itk-Itk^{act}-PIP_3 \rightarrow Itk-Itk + PIP_3$ | | 0.1 | | | | |
| $IP_4-Itk-Itk^{act}-PIP_3 \leftrightarrow Itk-Itk^{act}-PIP_3 + IP_4$ | $2.5 \times 10^{-3}$ | 0.1 | 40 | | | |
| $IP_4-Itk-Itk^{act}-PIP_3 \leftrightarrow Itk-Itk-IP_4 + PIP_3$ | 0.01 | 0.003 | 0.3 | | | |

**Table S11: Reactions and rate constants for model M3$^{lck}$.**

| Reactions | $k_{on}$ ($\mu M^{-1} s^{-1}$) | $k_{off}$ ($s^{-1}$) | $K_D$ ($\mu M$) | $k_{cat}$ ($\mu M^{-1} s^{-1}$) | $k_{act}$ ($s^{-1}$) | $k_{deact}$ ($s^{-1}$) |
|---|---|---|---|---|---|---|
| $Itk-Itk + PIP_3 \leftrightarrow Itk-Itk-PIP_3$ | $2.5 \times 10^{-4}$ | 0.1 | 400 | | | |
| $Itk-Itk + IP_4 \leftrightarrow Itk-Itk-IP_4$ | $2.5 \times 10^{-3}$ | 0.1 | 40 | | | |
| $PIP_3-Itk-Itk + IP4 \leftrightarrow PIP_3-Itk-Itk-IP_4$ | 0.1 | 0.003 | 0.03 | | | |

| | | 3 | | | | |
|---|---|---|---|---|---|---|
| $IP_4 - Itk - Itk + PIP_3 \leftrightarrow IP_4 - Itk - Itk - PIP_3$ | 0.01 | 0.003 | 0.3 | | | |
| $IP_4 - Itk - Itk + IP4 \leftrightarrow IP_4 - Itk - Itk - IP_4$ | 0.1 | 0.003 | 0.03 | | | |
| $IP_4 - Itk - Itk^{act} - PIP_3 + S \rightarrow IP_4 + IP_4 - Itk - Itk^{act} - PIP_3$ | | | | $1.5 \times 10^{-4}$ | | |
| $Itk - Itk^{act} - PIP_3 + S \rightarrow IP_4 + Itk - Itk^{act} - PIP_3$ | | | | $1.5 \times 10^{-4}$ | | |
| $Itk - Itk - PIP_3 \leftrightarrow Itk - Itk^{act} - PIP_3$ | | | | | 0.3 | 0.09 |
| $IP_4 - Itk - Itk - PIP_3 \leftrightarrow IP_4 - Itk - Itk^{act} - PIP_3$ | | | | | 0.3 | 0.09 |
| $Itk - Itk^{act} - PIP_3 \rightarrow Itk - Itk + PIP_3$ | | 0.1 | | | | |
| $IP_4 - Itk - Itk^{act} - PIP_3 \leftrightarrow Itk - Itk^{act} - PIP_3 + IP_4$ | 0.1 | 0.003 | 0.03 | | | |
| $IP_4 - Itk - Itk^{act} - PIP_3 \leftrightarrow Itk - Itk - IP_4 + PIP_3$ | 0.01 | 0.003 | 0.3 | | | |

**Table S12: Reactions and rate constants for model M4$^{lck}$.**

| Reactions | $k_{on}$ ($\mu M^{-1}s^{-1}$) | $k_{off}$ ($s^{-1}$) | $K_D$ ($\mu M$) | $k_{cat}$ ($\mu M^{-1}s^{-1}$) | $k_{act}$ ($s^{-1}$) | $k_{deact}$ ($s^{-1}$) |
|---|---|---|---|---|---|---|
| $Itk + PIP_3 \leftrightarrow Itk - PIP_3$ | $2.5 \times 10^{-4}$ | 0.1 | 400 | | | |
| $Itk + IP_4 \leftrightarrow Itk^* - IP_4$ | $2.5 \times 10^{-3}$ | 0.1 | 40 | | | |
| $Itk^* - IP_4 + PIP_3 \leftrightarrow Itk^* - PIP_3 + IP_4$ | 10 | | | | | |
| $Itk^*_{act} - PIP_3 + S \rightarrow IP_4 + Itk^*_{act} - PIP_3$ | | | | $1.5 \times 10^{-4}$ | | |
| $Itk_{act} - PIP_3 + S \rightarrow IP_4 + Itk_{act} - PIP_3$ | | | | $1.5 \times 10^{-4}$ | | |
| $Itk - PIP_3 \leftrightarrow Itk_{act} - PIP_3$ | | | | | 0.3 | 0.09 |
| $Itk^* - PIP_3 \leftrightarrow Itk^*_{act} - PIP_3$ | | | | | 0.3 | 0.09 |
| $Itk^*_{act} - PIP_3 + IP_4 \rightarrow Itk^* - IP_4 + PIP_3$ | 10 | | | | | |

**Table S13: Reactions and rate constants for model M5$^{lck}$.**

| Reactions | $k_{on}$ | $k_{off}$ | $K_D$ | $k_{cat}$ | $k_{ac}$ | $k_{deac}$ |
|---|---|---|---|---|---|---|

| | ($\mu M^{-1} s^{-1}$) | ($s^{-1}$) | ($\mu M$) | ($\mu M^{-1} s^{-1}$) | t ($s^{-1}$) | t ($s^{-1}$) |
|---|---|---|---|---|---|---|
| $Itk-Itk+PIP_3 \leftrightarrow Itk-Itk-PIP_3$ | $1.25 \times 10^{-4}$ | 0.05 | 400 | | | |
| $PIP_3-Itk-Itk+PIP_3 \leftrightarrow PIP_3-Itk-Itk-PIP_3$ | $1.25 \times 10^{-4}$ | 0.05 | 400 | | | |
| $Itk-Itk+IP_4 \leftrightarrow Itk-Itk-IP_4$ | $1.25 \times 10^{-3}$ | 0.05 | 40 | | | |
| $PIP_3-Itk-Itk+IP4 \leftrightarrow PIP_3-Itk-Itk-IP_4$ | $1.25 \times 10^{-3}$ | 0.05 | 40 | | | |
| $IP_4-Itk-Itk+PIP_3 \leftrightarrow IP_4-Itk-Itk-PIP_3$ | $1.25 \times 10^{-4}$ | 0.05 | 400 | | | |
| $IP_4-Itk-Itk+IP4 \leftrightarrow IP_4-Itk-Itk-IP_4$ | $1.25 \times 10^{-3}$ | 0.05 | 40 | | | |
| $IP_4-Itk-Itk^{act}-PIP_3+S \rightarrow IP_4+IP_4-Itk-Itk^{act}-PIP_3$ | | | | $1.5 \times 10^{-4}$ | | |
| $Itk-Itk^{act}-PIP_3+S \rightarrow IP_4+Itk-Itk^{act}-PIP_3$ | | | | $1.5 \times 10^{-4}$ | | |
| $PIP_3-Itk-Itk^{act}-PIP_3+S \rightarrow IP_4+PIP_3-Itk-Itk^{act}-PIP_3$ | | | | $1.5 \times 10^{-4}$ | | |
| $Itk-Itk-PIP_3 \leftrightarrow Itk-Itk^{act}-PIP_3$ | | | | | 0.3 | 0.09 |
| $IP_4-Itk-Itk-PIP_3 \leftrightarrow IP_4-Itk-Itk^{act}-PIP_3$ | | | | | 0.3 | 0.09 |
| $PIP_3-Itk-Itk-PIP_3 \leftrightarrow PIP_3-Itk-Itk^{act}-PIP_3$ | | | | | 0.3 | 0.09 |
| $Itk-Itk^{act}-PIP_3 \rightarrow Itk-Itk+PIP_3$ | | 0.05 | | | | |
| $IP_4-Itk-Itk^{act}-PIP_3 \leftrightarrow Itk-Itk^{act}-PIP_3+IP_4$ | $1.25 \times 10^{-3}$ | 0.05 | 40 | | | |
| $IP_4-Itk-Itk^{act}-PIP_3 \leftrightarrow Itk-Itk-IP_4+PIP_3$ | $1.25 \times 10^{-4}$ | 0.05 | 400 | | | |
| $PIP_3-Itk-Itk^{act}-PIP_3 \leftrightarrow Itk-Itk^{act}-PIP_3+PIP_3$ | $1.25 \times 10^{-4}$ | 0.05 | 400 | | | |

**Table S14: Reactions and rate constants for model M6$^{lck}$.**

| Reactions | $k_{on}$ ($\mu M^{-1} s^{-1}$) | $k_{off}$ ($s^{-1}$) | $K_D$ ($\mu M$) | $k_{cat}$ ($\mu M^{-1} s^{-1}$) | $k_{act}$ ($s^{-1}$) | $k_{deact}$ ($s^{-1}$) |
|---|---|---|---|---|---|---|
| $Itk + PIP_3 \leftrightarrow Itk - PIP_3$ | $1.25 \times 10^{-4}$ | 0.05 | 400 | | | |
| $Itk + IP_4 \leftrightarrow Itk - IP_4$ | $1.25 \times 10^{-3}$ | 0.05 | 40 | | | |
| $Itk - PIP_3 \leftrightarrow Itk^{act} - PIP_3$ | | | | | 0.3 | 0.09 |
| $Itk^{act} - PIP_3 \rightarrow Itk + PIP_3$ | | 0.05 | | | | |
| $Itk^{act} - PIP_3 + S \rightarrow IP_4 + Itk^{act} - PIP_3$ | | | | $1.5 \times 10^{-4}$ | | |

**Table S15: Reactions and rate constants for model M7$^{lck}$.**

| Reactions | $k_{on}$ ($\mu M^{-1} s^{-1}$) | $k_{off}$ ($s^{-1}$) | $K_D$ ($\mu M$) | $k_{cat}$ ($\mu M^{-1} s^{-1}$) | $k_{act}$ ($s^{-1}$) | $k_{deact}$ ($s^{-1}$) |
|---|---|---|---|---|---|---|
| $Itk - Itk + PIP_3 \leftrightarrow Itk - Itk - PIP_3$ | $2.5 \times 10^{-4}$ | 0.01 | 40 | | | |
| $PIP_3 - Itk - Itk + PIP_3 \leftrightarrow PIP_3 - Itk - Itk - PIP_3$ | 0.01 | 0.1 | 10 | | | |
| $Itk - Itk + IP_4 \leftrightarrow Itk - Itk - IP_4$ | $2.5 \times 10^{-3}$ | 0.01 | 4 | | | |
| $PIP_3 - Itk - Itk + IP4 \leftrightarrow PIP_3 - Itk - Itk - IP_4$ | $2.5 \times 10^{-3}$ | 0.01 | 4 | | | |
| $IP_4 - Itk - Itk + PIP_3 \leftrightarrow IP_4 - Itk - Itk - PIP_3$ | 0.01 | 0.1 | 10 | | | |
| $IP_4 - Itk - Itk + IP4 \leftrightarrow IP_4 - Itk - Itk - IP_4$ | 0.1 | 0.1 | 1 | | | |
| $IP_4 - Itk - Itk^{act} - PIP_3 + S \rightarrow IP_4 + IP_4 - Itk - Itk^{act} - PIP_3$ | | | | $1.5 \times 10^{-4}$ | | |
| $Itk - Itk^{act} - PIP_3 + S \rightarrow IP_4 + Itk - Itk^{act} - PIP_3$ | | | | $1.5 \times 10^{-4}$ | | |
| $PIP_3 - Itk - Itk^{act} - PIP_3 + S \rightarrow IP_4 + PIP_3 - Itk - Itk^{act} - PIP_3$ | | | | $1.5 \times 10^{-4}$ | | |

| | | | | | 0.3 | 0.09 |
|---|---|---|---|---|---|---|
| $Itk - Itk - PIP_3 \leftrightarrow Itk - Itk^{act} - PIP_3$ | | | | | 0.3 | 0.09 |
| $IP_4 - Itk - Itk - PIP_3 \leftrightarrow IP_4 - Itk - Itk^{act} - PIP_3$ | | | | | 0.3 | 0.09 |
| $PIP_3 - Itk - Itk - PIP_3 \leftrightarrow PIP_3 - Itk - Itk^{act} - PIP_3$ | | | | | 0.3 | 0.09 |
| $Itk - Itk^{act} - PIP_3 \rightarrow Itk - Itk + PIP_3$ | | 0.1 | | | | |
| $IP_4 - Itk - Itk^{act} - PIP_3 \leftrightarrow Itk - Itk^{act} - PIP_3 + IP_4$ | $2.5 \times 10^{-3}$ | 0.01 | 4 | | | |
| $IP_4 - Itk - Itk^{act} - PIP_3 \leftrightarrow Itk - Itk - IP_4 + PIP_3$ | $2.5 \times 10^{-4}$ | 0.01 | 40 | | | |
| $PIP_3 - Itk - Itk^{act} - PIP_3 \leftrightarrow Itk - Itk^{act} - PIP_3 + PIP_3$ | $2.5 \times 10^{-4}$ | 0.01 | 40 | | | |

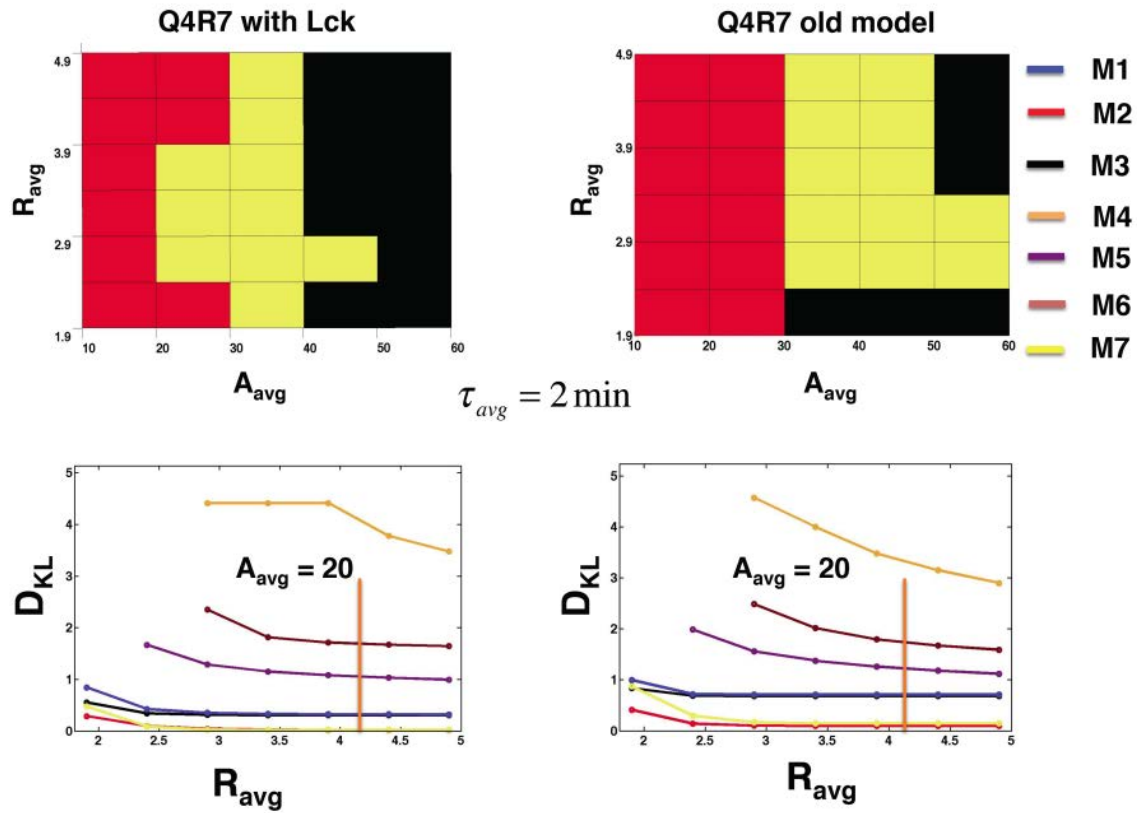

**Fig. S17: The effect of Lck mediated phosphorylation of Itk-PIP$_3$ on the relative robustness of M1-M7.** Upper panel (left most corner): For Itk$^0$ = 100 and PIP$_3^0$=370 the most robust models are shown as amplitude and the ratio of the Itk-PIP$_3$ kinetics are varied in presence of the Lck mediated phosphorylation of membrane recruited Itk at its Y511 residue. The average peak time is held at 2 mins. Upper panel (right most corner):

The same plot without any Lck mediated activation. Lower panel (left most corner): The relative robustness of the models M1-M7 for an amplitude average of 20 molecules in presence of Lck mediated activation of Itk. Lower panel (right most corner): Same plot without the explicit Lck mediated activation.

**Anti-CD3 and anti-CD3/CD4 antibody stimulation of polyclonal *MHC-DKO* thymocytes.**

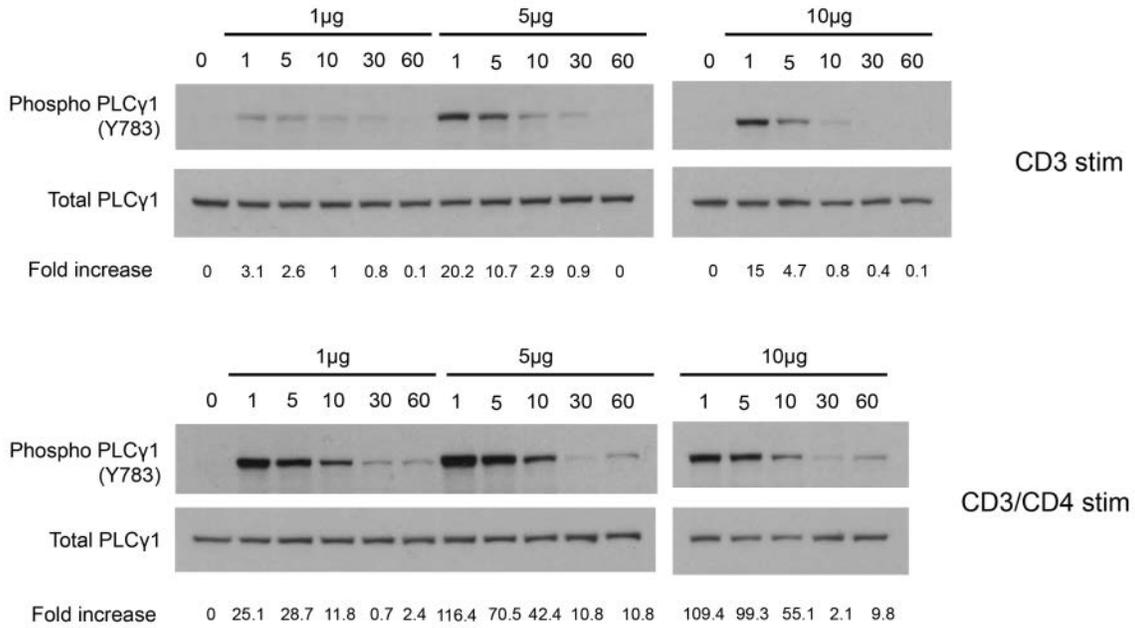

**Fig. S18: Kinetics of induction of PLCγ1 phosphorylation represented as the fold increase over non stimulated cells using total PLCγ1 protein as a loading control.**

Table S16: Values of peak time, peak width, and asymmetry ratio R calculated from the PLCγ1 activation kinetics in Fig. S18

| CD3 μg / mL | peak time ($\tau_p$) (mins.) | peak width ($\tau_w$) (mins.) | R |
|---|---|---|---|
| 1 | 1.0 | 8.0 | 8.0 |
| 5 | 1.0 | 5.5 | 5.5 |

| CD3 & CD4 μg / mL | peak time ($\tau_p$) (mins.) | peak width ($\tau_w$) (mins.) | R |
|---|---|---|---|
| 1 | 5.0 | 9.0 | 1.8 |
| 5 | 1.0 | 7.0 | 7.0 |

**Checkerboard plot of the most robust models as $R_{avg}$ and $A_{avg}$ are varied for different doses of anti-CD3 and anti-CD3/CD4 antibodies.**

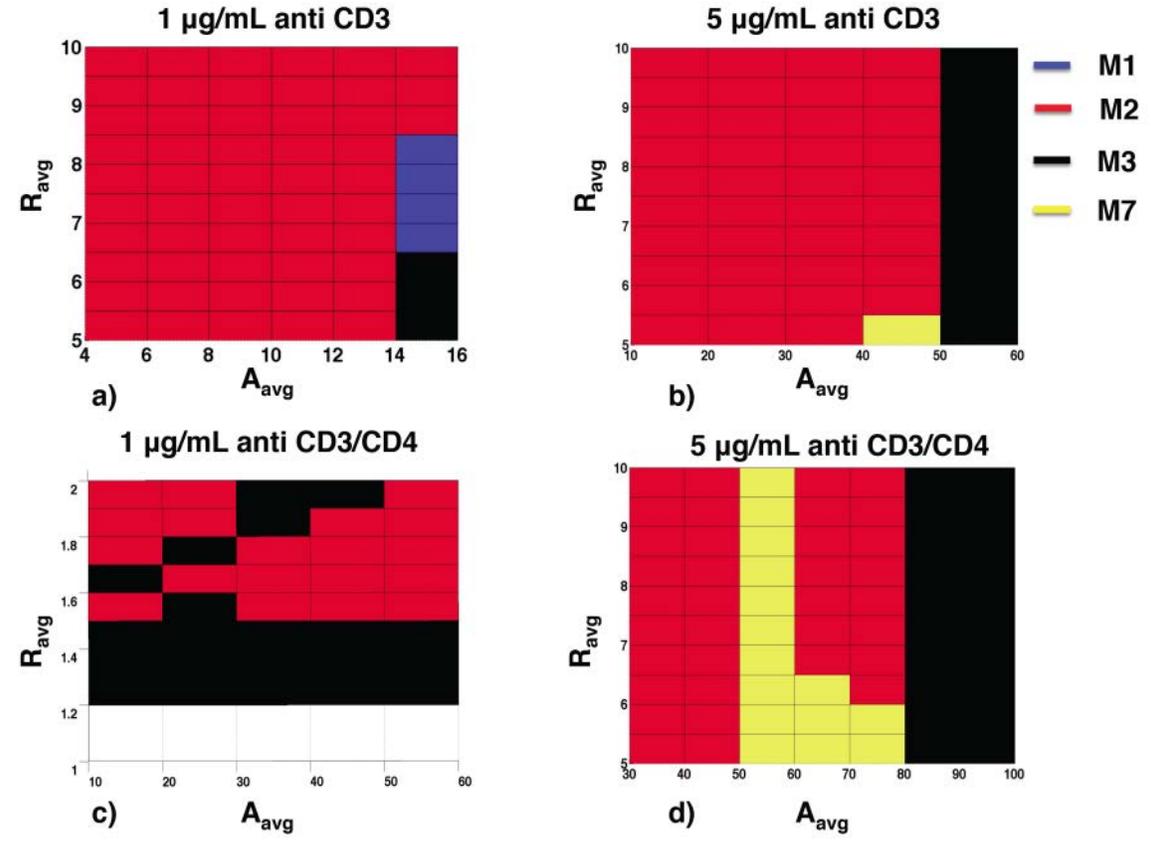

**Fig. S19: a)** $Itk^0 = 40$ and $PIP_3^0 = 130$ molecules are used to emulate the 1 μg/ mL anti CD3 stimulation. The $\tau_{avg}$ is held at 1 mins. The checkerboard diagram of the most robust models is shown as $R_{avg}$ and $A_{avg}$ are varied. **b)** Same as plot **a)** but $Itk^0 = 100$ and $PIP_3^0 = 370$ molecules are used as the initial concentrations. **c)** $Itk^0 = 100$ and $PIP_3^0 = 370$ molecules are used to emulate the 1 μg/ mL anti CD3/CD4 stimulation. The $\tau_{avg}$ is held at 5 mins. The checkerboard diagram of the most robust models is shown as $R_{avg}$ and $A_{avg}$ are varied. **d)** $Itk^0 = 140$ and $PIP_3^0 = 530$ molecules are used to emulate the 5 μg/ mL anti CD3/CD4 stimulation. The $\tau_{avg}$ is held at 1 mins. The checkerboard diagram of the most robust models is shown as $R_{avg}$ and $A_{avg}$ are varied.

**The plot of $D_{KL}$ for all the 7 models for a specific amplitude and different initial conditions for different doses of anti CD3 or anti CD3/CD4 antibodies.**

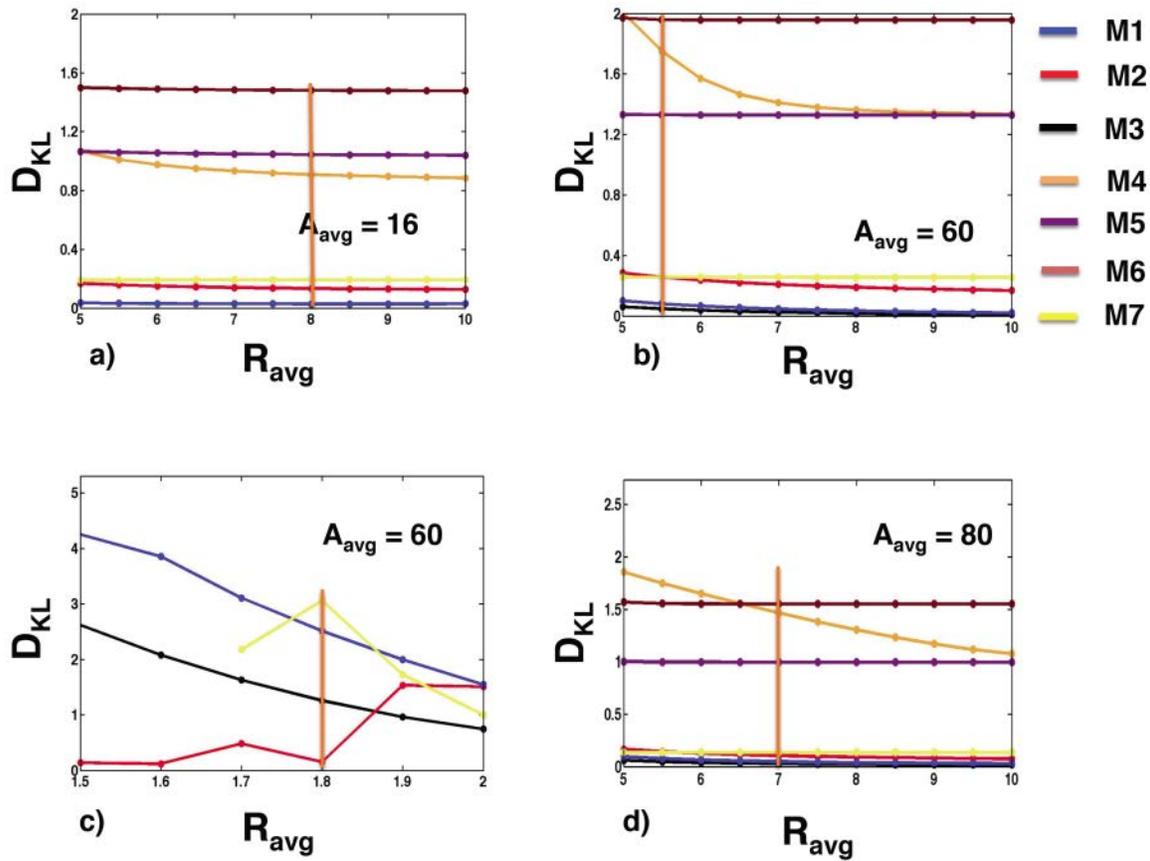

**Fig S20: a)** $Itk^0 = 40$ and $PIP_3^0 = 130$ molecules are used to emulate the 1 μg/ mL anti CD3 stimulation. The $\tau_{avg}$ is held at 1 mins. The $D_{KL}$ is shown for an $A_{avg} = 16$ molecules. **b)** Same as plot **a)** but $Itk^0 = 100$ and $PIP_3^0 = 370$ molecules are used as the initial concentrations and $A_{avg} = 60$ molecules. **c)** $Itk^0 = 100$ and $PIP_3^0 = 370$ molecules are used to emulate the 1 μg/ mL anti CD3/CD4 stimulation. The $\tau_{avg}$ is held at 5 mins. $A_{avg} = 60$ molecules. **d)** $Itk^0 = 140$ and $PIP_3^0 = 530$ molecules are used to emulate the 5 μg/ mL anti CD3/CD4 stimulation. The $\tau_{avg}$ is held at 1 mins and $A_{avg}$ is set equal to 80 molecules. The vertical orange bar shows the observed experimental values (Table S16).

### Dependence of $D_{KL}$ on parameters those weakly influence the Itk-PIP$_3$ kinetics.

Let us assume that the peak time $\tau_p$, the amplitude A and the ratio R depend on n parameters. If we add m new parameters, which do not in any way influence the outcome of the Itk-PIP$_3$ kinetics then the joint probability distribution that maximizes the entropy with the constraints becomes

$$p(k_1,...,k_{n+m}) = p(k_1,...,k_n) p(k_{n+1},...,k_{n+m})$$

Now as $k_{n+1},...,k_{n+m}$ are drawn from a uniform distribution and they do not contribute anything to the observables, $p(k_{n+1},...,k_{n+m}) = q(k_{n+1},...,k_{n+m})$, where $q$ is a uniform distribution. Therefore

$$D_{KL}^{new} = \int d^{n+m}k \ p(k_1,...,k_{n+m}) \ln \frac{p(k_1,...,k_{n+m})}{q(k_1,...,k_{n+m})} = \int d^n k \ p(k_1,...,k_n) \ln \frac{p(k_1,...,k_n)}{q(k_1,...,k_n)} \int d^m k \ p(k_{n+1},...,k_{n+m})$$

$$= \int d^n k \ p(k_1,...,k_n) \ln \frac{p(k_1,...,k_n)}{q(k_1,...,k_n)} = D_{KL}^{old} \ \left( \because \int d^m k \ q(k_{n+1},...,k_{n+m}) = 1 \right)$$

In order to probe the effect of parameters those weakly influence the kinetics of Itk-PIP$_3$ we have carried out a simulation for M3 with three added reactions. Instead of approximating the production of IP$_4$ from PIP$_2$ by a simple one step reaction, we have incorporated the fact that the membrane bound Itk phosphorylates PLCγ which in turn hydrolyses PIP$_2$ to form membrane bound DAG and soluble IP$_3$. IP$_3$ then gets converted into IP$_4$. In an effort to render the newly added parameters weak, we have chosen the rate constants in such a way that the PLCγ kinetics follow the Itk-PIP$_3$ kinetics and the turnover of IP$_3$ to IP$_4$ happens very quickly. The variations of these new parameters are confined within a two folds range. The details of the reactions are given in table S17. From Fig S21 we find that the $D_{KL}^{new}$ and $D_{KL}^{Old}$ are very similar, differing only in the second place of decimal.

**New reactions added to M3**
**Table S17**

| Reactions | $k_{forward}$) | $k_{back}$ (s$^{-1}$) | $k_{cat}$ |
|---|---|---|---|
| $Itk - PIP_3 + PLC\gamma \rightarrow Itk - PIP_3 + PLC\gamma^*$ | 5.0 μM$^{-1}$s$^{-1}$ | | |
| $IP_4 - Itk - PIP_3 + PLC\gamma \rightarrow IP_4 - Itk - PIP_3 + PLC\gamma^*$ | 5.0 μM$^{-1}$s$^{-1}$ | | |
| $PLC\gamma^* \rightarrow PLC\gamma$ | | 8.0 s$^{-1}$ | |
| $PLC\gamma^* + S \rightarrow IP_3 + PLC\gamma^*$ | | | $1.5 \times 10^{-4}$ μM$^{-1}$s$^{-1}$ |
| $IP_3 \rightarrow IP_4$ | | | 0.7 s$^{-1}$ |

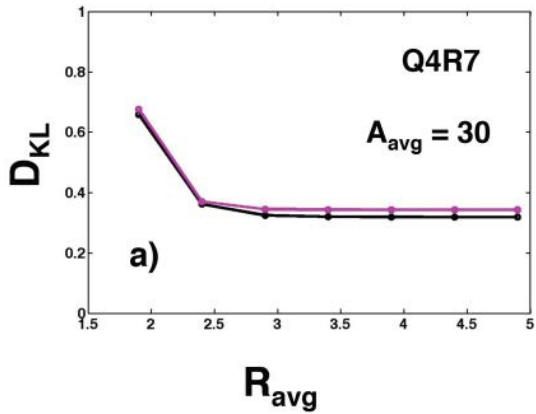
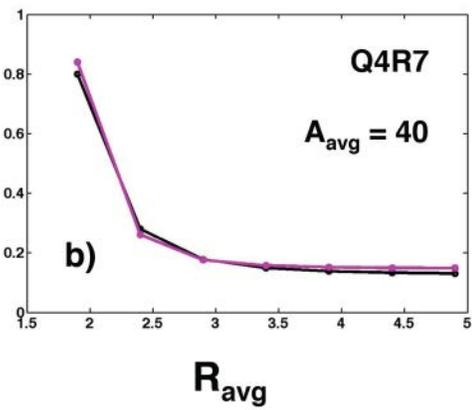

**Fig. S21: Addition of parameters which weakly affect the Itk-PIP$_3$ kinetics, do not lead to any significant difference in the D$_{KL}$:** For Itk$^0$ = 100 and PIP$_3^0$ = 370, **a)** we have looked at the relative difference in the D$_{KL}$ of our old M3 (black) and M3 with the added reactions (magenta) for an amplitude average of 30 molecules and peak time average of 2 mins. **b)** We have looked at the relative difference in the D$_{KL}$ of our old M3 (black) and M3 with the added reactions (magenta) for an amplitude average of 40 molecules and peak time average of 2 mins.

## Convergence of the D$_{KL}$ with the sample size

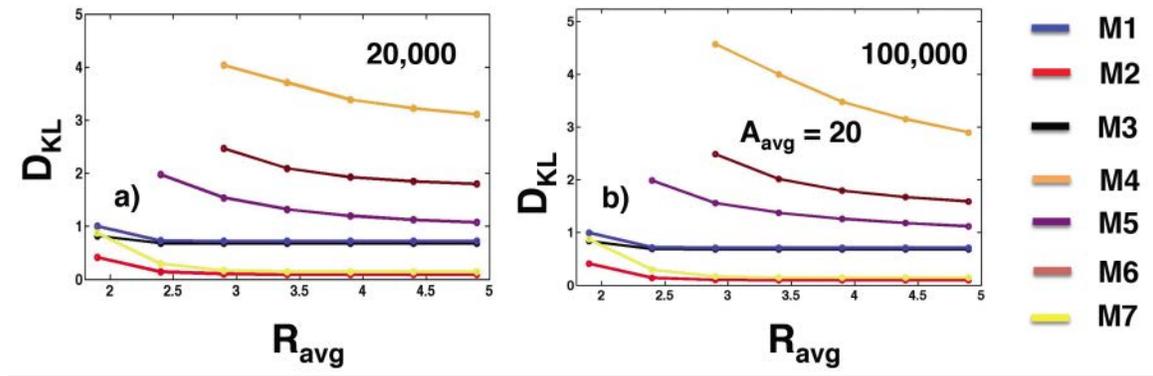

**Fig S22: The sample set of 100,000 is a good sample size:** We show the D$_{KL}$ of M1-M7 for Itk$^0$ = 100 and PIP$_3^0$ = 370 for a) 20,000 realizations and b) 100,000 realizations when the amplitude average is 20 molecules and the peak time average is 2 mins. The KL distances are identical.

## D$_{KL}$ without the constraint on amplitude

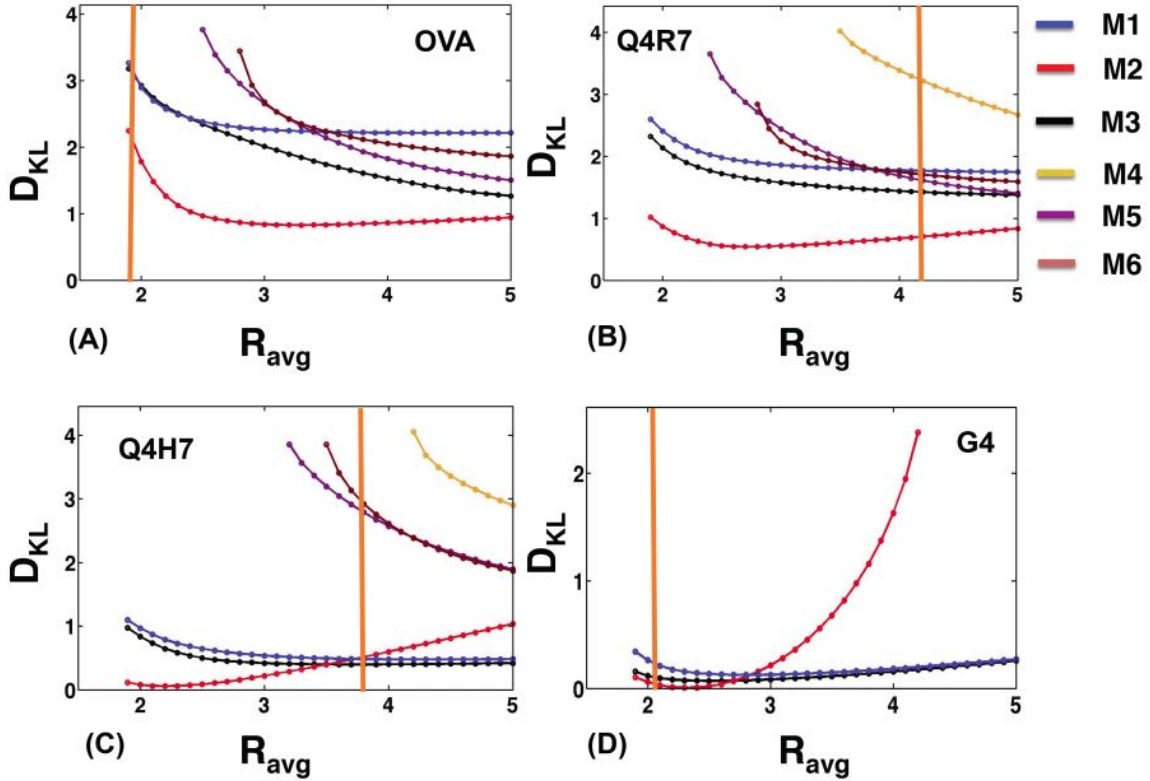

**Fig S23: Model robustness, quantified by the Kullback-Leibler distance ($D_{KL}$) as a function of the average asymmetry ratio $R_{avg}$.** Lower $D_{KL}$ values (shown in $\log_{10}$ scale) denote higher robustness for any given $R_{avg}$. Based on the data in Fig. 4, the average peak time was fixed at 2 mins in all cases. Experimentally measured $R_{avg}$ values are indicated by vertical orange lines. (A) Robustness for models M1-M3 and M5-M6 at high initial Itk ($Itk_0$=140 molecules) and $PIP_3$ concentrations ($PIP_{30}$=530 molecules), simulating high-affinity OVA stimulation. M2 appears most robust in the experimentally observed $R_{ave}$ range. M4 fails produce any R value in the range investigated here. (B) M2 shows maximal robustness for moderate concentrations of initial Itk (=100 molecules) and $PIP_3$ (=370 molecules), simulating Q4R7 stimulation. (C) For lower values of $Itk_0$ (=40 molecules) and $PIP_{30}$ (=130 molecules), simulating Q4H7 stimulation, M1-M3 are most robust with similar $D_{KL}$ values in the experimentally observed $R_{ave}$ range. (D) For low initial concentrations of Itk ($Itk_0$=20 molecules) and $PIP_3$ ($PIP_{30}$=50 molecules), simulating stimulation by the low affinity peptide G4, M1-M3 are again most robust inthe experimentally observed $R_{avg}$ range. Models M4-M6 fail to produce any value of R in the range investigated here. Model M7 is not shown.